\documentclass[12pt, review]{elsarticle}

\pdfoutput=1

\usepackage{lineno,hyperref}
\usepackage{amsmath, amsthm, amssymb, calrsfs, wasysym, verbatim, bbm, color, graphics, geometry, setspace, caption, float,siunitx}
\usepackage{pdfpages}

\newfloat{Video}{H}{lop}
\floatname{Video}{Video}

\journal{Acta Materialia}

\begin{document}

\begin{frontmatter}

\title{Pseudo\mbox{-}4D view of the growth and form of locked eutectic colonies}


\author[ummseaffil]{Paul Chao}
\ead{pchao@umich.edu}

\author[ummseaffil]{George R. Lindemann}
\ead{lindemge@umich.edu}

\author[mc2affil]{\textcolor{black}{Allen H. Hunter}}
\ead{ahhunter@umich.edu}

\author[ummseaffil,umcheaffil]{Ashwin J. Shahani\corref{mycorrespondingauthor}}
\cortext[mycorrespondingauthor]{Corresponding author. Tel.: +1 (734) 764-5648}
\ead{shahani@umich.edu}

\address[ummseaffil]{Department of Materials Science and Engineering, University of Michigan, 2300 Hayward St., Ann Arbor, MI 48109, USA}
\address[mc2affil]{\textcolor{black}{Michigan Center for Materials Characterization, University of Michigan, 2800 Plymouth Rd., Ann Arbor, MI 48109, USA}}
\address[umcheaffil]{Department of Chemical Engineering, University of Michigan, 2300 Hayward St., Ann Arbor, MI 48109, USA}

\begin{abstract}

We investigate solidification of an Al\mbox{-}Cu alloy as a model system to understand the emergence of patterns (such as lamellar, rod and maze\mbox{-}like) within eutectic colonies. To uncover the morphological transitions \emph{in~situ} and in 3D, we introduce here a new synchrotron\mbox{-}based procedure termed pseudo-4D X-ray imaging. Our method simultaneously maximizes the temporal (200\mbox{ }ms) and spatial resolution (0.69\textsuperscript{2} \SI{} {\micro\meter}\textsuperscript{2}/pixel) over that of traditional imaging approaches. The wealth of information obtained from this procedure enables us to visualize the development of a crystallographically `locked' eutectic microstructure in the presence of thermosolutal convection. This data provides direct insight into the mechanism of the lamella\mbox{-}to\mbox{-}rod transition as the eutectic accommodates fluctuations in interfacial composition and growth velocity. We \textcolor{black}{offer evidence to show} that this transition is \textcolor{black}{diffusive. It is} brought about by impurity\mbox{-}driven forces acting on the solid\mbox{-}solid\mbox{-}liquid trijunction that must overcome the stiffness of the solid\mbox{-}solid interfaces. Our pseudo\mbox{-}4D imaging strategy holds broad appeal to the solidification science community, as it can overcome the space\mbox{-}time trade\mbox{-}off in conventional \emph{in~situ} X\mbox{-}ray microtomography. 
\end{abstract}


\end{frontmatter}


\section{Introduction}\label{introduction}

Eutectics exhibit outstanding mechanical and electrical properties because their polyphase microstructures act as natural or \textit{in~situ} composite materials \cite{tiwary2022five,glicksman_principles_2011,aikin_mechanical_1997}. The microstructures of eutectics oftentimes show a structural hierarchy in the nanometer to micrometer regime, which arises during solidification. As the eutectic grows into the undercooled liquid, the interdiffusion between the solid phases at the duplex growth front leads to spacing selection \cite{Jackson_Hunt_1966, kurz_fundamentals_1992}. Based on a stability analysis, Datye and Langer \cite{datye_stability_1981} showed that the eutectic will self\mbox{-}organize into a variety of steady\mbox{-}state, tilted, and oscillating structures. These solid phases may, in turn, be arranged in cells or colonies, a type of long wavelength instability. The length\mbox{-}scale of the cells is typically 10\mbox{-}100{\texttimes} that of the lamellar spacing \cite{plapp_eutectic_1999}.

Cells form due to the presence of a third chemical component (so\mbox{-}called impurity), which induces a Mullins\mbox{-}Sekerka\mbox{-}type instability at the growth front \cite{mullins_stability_1964,caroli_mullinssekerka_1986}. Provided the partition coefficient is below unity, the rejected impurity species will pile up at the eutectic\mbox{-}liquid interface. Above a critical ratio of $G/V$ where $G$ is the thermal gradient and $V$ is the growth velocity, the constitutional undercooling will render the interface unstable. In this scenario, the interface will form a protrusion leading to the rejection of impurity laterally and accumulation of impurity around the protrusion \cite{plapp_eutectic_1999, tiller_redistribution_1953,chadwick_eutectic_1963,trivedi_effect_2002,plapp_eutectic_2002}. Technologically, eutectic solidification usually occurs in the presence of trace elements or impurities \cite{hecht_multiphase_2004,sargin_postsolidification_2016, steinmetz_crystal_2018}; thus, solidification along a univariant pathway warrants further consideration. 

Based on observations of solidification in organic eutectics, Trivedi has developed a morphological taxonomy of this long\mbox{-}wavelength instability \cite{trivedi_nonplane_2008}. Two\mbox{-}phase cellular or dendritic microstructures are common microstructures observed during non\mbox{-}planar solidification, including that of eutectics. These morphologies can begin as shallow, low\mbox{-}amplitude cells, \textit{i.e.}, ``fan" or ``needle"\mbox{-}shaped colonies, at growth rates just above the critical velocity. The former shows a fanning of lamellae at the eutectic\mbox{-}liquid interfaces whereas the needle colonies are aligned in a specific direction even if there is a macroscopic eutectic\mbox{-}liquid interfacial curvature. At higher velocities, deep cells termed cellular dendrites can form \cite{han_stability_1995}. Such two\mbox{-}phase dendrites assume a nearly parabolic shape, which is similar to that of a single\mbox{-}phase dendrite tip, hence the name \cite{trivedi_nonplane_2008}. Somewhat akin to single\mbox{-}phase dendrites, the lamellae in eutectic dendrites may be arranged in primary and secondary branches, which are not collinear. These microstructures can be further understood as eutectic grains, where each grain is composed of crystals interlocked in a unique crystallographic orientation \cite{caroli_lamellar_1992,akamatsu_formation_2001}. The interphase anisotropy can influence colony microstructure such that the eutectic lamellae within a particular colony can become ``locked" on a preferred crystallographic orientation and thus tilted with respect to the thermal gradient \cite{akamatsu_lamellar_2012, ghosh_interphase_2015, ghosh_influence_2017, bottinrousseau_special_2018}.

The 3D structure of eutectic cells has been debated since the 1960s \cite{Jackson_Hunt_1966,chadwick_modification_1962,hunt_lamella_1966} to present day \cite{lahiri_eutectic_2017, aramanda_exotic_2020, aramanda_exotic_2021}. A confounding issue is that investigators lacked suitable (\textit{i.e.}, 3D space\mbox{-} and time\mbox{-}resolved) measurements on the growth dynamics of eutectic colonies, as will be elaborated on below. Generally, past studies report a lamellar or rod\mbox{-}like microstructure or a coexistence of these two morphologies within eutectic colonies. Perhaps the earliest proposal is that one morphology is favored based strictly on the volume fraction of the minor phase \cite{Jackson_Hunt_1966}. That is, minimization of the interphase boundary energy (\textit{i.e.}, between solid phases) leads to a critical volume fraction below which rods are expected. This model utilizes the competitive growth principle, wherein selection of morphology is determined by which has the higher interface temperature under steady\mbox{-}state conditions. Recent extensions to this approach account also for the possibility of a coexistence between rods and lamellae \cite{ liu_dynamic_2011} but do not predict \textit{where} in the eutectic microstructure rods or lamellae may be found nor \textit{how} the rods may transition into lamellae and \textit{vice versa}. According to Chadwick \cite{chadwick_modification_1962}, rods should be located preferentially at the edges of the eutectic colonies (\textbf{Fig.~\ref{F1}(a)}). From earlier, during growth, the cells in the eutectic\mbox{-}liquid interface reject impurity into the melt laterally and thus one would envision a higher concentration of impurity at the colony edges than center. Assuming the partition coefficients of the impurity in the solid phases are different (and below unity), the equilibrium temperatures of the two phases will be different and hence one phase \textcolor{black}{($\alpha$)} will grow ahead of the other \textcolor{black}{($\beta$)}. Let us assume also that the imposed $G/V$ is such that the depressed \textcolor{black}{$\beta$} phase is locally constitutionally supercooled. If, by chance, \textcolor{black}{a rod\mbox{-}like protrusion forms at the $\beta$\mbox{-}liquid interface}, \textcolor{black}{then this protrusion will be stabilized just like a cellular structure is stabilized~\cite{chadwick_modification_1962}.} \textcolor{black}{The protrusion} will reject solute radially, which then incorporates into the \textcolor{black}{$\alpha$} phase. \textcolor{black}{That is, the protrusion of $\beta$ will be surrounded by a matrix of $\alpha$.} Hunt \cite{hunt_lamella_1966} questioned this proposal, arguing that rods may not necessarily form at the edges of the colony. Instead, rods \textcolor{black}{may be} produced any time lamellae are forced to accommodate the curvature of the eutectic\mbox{-}liquid interfaces in a eutectic cell (\textbf{Fig.\mbox{ }\ref{F1}(b)}). That is, a lamellar structure is formed when the eutectic grows with a low interphase energy between the lamellae, and rods are produced when the local growth direction (indicated by the arrows in \textbf{Fig.\mbox{ }\ref{F1}(b)}) is furthest from this low energy, lamellar plane. The curvature of the cell causes rod formation away from the lamellae by Cahn's rule \cite{datye_stability_1981, plapp_eutectic_2002}. In this view, the area fraction of rods depends on the interfacial curvature and the volume ratio of the two eutectic phases. 

In order to make sense of these mechanisms, we rely on experiments that capture the evolution of microstructure in eutectic colonies. To this end, full\mbox{-}field synchrotron\mbox{-}based X\mbox{-}ray imaging has opened a paradigm shift in solidification science, allowing us to visualize transient microstructural dynamics in optically opaque materials \cite{mathiesen_timeresolved_2002,yasuda_direct_2004,nguyenthi_interest_2012,clarke_xray_2015,shahani_characterization_2020}. The traditional imaging strategy for dynamic \textit{in~situ} 4D (\textit{i.e.}, 3D space plus time) computed tomography (CT) experiments is to collect a series of forward transmitted X\mbox{-}ray projections whilst the sample is continuously rotated~\cite{patterson_data_2018}; a sequence of projections from a 180{\textdegree} rotation is used to reconstruct successive volumes that show the microstructure formation in 4D \cite{maire_quantitative_2014,withers_xray_2021}. In such experiments, there is a fundamental trade\mbox{-}off between the spatial and temporal resolutions. That is, improving temporal resolution by reducing the number of projections or the exposure time of the camera leads to severe aliasing artifacts in the 3D reconstructions. Until now, the solidification dynamics of a regular eutectic (on the scale of the lamellar spacing) has been difficult to probe with 4D CT. This is because of the rapid growth velocity (on the order of tens of \SI{}{\micro\meter}/s) and vanishingly small crystal dimensions (on the order of single \SI{}{\micro\meter}). Regarding the former, the eutectic\mbox{-}liquid interfaces would have advanced over \SI{600}{\micro\meter} within a time interval of 1~min, assuming a modest undercooling of 0.3~{\textdegree}C for an Al\mbox{-}Al\textsubscript{2}Cu eutectic. This growth velocity is prohibitively fast for any meaningful analysis of the solidification dynamics \textit{via} 4D CT in the conventional approach. Here, we implement a novel imaging strategy termed ``pseudo\mbox{-}4D imaging" that circumvents the above challenges. We achieve the highest reported temporal and spatial resolutions for 4D CT, see \textbf{Fig.~\ref{F2}}. We do this by digitally fusing data from two synchrotron\mbox{-}based experiments conducted in series, namely, X\mbox{-}radiography and X\mbox{-}ray tomography. \textit{In~situ} X\mbox{-}radiography (videomicroscopy) reveals the evolution of solid\mbox{-}liquid interfaces. Following solidification, \textit{ex~situ} X\mbox{-}ray tomography provides insight on the solid\mbox{-}solid interfaces in 3D.

In this paper, we invoke our new imaging procedure to gain insight on the detailed dynamics of pattern formation in metallic eutectic colonies. As alluded to above, our studies are focused on a eutectic between $\alpha$\mbox{-}Al and $\theta$\mbox{-}Al$_2$Cu, for which there is a test bed of information in the literature including the scaling relations between velocity, undercooling, and lamellar spacing \cite{Jackson_Hunt_1966, ourdjini_eutectic_1994, cadirli_directional_1999, walker_eutectic_2007, murphy_combined_2013, luo_research_2019, kakitani_thermal_2019}; the orientation relations, defined by a pair of parallel crystallographic planes and parallel directions within those planes \cite{bottinrousseau_special_2018,sutton_interfaces_1995,howe_interfaces_1997}; and the crystallographic anisotropy of the solid\mbox{-}solid interfaces \cite{kraft_crystallography_1962, bonnet_geometric_1973, cantor_growth_1974, hecht_coupled_2012, kokotin_molecular_2014, wang_characteristic_2018, hecht_phase_2019}. We show the dynamic microstructural adjustment mechanisms within Al\mbox{-}Al\textsubscript{2}Cu eutectic colonies under the influence of thermosolutal convection. Our real\mbox{-}time observations paint a clear picture on the mechanism of the lamella\mbox{-}to\mbox{-}rod transition in cells, as the eutectic accommodates changes in interfacial composition and growth velocity. With this information, we can test the predictions of existing theories. We observe a coplanar transition from a single lamella to multiple rods that is unlike previous models by Chadwick \cite{chadwick_eutectic_1963}, Hunt \cite{hunt_lamella_1966}, and others \cite{Jackson_Hunt_1966,tiller_redistribution_1953,liu_dynamic_2011}. The transition reflects a competition between two opposing factors, an impurity\mbox{-}induced motion of the solid\mbox{-}solid\mbox{-}liquid trijunctions and an immobility of particular solid\mbox{-}solid interfaces. Our results reveal also an additional orientation relationship (OR) and a particular solid\mbox{-}solid interface within this OR that remains ``locked" in eutectic solidification. We justify the existence of the latter by drawing upon Kraft's ``puckered" interface density description, which allows us to determine a minimum misfit between the two phases \cite{kraft_crystallography_1962}. Our results provide the benchmark data to validate simulations (\textit{e.g.}, phase field) of microstructure evolution along a univariant solidification pathway in the face of thermosolutal convection and an anisotropy in solid\mbox{-}solid interface energy. Both factors are the norm and not the exception in solidification of technologically relevant eutectics.

\section{Methods} \label{sec_methods}
For the subsequent experiments, we used an Al\mbox{-}33wt\%Cu alloy, corresponding to the eutectic composition in the Al\mbox{-}Cu system. The sample was cast at the Materials Preparation Center at Ames Laboratory (Ames, IA, USA) with high purity Al (99.99\%) and Cu (99.99\%). We confirmed the composition of the \textcolor{black}{cast ingot} \textit{via} inductively coupled plasma mass spectrometry (ICP\mbox{-}MS), see \textbf{Table~\ref{tableSI}}. \textcolor{black}{Further analysis \textit{via} atom probe tomography (APT) \textcolor{black}{on the sample solidified at the beamline} revealed the presence of trace B, see Sec.~\ref{sec_origin}.} We prepared the sample in the shape of a \SI{30}{\micro\meter} thin foil by mechanically sectioning and polishing with 1200 grit SiC metallurgical paper. The foils were \textcolor{black}{ultrasonically cleaned in isopropanol},  sandwiched between two \SI{0.2}{\milli\meter} quartz slides, and \textcolor{black}{then} secured with boron nitride paste following the approach used in Refs.~\cite{liotti_crystal_2018, wang_insitu_2021, wang_situ_2022}, see \textbf{Fig.~\ref{SI_1}(a-b)} for additional details.

We conducted our experiments at sector 2\mbox{-}BM of Argonne National Laboratory's Advanced Photon Source (Lemont, IL, USA). The synchrotron beamline is equipped with a resistive furnace, described elsewhere \cite{wang_insitu_2021, wang_situ_2022,han2019side, moniri_mechanism_2019, moniri_singlytwinned_2020, reese_realtime_2021}. We calibrated the temperature at the sample position with a thermocouple prior to the experiment. We then raised the furnace temperature to 560~{\textdegree}C (above the eutectic temperature of 548~{\textdegree}C) and held it there for 5~min. before cooling at a rate of 0.5~{\textdegree}C/min.~for a total scan duration of 30~min. 

As mentioned in Sec.~\ref{introduction}, we first conducted an \textit{in~situ} X\mbox{-}radiography scan (at temperature) to retrieve the position of the eutectic\mbox{-}liquid interface as solidification progresses across the field\mbox{-}of\mbox{-}view (FOV). Then, we performed an \textit{ex~situ} CT scan to resolve the solid\mbox{-}solid interfaces in 3D. Refs.~\cite{deville_situ_2009_1,deville_situ_2009_2} have used these techniques in series to estimate interfacial velocities and determine bulk microstructure zones in ceramic freezing experiments; the distinguishing feature in our work is that we have fused the datasets to obtain a cohesive picture of the solidification process (\textbf{Fig.~\ref{F2}}). A monochromatic X\mbox{-}ray beam operating at 20\mbox{ }keV was focused onto our sample and X\mbox{-}rays were converted to visible light using a \SI{25}{\micro\meter} thick LuAG:Ce scintillator. These digital images were collected using a FLIR Oryx CCD with a 5{\texttimes} magnifying objective lens, yielding pixel sizes of 0.69\textsuperscript{2} \SI{}{\micro\meter}\textsuperscript{2} within a FOV of 1689\texttimes \SI{1413} {\micro\meter}\textsuperscript{2}. X\mbox{-}radiographs (or projection images) were acquired for approximately 30~min.~with an exposure of 200 ms, see \textbf{Video~\ref{V_S1}} for the time interval spanning eutectic growth. The collected images were processed in MATLAB \cite{MATLAB} to enhance the attenuation contrast between the solid and liquid phases, and hence delineate the eutectic\mbox{-}liquid interface. More specifically, we first normalized each image (corresponding to a distinct timestep) to remove detector artifacts, \textcolor{black}{also referred to as flat\mbox{-}field correction \cite{buffet2010measurement,soltani2020impact}}. This was accomplished by using a pixel\mbox{-}by\mbox{-}pixel division operation between two successive images, namely, a frame of interest and a background frame \cite{wang_insitu_2021, wang_situ_2022,tandjaoui_twinning_2013}. Then, we limited the range of the pixel intensity values to four standard deviations from the mean to adjust for random dead pixels on the detector. Direct segmentation of these simply processed radiographs was nontrivial due to random noise and minimal contrast between phases. After some trial\mbox{-}and\mbox{-}error, we found success by performing two additional operations: we applied a non\mbox{-}local means filter \cite{MATLAB} to the normalized frame to remove speckle noise while preserving edges; furthermore, we summed 100 normalized and filtered images to yield a single image (a technique commonly employed in visible light photography to reduce noise in astroimages). Stated differently, each pixel in the resulting, stacked image can be described as the sum of the pixels in the same spatial location within each image in the set of 100 images. This procedure improved image quality significantly (refer to \textbf{Video~\ref{V_S2}}). We then utilized Zeiss Zen Blue 3.1 with the Intellesis deep learning module \cite{andrew_usage_2017,volkenandt_machine_2018} to perform semantic segmentation, a pixel\mbox{-}based machine learning method, to identify the eutectic\mbox{-}liquid interface in the processed images. For this purpose, we supplied one dozen carefully and manually segmented images as training data. After training, the model was capable of tracing the eutectic\mbox{-}liquid interface, albeit with some residual noise (\textit{i.e.}, the interface was jagged and it also overlapped with interfaces in neighboring frames, two features that are unphysical). We overcame this lack of definition by applying an average filter along both spatial axes of all eutectic\mbox{-}liquid interfaces in combination to create smooth, non\mbox{-}overlapping interfaces at each timestep. \textbf{Fig.~\ref{SI_1}}(c-e) summarizes the data processing steps.

For the same foil sample, following eutectic solidification, we acquired an \textit{ex~situ} X\mbox{-}ray tomogram. We imaged the same FOV as in the above X\mbox{-}radiography experiment over a full 360\textdegree~rotation, with a pixel size of 0.69\textsuperscript{2} \SI{} {\micro\meter}\textsuperscript{2}, a total of 3000 projections, and an exposure of 200~ms. We used TomoPy \cite{gursoy_tomopy_2014}, a Python-based open source framework for the analysis of tomographic data, to reconstruct the volume. Although it may appear to be redundant, collecting images over a span of 360\textdegree~provided clarity on the as\mbox{-}solidified microstructure: the pairs of mirror images doubled the supplied projections for reconstruction and hence reduced detector noise. Within TomoPy, we first normalized the X\mbox{-}ray projections by the dark\mbox{-}field and white\mbox{-}field images. Normalization alone was not sufficient to correct for ``ring"\mbox{-}shaped artifacts, which are typically caused by a combination of dead pixels in the CCD as well as beam instabilities. Such artifacts were corrected here \textit{via} a combination of algorithms in Ref.~\cite{vo_superior_2018}. After normalization and artifact removal, the data were reconstructed \textit{via} the Gridrec algorithm \cite{dowd_developments_1999}, which is a direct Fourier\mbox{-}based method. Further details can be found in Ref.~\cite{gursoy_tomopy_2014} and the references therein.

We digitally fused the two results to obtain a pseudo\mbox{-}4D view of the solidification process, see \textbf{Fig.~\ref{F3}} for a sketch. We interpolated the spatial location of the eutectic\mbox{-}liquid interface onto the tomographic data, assuming there is no tilt of the interface in the direction of the X\mbox{-}ray beam (discussed later in Sec.~\ref{sec_outlook}). \textbf{Video~\ref{V_S3}} shows the resulting transverse cross\mbox{-}sectional views. To our advantage, the two sample frames of reference are nearly equivalent such that this mapping is straightforward. We confirmed equivalence of the FOV between the two experiments by registering the data \cite{MATLAB} and determining a \textless \SI{2}{\micro\meter} translation in the plane of the detector. Our procedure is a significant improvement over traditional 4D X\mbox{-}ray tomography (\textbf{Fig.~\ref{F2}}). In principle, the temporal resolution in pseudo\mbox{-}4D imaging is limited only by the frame rate of the camera. In addition, our imaging strategy eliminates the influence of convection caused by a centripetal acceleration of the sample during continuous rotation.

After the synchrotron experiment, we sought to analyze the same sample further \textit{via} electron microscopy at our home institution, with a focus on the crystallographic orientation of the eutectic solid\mbox{-}solid interfaces. We removed the sample from the quartz slides and mechanically polished it with 1200 grit SiC metallurgical paper; next, we ion polished with Xe plasma to achieve a smooth surface for electron backscatter diffraction (EBSD). We collected Kikuchi patterns from a 196{\texttimes}\SI{196} {\micro\meter}\textsuperscript{2} region of the as\mbox{-}solidified microstructure, using a step size of \SI{0.5} {\micro\meter} in a hexagonal grid. To align the EBSD frame of reference with the X\mbox{-}ray frame of reference, we used an affine transformation such that the lamellae in both sets of images are oriented similarly, see \textbf{Fig.~\ref{SI_2}}. 

By combining the X\mbox{-}ray and EBSD data, we can transform orientational data from the specimen ($s$) frame to the crystallographic ($c$) frame. Denoting the crystal symmetry operators as $T$ (obtained from Refs.~\cite{kocks_texture_1998, engler_introduction_2009,bachmann_texture_2010}), and the orientation matrix of a given phase as $g$ (determined \textit{via} EBSD), we can rotate the normal vectors $\hat{n}$ of the solid\mbox{-}solid interfaces in $s$ (calculated from the tomographic data) to crystal directions in $c$ according to $\hat{n}_c\,=\,T\, g\, \hat{n}_s$ \cite{rohrer_distribution_2004,rowenhorst_3d_2006}. This analysis is critical to deduce the crystal orientation of the solid\mbox{-}solid interfaces. That said, it is only applicable for a single crystal, in which case $g$ takes on a unique form. We will later prove this to be true in Section \ref{sec_crystal}, \textit{i.e.}, the Al\mbox{-}Al\textsubscript{2}Cu eutectic consists of two interpenetrating single crystals.

\textcolor{black}{We further deduced the interfacial chemistry of the Al\mbox{-}Al\textsubscript{2}Cu phase boundary with the aid of atom probe tomography (APT) \cite{larson2013local}.  The APT tips were prepared by focused ion beam (FIB) milling using a Helios G4 Plasma FIB UXe (ThermoFisher Scientific, Hillsboro, OR). The region of interest \textcolor{black}{was selected from the foil after solidification at the beamline to be near an area of larger average lamellar spacing (corresponding to lower growth velocity as discussed later in Sec.~\ref{sec_origin}); it} was first protected from FIB damage by capping the surface with several hundred nm thick Pt deposited using the electron beam at 2~kV accelerating voltage. A thicker layer of W was deposited on top of the Pt layer using the Xe ion beam at 12~kV accelerating voltage. FIB cuts were made at $\pm$30$^{\circ}$~normal to the surface using 30 kV Xe ions to produce a triangle shaped bar. The bar was extracted from the sample surface using a W manipulator needle and transferred to Mo posts.  The Mo posts were cut into a standard 3~mm Mo TEM half\mbox{-}grid by FIB milling.  The pieces of the bar were attached to the tops of the Mo posts using W deposited at 30~kV accelerating voltage.  The posts were sharpened to points using a series of annular milling patterns at 30~kV accelerating voltage.  Final tip polishing was performed with a circular pattern at 5~kV accelerating voltage.  The atom probe tips were analyzed in a LEAP 5000X HR local electrode atom probe (LEAP) (Cameca Instruments Inc., Madison, WI).  The specimen tip was run in laser pulsed mode using a pulse energy of 100~pJ per pulse at 125~kHz pulse repetition rate.  The tip temperature was 25~K and the average detection rate for the experiment was 0.5\%~ions/pulse.  The analyzed volume was reconstructed using the IVAS module within APSuite 6.1.0.26 (Cameca Instruments Inc., Madison, WI).  The tip profile reconstruction method was used to reconstruct the dataset according to a high resolution SEM image of the sharpened tip acquired just prior to collecting the dataset. See \textbf{Fig.~\ref{SI_3}} for additional details.} 

\section{Results and discussion}

\subsection{Analysis of pseudo\mbox{-}4D data}\label{analysis}

By examining the X\mbox{-}radiography data, we see the eutectic\mbox{-}liquid interface exhibits an oscillatory velocity. See \textbf{Fig.~\ref{F4}(a)}, \textbf{Fig.~\ref{F5}(a)}, and \textbf{Video~\ref{V_S1}}. A wealth of information can be quantified from this image stream. We begin with the velocity $V$ of the eutectic\mbox{-}liquid interface as it propagates along the positive $y$ direction (\textit{i.e.}, antiparallel to gravity). We calculated $V$ by fitting a second degree polynomial to each eutectic\mbox{-}liquid interface, and taking the derivative of the interfacial displacement in the $y$ direction. The average velocity of the eutectic\mbox{-}liquid interface oscillates between 3 to 9 \SI{}{\micro\meter/\second} for upwards of 4 periods within the FOV. The uncertainty in \textbf{Fig.~\ref{F5}(a)} represents the standard deviation of the calculated velocity. We also quantified the maximum curvature $\kappa$ within the projected eutectic\mbox{-}liquid interface. We compute $\kappa$ in a somewhat similar manner to $V$, by fitting polygons to the points on the line, and then calculating the analytical curvature from the polygons \cite{mjaavatten_curvature}. The maximum curvature for a given time\mbox{-}step varies from 0.04 to 0.08 \SI{}{\micro\meter}\textsuperscript{$-$1} (absolute values, see \textbf{Fig.\mbox{ }\ref{F5}(b)}). We take the error range as the 10 to 90 percentile. Finally, we determined the thermal gradient along the $y$ direction, by taking into account the cooling rate and the amount of time for the growth front to propagate vertically across the FOV, see also Refs.~\cite{mathiesen_xray_2011,reinhart_impact_2020}. Ultimately, we find the thermal gradient is on the order of 1\mbox{ }K/mm. 

As noted in Sec.~\ref{sec_methods}, the spatial location of growth front can be mapped onto the tomographic volume to capture the development of two\mbox{-}phase microstructure within the eutectic\mbox{-}liquid interface (\textbf{Fig.~\ref{F4}(b)}). We now have two complementary perspectives of the solidification process, one from a side view (corresponding to the detector plane) and the other from a bird’s eye view (corresponding to the transverse cross\mbox{-}section). This provides access into a number of key microstructural attributes by way of the pseudo\mbox{-}4D reconstruction, beyond the curvatures and velocities of the eutectic\mbox{-}liquid interfaces. They include the lamellar spacing, area fraction of phases within the eutectic\mbox{-}liquid interface, and area fraction of Al rods. We determined the time\mbox{-}dependant average eutectic spacing (\textbf{Fig.\mbox{ }\ref{F5}(c)}) \textcolor{black}{from the image autocorrelation} on these transverse cross\mbox{-}sections~\cite{kalidindi_3_2015, cecen_new_2018, aramanda_exotic_2020}, see \textbf{Fig.~\ref{SI_4}}. Peaks in the \textcolor{black}{image autocorrelation are diffuse and provide real\mbox{-}space information on periodicity of microstructure. We derived so\mbox{-}called ``primary" and ``secondary" eutectic spacings as the next\mbox{-}nearest peak from the center along two corresponding lamellae directions and approximate the error as \SI{0.5} {\micro\meter}.} We determined also the area fraction of the eutectic phases after segmentation (\textbf{Fig.~\ref{F5}(d)}). We selected an adaptive threshold \cite{MATLAB} with sensitivity of 0.57, which yielded a reasonable segmentation upon visual inspection and also the expected phase fractions for this alloy (47\% Al$_2$Cu, by volume) \cite{walker_eutectic_2007}. Finally, we found the area fraction of Al rods (\textbf{Fig.~\ref{F5}(e)}) by applying a morphological criteria based on eccentricity and solidity \cite{MATLAB} of connected components in the segmented, transverse cross\mbox{-}section. We selected a threshold of 0.9 for eccentricity (note line segments have an eccentricity of 1 \textcolor{black}{while circles have an eccentricity of 0}) and a threshold of 0.8 for solidity (convex shapes like circles have a solidity of 1). A connected component on the eutectic\mbox{-}liquid interface must have eccentricity less than and solidity greater than these values to classify as a rod. The reported fractional value represents the area fraction of Al that satisfies the criteria. \textcolor{black}{We quantified the variation in area fraction and area fraction of Al rods across the transverse cross\mbox{-}sections using a series of overlapping windowed subsections: each window is \SI{69} {\micro\meter} wide with an incremental step size of \SI{6.9} {\micro\meter} along $x$ for a total of 180 windows.} \textcolor{black}{Despite the fact that the two eutectic phases have nearly equal area fractions (Fig.~\ref{F5}(d)), there is still a significant fraction of rods. This result is consistent with phase field simulations that depict a family of periodic and disordered configurations ``in between" rods and lamellae \cite{parisi_defects_2010}.} We report in \textbf{Fig.\mbox{ }\ref{F5}(d\mbox{-}e)} the maximum and minimum average values of the windowed subsections as the corresponding error (grey bands). All results shown in \textbf{Fig.\mbox{ }\ref{F5}} have been smoothed with windowed polynomial fits for clarity. 

The velocity evolves out\mbox{-}of\mbox{-}phase with other measured attributes. That is, an increase in eutectic\mbox{-}liquid velocity corresponds to a decrease in the maximum eutectic\mbox{-}liquid curvature, average lamellar spacing, area fraction of Al\textsubscript{2}Cu, and area fraction of Al rods. We observe the converse for decreasing velocity. We next discuss the underlying source of these trends.

\subsection{Origin of cells and their oscillatory velocity} \label{sec_origin}

\textcolor{black}{We begin by analyzing the APT dataset, which consisted of approximately 10 million ions and contained an Al\mbox{-}Al\textsubscript{2}Cu interface.  The atoms detected within the reconstruction were Al, Cu, O, N, B, Xe.  H was detected in the reconstructed volume but will not be considered in this analysis due to the uncertainty of the true H concentration in the sample compared to background levels in the UHV vacuum chamber.  The Xe ions likely originated from implantation during FIB milling of the tip and are also not considered for futher analysis. The interface separating the Al from the Al\textsubscript{2}Cu phase was identified using a 10.0 at\% Cu iso\mbox{-}concentration surface.  The reconstructed view of the interface is shown in  \textbf{Fig. \ref{F13}(a)}.  The composition profile across the interface was measured using a proximity histogram \cite{hellman2000analysis} based on the Cu iso\mbox{-}concentration surface.  A step size of 0.01 nm was used to generate the profiles which were then smoothed using a 0.5 nm moving average function. The resulting composition profiles for Al, Cu, and B are shown in  \textbf{Fig. \ref{F13}(b)}.  The compositions profiles of the Al and Al\textsubscript{2}Cu are not flat. Within the Al phase, the Al content becomes enriched closer to the interface and the Cu is correspondingly depleted.  Within the Al\textsubscript{2}Cu phase, the Al profile shows a depletion near the interface and a corresponding increase in Cu concentration.  These profiles suggest some solid\mbox{-}state growth of the Al\textsubscript{2}Cu phase following eutectic solidification. The exact shape of the composition profile is likely due to a combination of diffusion\mbox{-}controlled growth and varying equilibrium phase fractions upon cooling, see \textbf{Fig.~\ref{SI_5}}. Examination of the B concentration shows a peak located approximately 10~nm from the Al/Al\textsubscript{2}Cu interface with almost zero B detected on the Al side of the interface.  This can be explained by the B being segregated at the Al\mbox{-}Al\textsubscript{2}Cu interface during eutectic solidification and then becoming trapped while the Al\textsubscript{2}Cu phase boundary shifted during diffusional growth.  N and O were also measured in the composition profile and both N and O were observed to partition slightly to the Al\textsubscript{2}Cu phase, however interfacial segregation was not apparent.}

The observed eutectic cells \textcolor{black}{in \textbf{Figs.~\ref{F3}-\ref{F4}}} are evidence of the morphological instability induced by the constitutional undercooling of an impurity species, \textcolor{black}{which we determine as B from the above analysis of the interfacial chemistry. The suspected B likely originated from the BN paste used to secure the foil and quartz assembly\textcolor{black}{, where the BN and molten Al will react to liberate B into the melt} \cite{fujii1993interfacial}. \textcolor{black}{At 548~\textdegree C, it is expected that B would diffuse quickly through the thin, molten sample (the characteristic diffusion time is only 5~s in a sample of thickness \SI{30} {\micro\meter}), relative to the 30~min.~duration of the experiment, following a similar analysis done in Ref.~\cite{reese_realtime_2021}. }As a cross\mbox{-}check, we solidified Al\mbox{-}Al\textsubscript{2}Cu eutectics with and without the BN; the results in \textbf{Fig.~\ref{SI_6}} show dramatically different microstructures, which supports the idea of an underlying chemical effect.} Accordingly, a planar interface will break down into an array of cells when the eutectic\mbox{-}liquid velocity exceeds a critical velocity, $V_c$. The latter can be calculated using the following relationship \cite{tiller_redistribution_1953, hecht_multiphase_2004}, $V_c = G k_c D / m_l C_o (k_c -1) $. In our experiments, we take the values of thermal gradient ($G$), solute \textcolor{black}{(B)} partition coefficient ($k_c$), solute diffusivity in the liquid ($D$), slope of the liquidus line ($m_l$), and solute composition ($C_o$) to be \SI{0.00167} {\degree C/\micro\meter}, \textcolor{black}{0.08, \SI{4500} {\micro\meter ^2/s} \cite{du_diffusion_2003}, \SI{-13.6} {~\degree C/at\%}, and 10$^{-1}$\mbox{-}10$^{-2}$\mbox{ }at\%, respectively.\footnote{The quantities $k_c$, $m_l$, and $D$ are not known for the ternary Al\mbox{-}Al\textsubscript{2}Cu\mbox{-}B system. Thus, in a manner similar to Ref.~\cite{akamatsu_traveling_2000}, we assume they are equal to the known quantities of the binary Al\mbox{-}B system.}} We find $V_c$ \textcolor{black}{lies in the range \SI{0.5}\mbox{-}\SI{5}  {\micro\meter/s}}, which is in line with \textcolor{black}{the order-of-magnitude of} our experimentally observed velocities (\textbf{Fig.\mbox{ }\ref{F5}(a)}). Technically speaking, the above equation applies for an established and steady impurity boundary layer, while in our case transient conditions prevail. Nevertheless, others \cite{bogno_situ_2011} have used the above equation to compute $V_c$ during an initial transient, without loss of generality. 

The observed variation of the eutectic\mbox{-}liquid interface velocity in time suggests an unsteady impurity concentration field ahead of the solidification front. \textcolor{black}{Note that fluid flow is otherwise negligible in \textit{binary} eutectic alloys since the solute boundary layer, on the order of the lamellar spacing, is smaller than the fluid boundary layer, as noted in Ref.~\cite{lee_effect_2005}. Instead, t}he instability arises due to thermosolutal convection of trace \textcolor{black}{B in univariant solidification}, and namely, cycles of \textcolor{black}{B} pile\mbox{-}up and convection away from the eutectic\mbox{-}liquid interfaces. The layer of rejected \textcolor{black}{B} at the eutectic\mbox{-}liquid interface (for $k_c<1$, see above) is unstable and subject to buoyancy forces (\textcolor{black}{B} is lighter than the Al\mbox{-}Cu melt). This buoyancy force will overcome the fluid inertia and result in thermosolutal convection of \textcolor{black}{B} upwards (\textit{i.e.}, antiparallel to gravity) thereby disturbing the impurity boundary layer. \textcolor{black}{In this configuration, the temperature gradient will have a stabilizing influence only when the thermal Rayleigh number (denoted $Ra$, with a length\mbox{-}scale dependence based on $D/V$) is much greater than 10, according to a theoretical treatment~\cite{mcfadden1984thermosolutal}. We calculate $Ra = (g \beta G/\nu \alpha)(D/V)^3 \approx 0.001$ where $g$ is the acceleration of gravity, $\beta$ is the thermal coefficient of expansion of the melt, $\nu$ is kinematic viscosity, and $\alpha$ is the thermal diffusivity of the melt. The values of the last three parameters are given in Ref.~\cite{carlberg_effect_1984} as \SI{1.4e-4}{1/\degree C}, \SI{8e5}{\micro\meter ^2/s}, and \SI{3e7}{\micro\meter ^2/s}, respectively. This analysis indicates that the thermal field plays little role in restraining fluid flow. Consequently, the} process of pile\mbox{-}up and ejection occurs with a characteristic frequency in time \textbf{(Fig.~\ref{F4})}. Ref.~\cite{de_wilde_solutal_2005} \textcolor{black}{also} suggested an unstable solute field when an Al\mbox{-}Al\textsubscript{2}Cu eutectic is solidified in an upwards direction and in the presence of trace Si, although they offer no microstructural data for this particular case. Among the alternative scenarios that we have considered, none can satisfactorily explain this behavior: poor furnace control is not responsible because we do not see any such systematic variation in eutectic\mbox{-}liquid interfacial velocity in our other \textit{in~situ} solidification experiments \cite{wang_insitu_2021,wang_situ_2022}. Marangoni convection is also not plausible because we observed no bubbles in the FOV nor in the sample overall (likely because it was degassed during casting). Furthermore, magnetohydrodynamic convection is unlikely in the presence of a small magnetic field (estimated as $\ll$\SI{1}{T}) generated by the heating elements in the furnace. \textcolor{black}{In comparison, Refs.~\cite{li2007spiral,li2010morphological} have demonstrated that a relatively high magnetic field of at least 2~T is necessary to modify the Al\mbox{-}Al\textsubscript{2}Cu eutectic microstructure during directional solidification}\textcolor{black}{, although further research is necessary to ascertain the effect of magnetohydrodynamics on fluid flow.}

Recently, Ref.~\cite{reinhart_impact_2020} reported, through X\mbox{-}radiography observations, an oscillatory velocity for single\mbox{-}phase dendritic growth in Ni\mbox{-}based alloys. We attempted to follow a similar approach outlined in that paper by computing the impurity concentration \textcolor{black}{ahead of} the eutectic\mbox{-}liquid interfaces by application of the Beer\mbox{-}Lambert law. Unfortunately, however, we did not see direct evidence of \textcolor{black}{impurity concentration variations in our} X\mbox{-}radiography data. This is likely a result of the low concentration of \textcolor{black}{B} in the alloy. Given a sample thickness of \SI{30}{\micro\meter}, a beam energy centered on 20 keV, and assuming that \textcolor{black}{B} substitutes for Cu, the normalized intensity of X\mbox{-}rays on the detector would be \textcolor{black}{$<$}0.005\% different from that of pure Al\mbox{-}33wt\%Cu. In comparison, the normalized standard deviation of intensities in the bulk liquid is of the order 10\%. Thus, calculation of \textcolor{black}{the impurity} composition profiles in the melt is beneath the detection limit of the instrument and thus may not be directly measurable by X\mbox{-}radiography. Nevertheless, the numerous macroscopic quantities measured in the fused 4D data reconstruction\textcolor{black}{, shown in \textbf{Fig.~\ref{F5}}, and the B identified by APT, shown in \textbf{Fig.~\ref{F13}},} provide ample evidence of impurity\mbox{-}induced microstructural changes. These results are self\mbox{-}consistent with our description of thermosolutal convection that induces oscillations in eutectic\mbox{-}liquid interface velocity. Firstly, we consider the coupling between interfacial velocity and interfacial curvature (\textbf{Fig.\mbox{ }\ref{F5}(a\mbox{-}b)}). The formation of cellular protrusions causes \textcolor{black}{the impurity} to be rejected laterally, piling up at the roots of the protrusion. This would \textcolor{black}{necessarily} lower the interface temperature \textcolor{black}{(by Alkamede's theorem~\cite{van1893graphical})}, causing the formation of depressions. Depressions are associated with a maximal curvature of eutectic\mbox{-}liquid interfaces (refer to \textbf{Fig.\mbox{ }\ref{F5}(b)}). Conversely, increasing velocity leads to a decrease in maximum curvature of the eutectic\mbox{-}liquid interface: when the impurity flows upwards, the solid\mbox{-}liquid interface temperature will rise at the roots/depressions of the perturbation, causing the interface to accelerate into the liquid (thereby decreasing the curvature). \textcolor{black}{These results are consistent with those of Ref.~\cite{rumball1968cellular} who showed that the cell size tends to decrease with an increased impurity content, holding all else constant; that is, the higher degree of constitutional undercooling can only be relieved by a shorter lateral diffusion distance.} Secondly, we observe that changes in velocity are coupled also to changes in area fraction of eutectic phases (\textbf{Fig.\mbox{ }\ref{F5}(d)}). It follows that the impurity species likely provokes a change in the diffusivity, solid\mbox{-}liquid interfacial energies, \textit{etc.}, which would, in turn, cause an increase in the undercooling of one particular phase. To maintain an isothermal eutectic liquid interface (as we see, on local, lamellar scales), the solid\mbox{-}liquid interfacial curvature would need to be reduced accordingly \cite{lahiri_eutectic_2017, lahiri_effect_2015,lahiri_revisiting_2017}. This leads to a departure of contact angles at the solid\mbox{-}solid\mbox{-}liquid trijunction, resulting in a driving force on the trijunction to restore the equilibrium contact angles. The volume fraction adjustment (observed experimentally in \textbf{Fig.~\ref{F5}(d)}) driven by trijunction movement regulates the boundary layer composition to bring the contact angles to their equilibrium values. The latter observation contradicts also the idea that the velocity oscillation is due to poor furnace control (\textit{vide supra}, Sec.~\ref{analysis}). If that were the case, then we would not expect to see the phase fractions oscillate in phase with the other measured attributes; they should be constant throughout (as in Ref. \cite{carlberg_effect_1984}), and only the lamellar spacing would adapt to the evolving interfacial velocity.	These measured attributes are instead evidence of the microstructural adjustment to changes in interface velocity brought about by thermosolutal convection. It is only through a 4D analysis that they could be monitored in time. At this point, it is unclear if the increasing area fraction of Al rods is brought about by an increasing eutectic\mbox{-}liquid interfacial curvature (by Hunt’s proposal~\cite{hunt_lamella_1966}) or by increasing the volume fraction of Al\textsubscript{2}Cu locally. There is support for both ideas from the data (\textbf{Fig.\mbox{ }\ref{F5}}). A closer look at the microstructure (below, Sec.~\ref{sec_struct}) can help us deduce the underlying mechanism. 

\subsection{Structural consequences of an oscillatory velocity} \label{sec_struct}
The oscillation in velocity give us a unique opportunity to examine how the eutectic microstructure adapts to changes in growth conditions. We begin by placing our work on the same plane of analysis as others do in the realm of eutectic solidification, by measuring the relationship between eutectic\mbox{-}liquid interface velocity, $V$, and lamellar spacing, $\lambda$, see \textbf{Fig.~\ref{F6}}. Our results show modest agreement with previous studies \cite{ourdjini_eutectic_1994, cadirli_directional_1999, walker_eutectic_2007, murphy_combined_2013, luo_research_2019, kakitani_thermal_2019} and Jackson\mbox{-}Hunt scaling behavior $\lambda \propto V^{-1/2}$~\cite{Jackson_Hunt_1966} (which applies also to univariant eutectics~\cite{de2005two}), despite the fact that solidification proceeds under non\mbox{-}steady state conditions and with smoothly curved eutectic\mbox{-}liquid interfaces. The two preferred orientations of the solid\mbox{-}solid interfaces (termed ``primary" and ``secondary," \textit{vide infra}) invite a deeper inquiry into the spacing selection. It is clear the primary and secondary lamellae possess distinct spacings\textcolor{black}{, see \textbf{Fig.~\ref{F5} and \ref{SI_4}}}. One contributing factor is a different crystallographic orientation between the two types of solid\mbox{-}solid interfaces; the crystallographic orientations of both lamellar planes are examined later, in Sec.~\ref{sec_crystal}.	The hysteresis in the $V$ \textit{vs.}~$\lambda$ plane is not as pronounced as in recent work by Hecht \cite{witusiewicz_insitu_2021}, despite the higher eutectic\mbox{-}liquid interface acceleration in our work (on the order of \SI{0.1}{\micro\meter/s^2} here, compared to 10\textsuperscript{$-$6} \SI{}{\micro\meter/s^2} there). 

Altogether, these results offer new evidence on the transient structures within eutectic colonies, that do not satisfy previous theories. \textcolor{black}{One is that the lamella\mbox{-}to\mbox{-}rod transition is  provoked by a change in velocity~\cite{tiller_redistribution_1953}.  If that were the case, we would expect rods to form at high velocity, as seen in other dynamic experiments~\cite{cserefoglu2012dynamics}; yet we see the opposite here (\textit{cf.}~Figs.~\ref{F5}(a,e)). Furthermore,} Chadwick's model \cite{chadwick_modification_1962} does not hold merit for two reasons: Firstly, we would expect (by \textcolor{black}{the logic} in Ref.~\cite{chadwick_modification_1962}) that the \textcolor{black}{Al\textsubscript{2}Cu\mbox{-}liquid} interface extends ahead of the \textcolor{black}{Al\mbox{-}liquid} interface (or \textit{vice versa}). We do not see one phase grow substantially ahead of the other, at least at the resolution of the X\mbox{-}radiographs (0.69\textsuperscript{2} \SI{}{\micro\meter}\textsuperscript{2}/pixel). Secondly, his proposal violates Young's law at the trijunction. Instead, and as discussed previously (Sec.~\ref{sec_origin}), the two phases adjust volume fractions to maintain an isothermal front. According to Hunt's model \cite{hunt_lamella_1966}, the eutectic is forced to accommodate the curvature of solid\mbox{-}liquid interface. Consequently, we would predict the growth of rods in a direction that is away from the preferred, low\mbox{-}energy lamellar plane, \textit{i.e.}, rods and lamellae are oriented differently. Instead, our pesudo\mbox{-}4D observations of a single colony in \textbf{Fig.~\ref{F7}} and \textbf{Video~\ref{V_S4}} paint a different picture, that of a coplanar transition. This can be considered as microstructural accommodation following a ``locked" crystallographic growth direction (characterized below). Ultimately, we find that the selection of eutectic morphology (lamellae or rods) is a delicate balance between an impurity\mbox{-}induced change in solid volume fractions (stimulating a change in eutectic morphology) and a strong anisotropy of solid\mbox{-}solid interfacial energy (which tends to resist such a change). In fact, we see the latter effect constrains the lamella\mbox{-}to\mbox{-}rod transition to the lamellar plane, preserving the mirror symmetry of the pattern (unlike, \textit{e.g.}, a zigzag instability~\cite{parisi_stability_2008}). The volume fractions may become highly unequal on local scales:~for example, at the early stages in \textbf{Fig.\mbox{ }\ref{F7}(b)}, the Al phase occupies 75\% of the growth front (by area) for a local region of 6.9~\SI{}{\micro\meter} along $x$ (see boxed region); at the later stages, it takes up only 55\% for this same patch of interface. Ultimately, the volume fractions will approach the global trends (\textbf{Fig.~\ref{F5}(d)}) for larger regions containing more than one lamella. 

\textcolor{black}{It remains to be determined the physical origin of the rod spacing that is selected in the lamella\mbox{-}to\mbox{-}rod transition. If the breakdown of the lamella is due to a capillary effect alone, \textit{i.e.}, a Rayleigh-Plateau instability~\cite{rayleigh1878instability}, then the separation $\lambda^*$ between rods at the transition should scale as the lamellar thickness $R$. Instead, the instability may be governed by an interplay between capillary and diffusive effects, \textit{i.e.}, a Mullins\mbox{-}Sekerka instability~\cite{mullins_stability_1964}, in which case $\lambda^*$ should scale as the geometric mean of capillary and diffusion lengths (equivalent to the Jackson\mbox{-}Hunt spacing, a \textit{constant} for a given velocity, \textit{cf.} Fig.~\ref{F6}). In Fig.~\ref{FX} we plot $\lambda^*$ against $R$ for a particular 2~s window in time, during which we observed 14 lamellae of various thicknesses $R$ and orientations transforming into rods by the same mechanism shown in Fig.~\ref{F7}(c). It can be seen that $\lambda^*$ is nearly invariant with respect to $R$, which points not to a Rayleigh instability but to a diffusive transformation. \textcolor{black}{We find} the average rod separation in Fig.~\ref{FX} of 5.1~\textpm~0.3~\SI{}{\micro\meter} matches reasonably well the smallest stable eutectic spacing of \SI{4.9}{\micro\meter} for an interfacial velocity of \SI{3.8}{\micro\meter/s} \cite{walker_eutectic_2007}, determined using the Jackson\mbox{-}Hunt model~\cite{Jackson_Hunt_1966}. This finding is broadly in agreement with 3D phase field simulations~\cite{parisi_stability_2008}}.

We note also that the morphological transition is not reversible. As the volume fractions of the two phases become comparable on local scales, the same Al rods will not coalesce to form a lamella. Instead, the rods will flatten in cross\mbox{-}section and merge with other lamellar branches, a process that involves the collective interactions of many lamellae (not pictured). 

Our observations in \textbf{Fig.~\ref{F7}(c)} and \textbf{Video~\ref{V_S5}} bear some similarity to that of Refs.~\cite{witusiewicz_insitu_2021,bottin-rousseau_coexistence_2022}, wherein transverse thermal gradients \cite{bottin-rousseau_coexistence_2022} and velocity changes \cite{witusiewicz_insitu_2021} destabilize the steady-pattern formation in transparent eutectics. In Ref.~\cite{bottin-rousseau_coexistence_2022}, the authors observed a varicose instability that stimulates a coplanar transition from a lamella to rods. However, there are two key differences between their work and ours: firstly, in their case, lamellae are stabilized only by a wall/boundary effect while rods are ordinarily formed in their system\mbox{-}of\mbox{-}interest, succinonitrile(D),camphor. Meanwhile, the morphological transition in our case is provoked by an impurity effect. Secondly, anisotropy effects in succinonitrile(D),camphor (SCN-DC) are significantly weaker than in Al\mbox{-}Al\textsubscript{2}Cu. \textcolor{black}{That is, the SCN- and DC\mbox{-}liquid interfaces are not only nonfaceted but also weakly anisotropic; furthermore, there exist no orientation relationship between SCN and DC~\cite{akamatsu2007real}. In comparison, Al and Al\textsubscript{2}Cu form ``epitaxial" eutectic grains during directional solidification~\cite{kraft_crystallography_1962, bonnet_geometric_1973, cantor_growth_1974, hecht_coupled_2012, wang_characteristic_2018, hecht_phase_2019}. Thus, o}ur rods are not circular in cross section (eccentricity of 0) but do show a slight elongation or ellipsoidal cross section (eccentricity of $\sim$0.7)\textcolor{black}{, see \textbf{Fig.~\ref{SI_7}}}.{\footnote{\textcolor{black}{It follows from Fig.~\ref{FX} that if $\lambda^* \neq R$, the rods will necessarily show an elongated shape when viewed cross\mbox{-}sectionally (trivial case). Yet even those few rods for which $\lambda^* \simeq R \simeq 5$~\SI{}{\micro\meter} show a partially faceted shape, supporting the idea of an underlying anisotropy of solid\mbox{-}solid interfacial free energy.}} This geometry hints at a resistance to morphological instability due to a low\mbox{-}energy solid\mbox{-}solid boundary, an effect that is seen also in the capillary\mbox{-}driven evolution of pore channels in sapphire during annealing \cite{santala_surfaceenergyanisotropyinduced_2006}. 

\subsection{Crystallography of `locked’ eutectic grains} \label{sec_crystal}
Why do the lamellae follow specific orientations during growth? Why do they not `fan out’ as in past studies \cite{trivedi_nonplane_2008,han_stability_1995} of eutectic cells? We suppose that there exists a preferred growth crystallography between Al and Al\textsubscript{2}Cu phases. We begin by characterizing the OR between the two phases. The orientation relationship in the sample deviates from the commonly observed Beta\mbox{-}6 and Alpha\mbox{-}4 orientation relationships reported in \cite{kraft_crystallography_1962, bonnet_geometric_1973, cantor_growth_1974, hecht_coupled_2012, wang_characteristic_2018, hecht_phase_2019} (See \textbf{Fig.\mbox{ }\ref{F8}}). Instead, we identify a new OR based on the EBSD data collected on the same specimen, following the method described in Ref.~\cite{bottinrousseau_special_2018}. We calculated the mismatch angle $\Delta p$ between $\alpha$\mbox{-}Al and $\theta$\mbox{-}Al\textsubscript{2}Cu lattice planes, and the angle $\Delta d$ between directions belonging to those planes \cite{bachmann_texture_2010}. We considered planes and directions with Miller indices lower than 5. We dismissed pairs of common planes and common directions with $\Delta p$ and $\Delta d$ angles larger than a threshold value of 8\textdegree. On this basis, we designate the new OR as $(211)_{\textrm{Al}} // (\bar{1}10)_{\textrm{Al\textsubscript{2}Cu}}$ with $\Delta p$ of 3.34\textdegree and $[\bar{1}02]_{\textrm{Al}} // [332]_{\textrm{Al\textsubscript{2}Cu}}$ with $\Delta d$ of 6.88\textdegree. 

By aligning the EBSD and X\mbox{-}ray tomography frames of reference, we can achieve a crystallographic description of the eutectic solid\mbox{-}solid interfaces (and specifically, the normal vectors along these interfaces), see \textbf{Fig.~\ref{F9}}. We first examine the orientation of microstructural features relative to each other (\textbf{Fig.~\ref{F10}(a)}). The angle between the normal vectors of primary and secondary lamellae is $80 \pm 2^{\circ}$. Intriguingly, the growth direction (GD) of primary rods (here, ``primary rod" indicates that the rods emerge from a primary lamella during solidification) is orthogonal to both lamellar normals. The growth direction of secondary rods is inclined around 30\textdegree~to the primary rod growth direction. By this nomenclature, the primary rods are within the primary lamellar plane, and likewise for the secondary rods. 

Next, we invoke the textural data from EBSD to transform the solid\mbox{-}solid orientations from specimen to crystallographic frame (as described in Sec.~\ref{sec_methods} and depicted in \textbf{Fig.~\ref{F9}}). The Al and Al\textsubscript{2}Cu inverse pole figures (IPFs) in \textbf{Fig.~\ref{F10}(b\mbox{-}c)} represent the crystallographic orientations of the primary and secondary lamella normals. With the aid of these results, we are able to determine completely the interfacial bi\mbox{-}crystallography. The habit planes of Al and Al\textsubscript{2}Cu are not the expected dense, low index planes \cite{bottinrousseau_lockedlamellar_2021}. Instead, we find the primary lamella habit plane is oriented along $(011)_{\textrm{Al}}$ // $(433)_{\textrm{Al\textsubscript{2}Cu}}$ and the secondary lamella habit plane is $(223)_{\textrm{Al}}$ // $(\bar{4}51)_{\textrm{Al\textsubscript{2}Cu}}$. In the crystal frame, the angle between the two habit planes is $82^{\circ}$, which falls within the error margins of our finding from above.

This interface between Al and Al\textsubscript{2}Cu is semicoherent \cite{sutton_interfaces_1995, howe_interfaces_1997,garmong_interfacial_1974,garmong_structure_1974}. Generally, the minimum in interfacial energy for a semicoherent interface should correspond to a low misfit and hence a reduced lattice strain, not accounting for chemical or thermal contributions \cite{howe_interfaces_1997}. We define the lattice misfit as $\Delta = 2|\rho_\alpha - \rho_\theta | / (\rho_\alpha + \rho_\theta$), where subscripts indicate the two eutectic phases that meet at a solid\mbox{-}solid interface \cite{sutton_interfaces_1995, howe_interfaces_1997}. The atomic density of a lattice plane is $\rho = n_{2D} d / \Omega$, where $n_{2D}$ is the number of atoms per unit cell in the plane, $d$ the interplane spacing, and $\Omega$ the volume of the 3D unit cell. Unfortunately, this definition forces the selection of a specific atomic layer which may have variable atomic density \cite{bottinrousseau_special_2018, bottinrousseau_lockedlamellar_2021}. We bypass this limitation by following Kraft’s description of a ``puckered" interface to calculate the atomic planar density \cite{kraft_crystallography_1962}. He describes a ``puckered" atomic plane as one that accommodates additional atoms above and below it. By accounting for atoms that are not strictly on the plane but close to it, the density difference between the planes can be reduced (\textit{vide infra}). In this way, we determine a minimum misfit across the primary lamellar interface as 0.01\% for 8 atoms/unit area of Al\textsubscript{2}Cu and 1 atom/unit area of Al; for the the secondary lamellar interface, we find a misfit of 4.57\% for 3 atoms/unit area of Al\textsubscript{2}Cu and 1 atom/unit area of Al. \textbf{Figs. \ref{F11}(a\mbox{-}b)} depict the corresponding puckered densities for the lamellar habit planes for increasing thicknesses. We have also included the atomic structure of these planes as a visual aid similar to Kraft’s original sketch \cite{kraft_crystallography_1962}. These lattices are oriented according to the OR determined by EBSD. Overlaying both atomic lattices can yield additional details such as 2D lattice strain tensors and coincident site lattice descriptions for these interfaces, as described elsewhere \cite{wang_characteristic_2018}. Though the puckered interface density has been used to explain ORs in Al\mbox{-}Al\textsubscript{2}Cu \cite{kraft_crystallography_1962}, the model may not necessarily apply to all eutectic systems \cite{kaya_transmission_1992}. 

The aforementioned theory holds for pristine or chemically pure Al\mbox{-}Al\textsubscript{2}Cu interfaces. In univariant solidification, however, \textcolor{black}{the impurity species} may segregate to the semicoherent solid\mbox{-}solid interfaces \textcolor{black}{(as in \textbf{Fig.~\ref{F13}})}. \textcolor{black}{In a similar vein, APT} shows that \textcolor{black}{trace} Si chemically modifies the semicoherent interface structure and lowers the solid\mbox{-}solid interfacial Gibbs free energy five times more than that of a coherent interface in $\theta'$ precipitates embedded in the Al matrix \cite{biswas_simultaneous_2010}. The sequence of volume renders in \textbf{Fig.~\ref{SI_8}} indicates that primary lamellae predominate at low velocity (corresponding to a greater impurity concentration in the boundary layer and presumably also a greater \textcolor{black}{impurity} segregation to the lamellar interfaces). It stands to reason, then, that \textcolor{black}{B} may stabilize this particular primary interface over that of the secondary lamellae (\textit{via} Gibbs adsorption \cite{biswas_simultaneous_2010}). A sequence of renders demonstrating such lamella\mbox{-}to\mbox{-}lamella transitions are included in \textbf{Fig.~\ref{SI_8}}. We await confirmation of this idea \textit{via} atomistic modeling of the interfacial structures in eutectics.


Based on rotating directional furnace experiments \cite{akamatsu_lamellar_2012}, large tilted domains can persist due to interfacial anisotropy. In many cases, the growth direction of the monocrystalline eutectic is the result of the initial seed crystal \cite{bottinrousseau_lockedlamellar_2021}, not pictured in our experiments. Another consequence of a crystalline anisotropy is that the eutectic cells will appear to drift across the imaged FOV, as we see here in \textbf{Fig.\mbox{ }\ref{F4}}. A similar behavior is reported also in phase field simulations of locked eutectic grains \cite{lahiri_eutectic_2017}.

\subsection{Outlook on pseudo\mbox{-}4D tomography} \label{sec_outlook}
The imaging strategy introduced in this study provides access to temporal and spatial resolutions scales that are not accessible in conventional 4D CT. Results presented here can lay the groundwork for future \textit{in~situ} tomography campaigns. Even so, one must bear in mind that there is one assumption that enables the pseudo\mbox{-}4D visualization with data fusion: when we extract and interpolate the 3D shape of the growth front from a projected view of the solidification process, we assumed that the eutectic\mbox{-}liquid interface is parallel to the beam direction. We do not know the `tilt’ of the growth front \textit{a priori} but set it to be oriented along the beam direction for sake of simplicity. This assumption is strictly valid only in the limit that the sample thickness is infinitely reduced, corresponding to pseudo\mbox{-}2D solidification. With higher\mbox{-}resolution imaging, \textit{e.g.}, X\mbox{-}ray nanotomography \cite{withers_xray_2021}, this assumption can be lifted:~one can reproject the 3D microstructure to the detector plane and compare the solid\mbox{-}liquid interface against that of the actual X\textbf{-}radiograph. If the interface is tilted by an angle of $\theta$ in the direction of the X\mbox{-}ray beam, it should blur in the reprojected image by an amount $l\mbox{tan} \theta$, where $l$ is the sample thickness. By incrementally changing $\theta$, one can find the tilt angle that results in a projection that is most similar in contrast to the X\mbox{-}radiograph, taken as ground truth. Efforts are underway to implement this approach. 

Thin samples are otherwise beneficial for X\mbox{-}radiography, since it is difficult to interpret signals in projection images of thick ($\sim$1 mm) samples. Yet the presence of surfaces leads to no\mbox{-}flux boundary conditions in solidification, which may influence solidification dynamics for sample sizes of less than 10\mbox{-}100\texttimes that of the eutectic lamellar spacing \cite{plapp_eutectic_2002,asta_solidification_2009}. \textcolor{black}{For example, the lamella\mbox{-}to\mbox{-}rod transition may be restrained when the sample thickness is less than the wavelength of any instability threshold, thereby stabilizing lamellae~\cite{cserefoglu2012dynamics}.} 

Nonetheless, this technique is especially valuable when the final, solidified structure is the one that is growing during the \textit{in~situ} scan. Conversely, it is not applicable for solidification of metastable phases or transient states, which give way to stable ones during cooling at low rates. For this reason, pseudo\mbox{-}4D tomography is not suitable for capturing the peritectic transition, wherein the primary phase is consumed to yield a peritectic phase (a ``perfect crime" since the latter ``kills" the former phase in solidification \cite{rappaz_thermodynamic_2015}). 

\section*{Conclusion}
This study explores the evolution of Al\mbox{-}Al\textsubscript{2}Cu eutectic patterns along a univariant solidification pathway in the face of thermosolutal convection. To capture the transient microstructural details, we developed a new pseudo\mbox{-}4D approach that combines \textit{in~situ} X\mbox{-}radiography and \textit{postmortem} X\mbox{-}ray microtomography. With this method, we achieve the highest reported temporal and spatial resolution for absorption\mbox{-}contrast X\mbox{-}ray imaging \textcolor{black}{and gain unique insight into the solidification dynamics. Several conclusions can be drawn from this study:}
\begin{enumerate}
  \color{black}{\item The influence of impurities (such as B, $\sim$0.01at\%), is two\mbox{-}fold. It bring about a constitutional undercooling that destabilizes the eutectic\mbox{-}liquid interfaces at long wavelengths ($\gtrsim$~10~\SI{}{\micro\meter}), giving rise to shallow cells or colonies. On shorter wavelengths, it changes the undercooling of one eutectic phase with respect to the other, which in turn changes the phase fractions and hence the eutectic morphology. Our time\mbox{-}resolved analysis provides evidence of these trends in phase fraction, morphology, and their relation.
  \item The eutectic accommodates the changes in phase fractions through a co-planar lamella\mbox{-}to\mbox{-}rod transition despite the resistance of `locked' interfaces. The partially faceted morphology of the rods so\mbox{-}produced hints at a low\mbox{-}energy solid\mbox{-}solid boundary. Furthermore, the spacing between the rods \textcolor{black}{is nearly invariant with the lamellar thickness}, signifying a \textcolor{black}{diffusive instability upon eutectic solidification}.
  \item A multimodal crystallographic analysis (linking EBSD and pseudo\mbox{-}4D imaging) provides a complementary view of the microstructure. We detect a new OR and unexpectedly high index (yet low misfit) lamellar habit planes. As a result of this crystalline anisotropy, the eutectic colonies drift across the FOV, following the inclination of the habit planes with respect to the macroscopic growth direction.}
\end{enumerate}

Altogether, these results provides a fresh picture on the selection of eutectic patterns in systems with a strong anisotropy in solid\mbox{-}solid interfacial energy. Chemical effects give rise to a complexity in the dynamics of pattern formation that is not seen in nonvariant systems. We expect that pseudo\mbox{-}4D imaging may open other doors in solidification science, enabling researchers to test the predictions of theory with high precision.

\section*{Acknowledgements}
We gratefully acknowledge financial support from the National Science Foundation (NSF) CAREER program under Award No.~1847855. We thank Dr.~Caleb Reese, Aaron Gladstein, Dr.~Francesco De Carlo, and Pavel Shevshenko for assisting in the synchrotron\mbox{-}based experiment. We also thank Drs.~Silvère Akamatsu and Sabine Bottin\mbox{-}Rousseau for fruitful discussions. This research used resources of the Advanced Photon Source, a U.S.~Department of Energy (DOE) Office of Science User Facility operated for the DOE Office of Science by Argonne National Laboratory under Contract No.~DE\mbox{-}AC02\mbox{-}06CH11357. We also \textcolor{black}{acknowledge the University of Michigan College of Engineering for financial support and} the Michigan Center for Materials Characterization for use of the instruments and \textcolor{black}{Bobby Kerns, Dr. Nancy Senabulya Muyanja, Dr. Tao Ma, and Dr. Haiping Sun for their} assistance.

\newpage

\begin{thebibliography}{113}
\expandafter\ifx\csname natexlab\endcsname\relax\def\natexlab#1{#1}\fi
\providecommand{\url}[1]{\texttt{#1}}
\providecommand{\href}[2]{#2}
\providecommand{\path}[1]{#1}
\providecommand{\DOIprefix}{doi:}
\providecommand{\ArXivprefix}{arXiv:}
\providecommand{\URLprefix}{URL: }
\providecommand{\Pubmedprefix}{pmid:}
\providecommand{\doi}[1]{\href{http://dx.doi.org/#1}{\path{#1}}}
\providecommand{\Pubmed}[1]{\href{pmid:#1}{\path{#1}}}
\providecommand{\bibinfo}[2]{#2}
\ifx\xfnm\relax \def\xfnm[#1]{\unskip,\space#1}\fi
\bibitem[{Tiwary et~al.(2022)Tiwary, Pandey, Sarkar, Das, Samal, Biswas, and
  Chattopadhyay}]{tiwary2022five}
\bibinfo{author}{C.~S. Tiwary}, \bibinfo{author}{P.~Pandey},
  \bibinfo{author}{S.~Sarkar}, \bibinfo{author}{R.~Das},
  \bibinfo{author}{S.~Samal}, \bibinfo{author}{K.~Biswas},
  \bibinfo{author}{K.~Chattopadhyay},
\newblock \bibinfo{title}{Five decades of research on the development of
  eutectic as engineering materials},
\newblock \bibinfo{journal}{Progress in Materials Science}
  \bibinfo{volume}{123} (\bibinfo{year}{2022}) \bibinfo{pages}{100793}.
\bibitem[{Glicksman(2011)}]{glicksman_principles_2011}
\bibinfo{author}{M.~E. Glicksman}, \bibinfo{title}{Principles of
  Solidification}, \bibinfo{publisher}{Springer New York},
  \bibinfo{year}{2011}. \DOIprefix\doi{10.1007/978-1-4419-7344-3}.
\bibitem[{Aikin(1997)}]{aikin_mechanical_1997}
\bibinfo{author}{R.~M. Aikin},
\newblock \bibinfo{title}{The mechanical properties of in-situ composites},
\newblock \bibinfo{journal}{{JOM}} \bibinfo{volume}{49} (\bibinfo{year}{1997})
  \bibinfo{pages}{35}.
\bibitem[{Jackson and Hunt(1966)}]{Jackson_Hunt_1966}
\bibinfo{author}{K.~Jackson}, \bibinfo{author}{J.~Hunt},
\newblock \bibinfo{title}{Lamellar and rod eutectic growth},
\newblock \bibinfo{journal}{Trans. Met. Soc. AIME} \bibinfo{volume}{236}
  (\bibinfo{year}{1966}).
\bibitem[{Kurz and Fisher(1992)}]{kurz_fundamentals_1992}
\bibinfo{author}{W.~Kurz}, \bibinfo{author}{D.~J. Fisher},
  \bibinfo{title}{Fundamentals of solidification}, \bibinfo{edition}{3. rev.
  ed., reprinted} ed., \bibinfo{publisher}{Trans Tech Publ},
  \bibinfo{year}{1992}.
\bibitem[{Datye and Langer(1981)}]{datye_stability_1981}
\bibinfo{author}{V.~Datye}, \bibinfo{author}{J.~S. Langer},
\newblock \bibinfo{title}{Stability of thin lamellar eutectic growth},
\newblock \bibinfo{journal}{Physical Review B} \bibinfo{volume}{24}
  (\bibinfo{year}{1981}) \bibinfo{pages}{4155--4169}.
\bibitem[{Plapp and Karma(1999)}]{plapp_eutectic_1999}
\bibinfo{author}{M.~Plapp}, \bibinfo{author}{A.~Karma},
\newblock \bibinfo{title}{Eutectic colony formation: a stability analysis},
\newblock \bibinfo{journal}{Physical Review. E, Statistical Physics, Plasmas,
  Fluids, and Related Interdisciplinary Topics} \bibinfo{volume}{60}
  (\bibinfo{year}{1999}) \bibinfo{pages}{6865--6889}.
\bibitem[{Mullins and Sekerka(1964)}]{mullins_stability_1964}
\bibinfo{author}{W.~W. Mullins}, \bibinfo{author}{R.~F. Sekerka},
\newblock \bibinfo{title}{Stability of a planar interface during solidification
  of a dilute binary alloy},
\newblock \bibinfo{journal}{Journal of Applied Physics} \bibinfo{volume}{35}
  (\bibinfo{year}{1964}) \bibinfo{pages}{444--451}.
\bibitem[{Caroli et~al.(1986)Caroli, Caroli, and
  Roulet}]{caroli_mullinssekerka_1986}
\bibinfo{author}{B.~Caroli}, \bibinfo{author}{C.~Caroli},
  \bibinfo{author}{B.~Roulet},
\newblock \bibinfo{title}{The {Mullins-Sekerka} instability in directional
  solidification of thin samples},
\newblock \bibinfo{journal}{Journal of Crystal Growth} \bibinfo{volume}{76}
  (\bibinfo{year}{1986}) \bibinfo{pages}{31--49}.
\bibitem[{Tiller(1991)}]{tiller_redistribution_1953}
\bibinfo{author}{W.~A. Tiller}, \bibinfo{title}{The science of crystallization:
  microscopic interfacial phenomena}, \bibinfo{publisher}{Cambridge University
  Press}, \bibinfo{year}{1991}.
\bibitem[{Chadwick(1963)}]{chadwick_eutectic_1963}
\bibinfo{author}{G.~Chadwick},
\newblock \bibinfo{title}{Eutectic alloy solidification},
\newblock \bibinfo{journal}{Progress in Materials Science} \bibinfo{volume}{12}
  (\bibinfo{year}{1963}) \bibinfo{pages}{99--182}.
\bibitem[{Trivedi et~al.(2002)Trivedi, Mazumder, and
  Tewari}]{trivedi_effect_2002}
\bibinfo{author}{R.~Trivedi}, \bibinfo{author}{P.~Mazumder},
  \bibinfo{author}{S.~N. Tewari},
\newblock \bibinfo{title}{The effect of convection on disorder in primary
  cellular and dendritic arrays},
\newblock \bibinfo{journal}{Metallurgical and Materials Transactions A}
  \bibinfo{volume}{33} (\bibinfo{year}{2002}) \bibinfo{pages}{3763--3775}.
\bibitem[{Plapp and Karma(2002)}]{plapp_eutectic_2002}
\bibinfo{author}{M.~Plapp}, \bibinfo{author}{A.~Karma},
\newblock \bibinfo{title}{Eutectic colony formation: A phase-field study},
\newblock \bibinfo{journal}{Physical Review E} \bibinfo{volume}{66}
  (\bibinfo{year}{2002}) \bibinfo{pages}{061608}.
\bibitem[{Hecht et~al.(2004)Hecht, Gránásy, Pusztai, Böttger, Apel,
  Witusiewicz, Ratke, De~Wilde, Froyen, Camel, Drevet, Faivre, Fries, Legendre,
  and Rex}]{hecht_multiphase_2004}
\bibinfo{author}{U.~Hecht}, \bibinfo{author}{L.~Gránásy},
  \bibinfo{author}{T.~Pusztai}, \bibinfo{author}{B.~Böttger},
  \bibinfo{author}{M.~Apel}, \bibinfo{author}{V.~Witusiewicz},
  \bibinfo{author}{L.~Ratke}, \bibinfo{author}{J.~De~Wilde},
  \bibinfo{author}{L.~Froyen}, \bibinfo{author}{D.~Camel},
  \bibinfo{author}{B.~Drevet}, \bibinfo{author}{G.~Faivre},
  \bibinfo{author}{S.~Fries}, \bibinfo{author}{B.~Legendre},
  \bibinfo{author}{S.~Rex},
\newblock \bibinfo{title}{Multiphase solidification in multicomponent alloys},
\newblock \bibinfo{journal}{Materials Science and Engineering: R: Reports}
  \bibinfo{volume}{46} (\bibinfo{year}{2004}) \bibinfo{pages}{1--49}.
\bibitem[{Sargin et~al.(2016)Sargin, Genau, and
  Napolitano}]{sargin_postsolidification_2016}
\bibinfo{author}{I.~Sargin}, \bibinfo{author}{A.~L. Genau},
  \bibinfo{author}{R.~E. Napolitano},
\newblock \bibinfo{title}{Post-solidification effects in directionally grown
  {Al-Ag\textsubscript{2}Al-Al\textsubscript{2}Cu} eutectics},
\newblock \bibinfo{journal}{Journal of Phase Equilibria and Diffusion}
  \bibinfo{volume}{37} (\bibinfo{year}{2016}) \bibinfo{pages}{75--85}.
\bibitem[{Steinmetz et~al.(2018)Steinmetz, Dennstedt, Şerefoğlu, Sargin,
  Genau, and Hecht}]{steinmetz_crystal_2018}
\bibinfo{author}{P.~Steinmetz}, \bibinfo{author}{A.~Dennstedt},
  \bibinfo{author}{M.~Şerefoğlu}, \bibinfo{author}{I.~Sargin},
  \bibinfo{author}{A.~Genau}, \bibinfo{author}{U.~Hecht},
\newblock \bibinfo{title}{Crystal orientation relationships in ternary eutectic
  {Al-Al$_2$Cu-Ag$_2$Al}},
\newblock \bibinfo{journal}{Acta Materialia} \bibinfo{volume}{157}
  (\bibinfo{year}{2018}) \bibinfo{pages}{96--105}.
\bibitem[{Trivedi and Sunseri(2008)}]{trivedi_nonplane_2008}
\bibinfo{author}{R.~Trivedi}, \bibinfo{author}{E.~Sunseri},
\newblock \bibinfo{title}{Non-plane front solidification},
\newblock in: \bibinfo{editor}{S.~Viswanathan}, \bibinfo{editor}{D.~Apelian},
  \bibinfo{editor}{R.~J. Donahue}, \bibinfo{editor}{B.~DasGupta},
  \bibinfo{editor}{M.~Gywn}, \bibinfo{editor}{J.~L. Jorstad},
  \bibinfo{editor}{R.~W. Monroe}, \bibinfo{editor}{M.~Sahoo},
  \bibinfo{editor}{T.~E. Prucha}, \bibinfo{editor}{D.~Twarog} (Eds.),
  \bibinfo{booktitle}{Casting}, \bibinfo{publisher}{ASM International},
  \bibinfo{year}{2008}, pp. \bibinfo{pages}{299--306}.
  \DOIprefix\doi{10.31399/asm.hb.v15.a0005210}.
\bibitem[{Han(1995)}]{han_stability_1995}
\bibinfo{author}{S.~H. Han}, \bibinfo{title}{Stability of a eutectic interface
  during directional solidification}, \bibinfo{year}{1995}.
  \DOIprefix\doi{10.31274/rtd-180813-10182}.
\bibitem[{Caroli et~al.(1992)Caroli, Caroli, Faivre, and
  Mergy}]{caroli_lamellar_1992}
\bibinfo{author}{B.~Caroli}, \bibinfo{author}{C.~Caroli},
  \bibinfo{author}{G.~Faivre}, \bibinfo{author}{J.~Mergy},
\newblock \bibinfo{title}{Lamellar eutectic growth of
  {CBr}\textsubscript{4}-{C}\textsubscript{2}{Cl}\textsubscript{6}: effect of
  crystal anisotropy on lamellar orientations and wavelength dispersion},
\newblock \bibinfo{journal}{Journal of Crystal Growth} \bibinfo{volume}{118}
  (\bibinfo{year}{1992}) \bibinfo{pages}{135--150}.
\bibitem[{Akamatsu et~al.(2001)Akamatsu, Faivre, and
  Moulinet}]{akamatsu_formation_2001}
\bibinfo{author}{S.~Akamatsu}, \bibinfo{author}{G.~Faivre},
  \bibinfo{author}{S.~Moulinet},
\newblock \bibinfo{title}{The formation of lamellar-eutectic grains in thin
  samples},
\newblock \bibinfo{journal}{Metallurgical and Materials Transactions A}
  \bibinfo{volume}{32} (\bibinfo{year}{2001}) \bibinfo{pages}{2039--2048}.
\bibitem[{Akamatsu et~al.(2012)Akamatsu, Bottin-Rousseau, Şerefoğlu, and
  Faivre}]{akamatsu_lamellar_2012}
\bibinfo{author}{S.~Akamatsu}, \bibinfo{author}{S.~Bottin-Rousseau},
  \bibinfo{author}{M.~Şerefoğlu}, \bibinfo{author}{G.~Faivre},
\newblock \bibinfo{title}{Lamellar eutectic growth with anisotropic interphase
  boundaries: Experimental study using the rotating directional solidification
  method},
\newblock \bibinfo{journal}{Acta Materialia} \bibinfo{volume}{60}
  (\bibinfo{year}{2012}) \bibinfo{pages}{3206--3214}.
\bibitem[{Ghosh et~al.(2015)Ghosh, Choudhury, Plapp, Bottin-Rousseau, Faivre,
  and Akamatsu}]{ghosh_interphase_2015}
\bibinfo{author}{S.~Ghosh}, \bibinfo{author}{A.~Choudhury},
  \bibinfo{author}{M.~Plapp}, \bibinfo{author}{S.~Bottin-Rousseau},
  \bibinfo{author}{G.~Faivre}, \bibinfo{author}{S.~Akamatsu},
\newblock \bibinfo{title}{Interphase anisotropy effects on lamellar eutectics:
  A numerical study},
\newblock \bibinfo{journal}{Physical Review E} \bibinfo{volume}{91}
  (\bibinfo{year}{2015}) \bibinfo{pages}{022407}.
\bibitem[{Ghosh and Plapp(2017)}]{ghosh_influence_2017}
\bibinfo{author}{S.~Ghosh}, \bibinfo{author}{M.~Plapp},
\newblock \bibinfo{title}{Influence of interphase boundary anisotropy on bulk
  eutectic solidification microstructures},
\newblock \bibinfo{journal}{Acta Materialia} \bibinfo{volume}{140}
  (\bibinfo{year}{2017}) \bibinfo{pages}{140--148}.
\bibitem[{Bottin-Rousseau et~al.(2018)Bottin-Rousseau, Senninger, Faivre, and
  Akamatsu}]{bottinrousseau_special_2018}
\bibinfo{author}{S.~Bottin-Rousseau}, \bibinfo{author}{O.~Senninger},
  \bibinfo{author}{G.~Faivre}, \bibinfo{author}{S.~Akamatsu},
\newblock \bibinfo{title}{Special interphase orientation relationships and
  locked lamellar growth in thin in-in2bi eutectics},
\newblock \bibinfo{journal}{Acta Materialia} \bibinfo{volume}{150}
  (\bibinfo{year}{2018}) \bibinfo{pages}{16--24}.
\bibitem[{Chadwick(1962)}]{chadwick_modification_1962}
\bibinfo{author}{G.~A. Chadwick},
\newblock \bibinfo{title}{Modification of lamellar eutectic structures},
\newblock \bibinfo{journal}{Journal of the Institute of Metals}
  \bibinfo{volume}{91} (\bibinfo{year}{1962}) \bibinfo{pages}{298--303}.
\bibitem[{Hunt(1966)}]{hunt_lamella_1966}
\bibinfo{author}{J.~D. Hunt},
\newblock \bibinfo{title}{The lamella -{\textgreater} rod transformation in
  eutectics},
\newblock \bibinfo{journal}{Journal of the Institute of Metals}
  \bibinfo{volume}{94} (\bibinfo{year}{1966}) \bibinfo{pages}{125--129}.
\bibitem[{Lahiri et~al.(2017)Lahiri, Tiwary, Chattopadhyay, and
  Choudhury}]{lahiri_eutectic_2017}
\bibinfo{author}{A.~Lahiri}, \bibinfo{author}{C.~Tiwary},
  \bibinfo{author}{K.~Chattopadhyay}, \bibinfo{author}{A.~Choudhury},
\newblock \bibinfo{title}{Eutectic colony formation in systems with interfacial
  energy anisotropy: A phase field study},
\newblock \bibinfo{journal}{Computational Materials Science}
  \bibinfo{volume}{130} (\bibinfo{year}{2017}) \bibinfo{pages}{109--120}.
\bibitem[{Aramanda et~al.(2020)Aramanda, Salapaka, Khanna, Chattopadhyay, and
  Choudhury}]{aramanda_exotic_2020}
\bibinfo{author}{S.~K. Aramanda}, \bibinfo{author}{S.~K. Salapaka},
  \bibinfo{author}{S.~Khanna}, \bibinfo{author}{K.~Chattopadhyay},
  \bibinfo{author}{A.~Choudhury},
\newblock \bibinfo{title}{Exotic colony formation in {Sn-Te} eutectic system},
\newblock \bibinfo{journal}{Acta Materialia} \bibinfo{volume}{197}
  (\bibinfo{year}{2020}) \bibinfo{pages}{108--121}.
\bibitem[{Aramanda et~al.(2021)Aramanda, Chattopadhyay, and
  Choudhury}]{aramanda_exotic_2021}
\bibinfo{author}{S.~K. Aramanda}, \bibinfo{author}{K.~Chattopadhyay},
  \bibinfo{author}{A.~Choudhury},
\newblock \bibinfo{title}{Exotic three-phase microstructures in the ternary
  {Ag-Cu-Sb} eutectic system},
\newblock \bibinfo{journal}{Acta Materialia} \bibinfo{volume}{221}
  (\bibinfo{year}{2021}) \bibinfo{pages}{117400}.
\bibitem[{Liu et~al.(2011)Liu, Lee, and Trivedi}]{liu_dynamic_2011}
\bibinfo{author}{S.~Liu}, \bibinfo{author}{J.~Lee},
  \bibinfo{author}{R.~Trivedi},
\newblock \bibinfo{title}{Dynamic effects in the lamellar–rod eutectic
  transition},
\newblock \bibinfo{journal}{Acta Materialia} \bibinfo{volume}{59}
  (\bibinfo{year}{2011}) \bibinfo{pages}{3102--3115}.
\bibitem[{Mathiesen et~al.(2002)Mathiesen, Arnberg, Ramsøskar, Weitkamp, Rau,
  and Snigirev}]{mathiesen_timeresolved_2002}
\bibinfo{author}{R.~H. Mathiesen}, \bibinfo{author}{L.~Arnberg},
  \bibinfo{author}{K.~Ramsøskar}, \bibinfo{author}{T.~Weitkamp},
  \bibinfo{author}{C.~Rau}, \bibinfo{author}{A.~Snigirev},
\newblock \bibinfo{title}{Time-resolved x-ray imaging of aluminum alloy
  solidification processes},
\newblock \bibinfo{journal}{Metallurgical and Materials Transactions B}
  \bibinfo{volume}{33} (\bibinfo{year}{2002}) \bibinfo{pages}{613--623}.
\bibitem[{Yasuda et~al.(2004)Yasuda, Ohnaka, Kawasaki, Sugiyama, Ohmichi,
  Iwane, and Umetani}]{yasuda_direct_2004}
\bibinfo{author}{H.~Yasuda}, \bibinfo{author}{I.~Ohnaka},
  \bibinfo{author}{K.~Kawasaki}, \bibinfo{author}{A.~Sugiyama},
  \bibinfo{author}{T.~Ohmichi}, \bibinfo{author}{J.~Iwane},
  \bibinfo{author}{K.~Umetani},
\newblock \bibinfo{title}{Direct observation of stray crystal formation in
  unidirectional solidification of {Sn–Bi} alloy by x-ray imaging},
\newblock \bibinfo{journal}{Journal of Crystal Growth} \bibinfo{volume}{262}
  (\bibinfo{year}{2004}) \bibinfo{pages}{645--652}.
\bibitem[{Nguyen-Thi et~al.(2012)Nguyen-Thi, Salvo, Mathiesen, Arnberg, Billia,
  Suery, and Reinhart}]{nguyenthi_interest_2012}
\bibinfo{author}{H.~Nguyen-Thi}, \bibinfo{author}{L.~Salvo},
  \bibinfo{author}{R.~H. Mathiesen}, \bibinfo{author}{L.~Arnberg},
  \bibinfo{author}{B.~Billia}, \bibinfo{author}{M.~Suery},
  \bibinfo{author}{G.~Reinhart},
\newblock \bibinfo{title}{On the interest of synchrotron x-ray imaging for the
  study of solidification in metallic alloys},
\newblock \bibinfo{journal}{Comptes Rendus Physique} \bibinfo{volume}{13}
  (\bibinfo{year}{2012}) \bibinfo{pages}{237--245}.
\bibitem[{Clarke et~al.(2015)Clarke, Tourret, Imhoff, Gibbs, Fezzaa, Cooley,
  Lee, Deriy, Patterson, Papin, Clarke, Field, and Smith}]{clarke_xray_2015}
\bibinfo{author}{A.~J. Clarke}, \bibinfo{author}{D.~Tourret},
  \bibinfo{author}{S.~D. Imhoff}, \bibinfo{author}{P.~J. Gibbs},
  \bibinfo{author}{K.~Fezzaa}, \bibinfo{author}{J.~C. Cooley},
  \bibinfo{author}{W.-K. Lee}, \bibinfo{author}{A.~Deriy},
  \bibinfo{author}{B.~M. Patterson}, \bibinfo{author}{P.~A. Papin},
  \bibinfo{author}{K.~D. Clarke}, \bibinfo{author}{R.~D. Field},
  \bibinfo{author}{J.~L. Smith},
\newblock \bibinfo{title}{X-ray imaging and controlled solidification of
  {Al-Cu} alloys toward microstructures by design},
\newblock \bibinfo{journal}{Advanced Engineering Materials}
  \bibinfo{volume}{17} (\bibinfo{year}{2015}) \bibinfo{pages}{454--459}.
\bibitem[{Shahani et~al.(2020)Shahani, Xiao, Lauridsen, and
  Voorhees}]{shahani_characterization_2020}
\bibinfo{author}{A.~J. Shahani}, \bibinfo{author}{X.~Xiao},
  \bibinfo{author}{E.~M. Lauridsen}, \bibinfo{author}{P.~W. Voorhees},
\newblock \bibinfo{title}{Characterization of metals in four dimensions},
\newblock \bibinfo{journal}{Materials Research Letters} \bibinfo{volume}{8}
  (\bibinfo{year}{2020}) \bibinfo{pages}{462--476}.
\bibitem[{Patterson et~al.(2018)Patterson, Cordes, Henderson, Xiao, and
  Chawla}]{patterson_data_2018}
\bibinfo{author}{B.~M. Patterson}, \bibinfo{author}{N.~L. Cordes},
  \bibinfo{author}{K.~Henderson}, \bibinfo{author}{X.~Xiao},
  \bibinfo{author}{N.~Chawla},
\newblock \bibinfo{title}{Data challenges of in situ x-ray tomography for
  materials discovery and characterization},
\newblock in: \bibinfo{editor}{T.~Lookman}, \bibinfo{editor}{S.~Eidenbenz},
  \bibinfo{editor}{F.~Alexander}, \bibinfo{editor}{C.~Barnes} (Eds.),
  \bibinfo{booktitle}{Materials Discovery and Design: By Means of Data Science
  and Optimal Learning}, Springer Series in Materials Science,
  \bibinfo{publisher}{Springer International Publishing}, \bibinfo{year}{2018},
  pp. \bibinfo{pages}{129--165}. \DOIprefix\doi{10.1007/978-3-319-99465-9_6}.
\bibitem[{Maire and Withers(2014)}]{maire_quantitative_2014}
\bibinfo{author}{E.~Maire}, \bibinfo{author}{P.~J. Withers},
\newblock \bibinfo{title}{Quantitative {X-ray} tomography},
\newblock \bibinfo{journal}{International Materials Reviews}
  \bibinfo{volume}{59} (\bibinfo{year}{2014}) \bibinfo{pages}{1--43}.
\bibitem[{Withers et~al.(2021)Withers, Bouman, Carmignato, Cnudde, Grimaldi,
  Hagen, Maire, Manley, Du~Plessis, and Stock}]{withers_xray_2021}
\bibinfo{author}{P.~J. Withers}, \bibinfo{author}{C.~Bouman},
  \bibinfo{author}{S.~Carmignato}, \bibinfo{author}{V.~Cnudde},
  \bibinfo{author}{D.~Grimaldi}, \bibinfo{author}{C.~K. Hagen},
  \bibinfo{author}{E.~Maire}, \bibinfo{author}{M.~Manley},
  \bibinfo{author}{A.~Du~Plessis}, \bibinfo{author}{S.~R. Stock},
\newblock \bibinfo{title}{{X-ray} computed tomography},
\newblock \bibinfo{journal}{Nature Reviews Methods Primers} \bibinfo{volume}{1}
  (\bibinfo{year}{2021}) \bibinfo{pages}{1--21}.
\bibitem[{Ourdjini et~al.(1994)Ourdjini, Liu, and
  Elliott}]{ourdjini_eutectic_1994}
\bibinfo{author}{A.~Ourdjini}, \bibinfo{author}{J.~Liu},
  \bibinfo{author}{R.~Elliott},
\newblock \bibinfo{title}{Eutectic spacing selection in {Al\mbox{–}Cu}
  system},
\newblock \bibinfo{journal}{Materials Science and Technology}
  \bibinfo{volume}{10} (\bibinfo{year}{1994}) \bibinfo{pages}{312--318}.
\bibitem[{Çadirli et~al.(1999)Çadirli, Ülgen, and
  Gündüz}]{cadirli_directional_1999}
\bibinfo{author}{E.~Çadirli}, \bibinfo{author}{A.~Ülgen},
  \bibinfo{author}{M.~Gündüz},
\newblock \bibinfo{title}{Directional solidification of the {Aluminium-Copper}
  eutectic alloy},
\newblock \bibinfo{journal}{Materials Transactions, {JIM}} \bibinfo{volume}{40}
  (\bibinfo{year}{1999}) \bibinfo{pages}{989--996}.
\bibitem[{Walker et~al.(2007)Walker, Liu, Lee, and
  Trivedi}]{walker_eutectic_2007}
\bibinfo{author}{H.~Walker}, \bibinfo{author}{S.~Liu}, \bibinfo{author}{J.~H.
  Lee}, \bibinfo{author}{R.~Trivedi},
\newblock \bibinfo{title}{Eutectic growth in three dimensions},
\newblock \bibinfo{journal}{Metallurgical and Materials Transactions A}
  \bibinfo{volume}{38} (\bibinfo{year}{2007}) \bibinfo{pages}{1417--1425}.
\bibitem[{Murphy et~al.(2013)Murphy, Browne, Mirihanage, and
  Mathiesen}]{murphy_combined_2013}
\bibinfo{author}{A.~G. Murphy}, \bibinfo{author}{D.~J. Browne},
  \bibinfo{author}{W.~U. Mirihanage}, \bibinfo{author}{R.~H. Mathiesen},
\newblock \bibinfo{title}{Combined in situ {X-ray} radiographic observations
  and post-solidification metallographic characterisation of eutectic
  transformations in {Al\mbox{-}Cu} alloy systems},
\newblock \bibinfo{journal}{Acta Materialia} \bibinfo{volume}{61}
  (\bibinfo{year}{2013}) \bibinfo{pages}{4559--4571}.
\bibitem[{Luo et~al.(2019)Luo, Wei, Li, and Wang}]{luo_research_2019}
\bibinfo{author}{L.~Luo}, \bibinfo{author}{C.~Wei}, \bibinfo{author}{X.-M. Li},
  \bibinfo{author}{X.~Wang},
\newblock \bibinfo{title}{Research on microstructure evolution of
  {Al\mbox{-}Al\textsubscript{2}Cu} eutectic by regional melting under
  directional solidification},
\newblock \bibinfo{journal}{Crystal Research and Technology}
  \bibinfo{volume}{54} (\bibinfo{year}{2019}) \bibinfo{pages}{1900108}.
\bibitem[{Kakitani et~al.(2019)Kakitani, Gouveia, Garcia, Cheung, and
  Spinelli}]{kakitani_thermal_2019}
\bibinfo{author}{R.~Kakitani}, \bibinfo{author}{G.~L.~d. Gouveia},
  \bibinfo{author}{A.~Garcia}, \bibinfo{author}{N.~Cheung},
  \bibinfo{author}{J.~E. Spinelli},
\newblock \bibinfo{title}{Thermal analysis during solidification of an
  {Al\mbox{-}Cu} eutectic alloy: interrelation of thermal parameters,
  microstructure and hardness},
\newblock \bibinfo{journal}{Journal of Thermal Analysis and Calorimetry}
  \bibinfo{volume}{137} (\bibinfo{year}{2019}) \bibinfo{pages}{983--996}.
\bibitem[{Sutton and Balluffi(1995)}]{sutton_interfaces_1995}
\bibinfo{author}{A.~P. Sutton}, \bibinfo{author}{R.~Balluffi},
  \bibinfo{title}{Interfaces in crystalline materials}, Monographs on the
  physics and chemistry of materials ;51, \bibinfo{publisher}{Clarendon Press ;
  Oxford University}, \bibinfo{year}{1995}.
\bibitem[{Howe(1997)}]{howe_interfaces_1997}
\bibinfo{author}{J.~M. Howe}, \bibinfo{title}{Interfaces in materials: atomic
  structure, thermodynamics and kinetics of solid-vapor, solid-liquid and
  solid-solid interfaces}, \bibinfo{publisher}{Wiley}, \bibinfo{year}{1997}.
\bibitem[{Kraft(1962)}]{kraft_crystallography_1962}
\bibinfo{author}{R.~W. Kraft},
\newblock \bibinfo{title}{Crystallography of equilibrium phase interfaces in
  {Al\mbox{-}CuAl\textsubscript{2}} eutectic alloys},
\newblock \bibinfo{journal}{Transactions of The Metallurgical Society of
  {AIME}} \bibinfo{volume}{224} (\bibinfo{year}{1962}) \bibinfo{pages}{65--75}.
\bibitem[{Bonnet and Durand(1973)}]{bonnet_geometric_1973}
\bibinfo{author}{R.~Bonnet}, \bibinfo{author}{F.~Durand},
\newblock \bibinfo{title}{Geometric discussion of the relationships between the
  phases {Al} and {CuAl\textsubscript{2}} for the eutectic and precipitates of
  {CuAl\textsubscript{2}}},
\newblock in: \bibinfo{booktitle}{Conference on in situ Composites}, volume
  \bibinfo{volume}{308-I} of \textit{\bibinfo{series}{Conference on in situ
  Composites}}, \bibinfo{year}{1973}, pp. \bibinfo{pages}{209--223}.
\bibitem[{Cantor and Chadwick(1974)}]{cantor_growth_1974}
\bibinfo{author}{B.~Cantor}, \bibinfo{author}{G.~A. Chadwick},
\newblock \bibinfo{title}{The growth crystallography of unidirectionally
  solidified {Al\mbox{-}Al\textsubscript{3}Ni} and
  {Al\mbox{-}Al\textsubscript{2}Cu} eutectics},
\newblock \bibinfo{journal}{Journal of Crystal Growth} \bibinfo{volume}{23}
  (\bibinfo{year}{1974}) \bibinfo{pages}{12--20}.
\bibitem[{Hecht et~al.(2012)Hecht, Witusiewicz, and
  Drevermann}]{hecht_coupled_2012}
\bibinfo{author}{U.~Hecht}, \bibinfo{author}{V.~Witusiewicz},
  \bibinfo{author}{A.~Drevermann},
\newblock \bibinfo{title}{Coupled growth of {Al\mbox{-}Al\textsubscript{2}Cu}
  eutectics in {Al\mbox{-}Cu\mbox{-}Ag} alloys},
\newblock \bibinfo{journal}{{IOP} Conference Series: Materials Science and
  Engineering} \bibinfo{volume}{27} (\bibinfo{year}{2012})
  \bibinfo{pages}{012029}.
\bibitem[{Kokotin and Hecht(2014)}]{kokotin_molecular_2014}
\bibinfo{author}{V.~Kokotin}, \bibinfo{author}{U.~Hecht},
\newblock \bibinfo{title}{Molecular dynamics simulations of
  {Al\mbox{-}Al\textsubscript{2}Cu} phase boundaries},
\newblock \bibinfo{journal}{Computational Materials Science}
  \bibinfo{volume}{86} (\bibinfo{year}{2014}) \bibinfo{pages}{30--37}.
\bibitem[{Wang et~al.(2018)Wang, Liu, Wang, and
  Misra}]{wang_characteristic_2018}
\bibinfo{author}{S.~J. Wang}, \bibinfo{author}{G.~Liu},
  \bibinfo{author}{J.~Wang}, \bibinfo{author}{A.~Misra},
\newblock \bibinfo{title}{Characteristic orientation relationships in nanoscale
  {Al\mbox{-}Al\textsubscript{2}Cu} eutectic},
\newblock \bibinfo{journal}{Materials Characterization} \bibinfo{volume}{142}
  (\bibinfo{year}{2018}) \bibinfo{pages}{170--178}.
\bibitem[{Hecht et~al.(2019)Hecht, Eiken, Akamatsu, and
  Bottin-Rousseau}]{hecht_phase_2019}
\bibinfo{author}{U.~Hecht}, \bibinfo{author}{J.~Eiken},
  \bibinfo{author}{S.~Akamatsu}, \bibinfo{author}{S.~Bottin-Rousseau},
\newblock \bibinfo{title}{Phase boundary anisotropy and its effects on the
  maze\mbox{-}to\mbox{-}lamellar transition in a directionally solidified
  {Al\mbox{-}Al\textsubscript{2}Cu} eutectic},
\newblock \bibinfo{journal}{Acta Materialia} \bibinfo{volume}{170}
  (\bibinfo{year}{2019}) \bibinfo{pages}{268--277}.
\bibitem[{Liotti et~al.(2018)Liotti, Arteta, Zisserman, Lui, Lempitsky, and
  Grant}]{liotti_crystal_2018}
\bibinfo{author}{E.~Liotti}, \bibinfo{author}{C.~Arteta},
  \bibinfo{author}{A.~Zisserman}, \bibinfo{author}{A.~Lui},
  \bibinfo{author}{V.~Lempitsky}, \bibinfo{author}{P.~S. Grant},
\newblock \bibinfo{title}{Crystal nucleation in metallic alloys using x-ray
  radiography and machine learning},
\newblock \bibinfo{journal}{Science Advances} \bibinfo{volume}{4}
  (\bibinfo{year}{2018}) \bibinfo{pages}{eaar4004}.
\bibitem[{Wang et~al.(2021)Wang, Gao, Chao, Muyanja, Mathiesen, and
  Shahani}]{wang_insitu_2021}
\bibinfo{author}{Y.~Wang}, \bibinfo{author}{J.~Gao}, \bibinfo{author}{P.~Chao},
  \bibinfo{author}{N.~S. Muyanja}, \bibinfo{author}{R.~H. Mathiesen},
  \bibinfo{author}{A.~J. Shahani},
\newblock \bibinfo{title}{{In-situ} evidence for impurity-induced formation of
  eutectic colonies in an interdendritic liquid},
\newblock \bibinfo{journal}{Materials Letters} \bibinfo{volume}{292}
  (\bibinfo{year}{2021}).
\bibitem[{Wang et~al.(2022)Wang, Gao, Sun, and Shahani}]{wang_situ_2022}
\bibinfo{author}{Y.~Wang}, \bibinfo{author}{J.~Gao}, \bibinfo{author}{W.~Sun},
  \bibinfo{author}{A.~J. Shahani},
\newblock \bibinfo{title}{In situ observation of faceted growth and
  morphological instability of a complex-regular eutectic in {Zn-Mg-Al}
  system},
\newblock \bibinfo{journal}{Scripta Materialia} \bibinfo{volume}{206}
  (\bibinfo{year}{2022}) \bibinfo{pages}{114224}.
\bibitem[{Han et~al.(2019)Han, Xiao, Sun, and Shahani}]{han2019side}
\bibinfo{author}{I.~Han}, \bibinfo{author}{X.~Xiao}, \bibinfo{author}{H.~Sun},
  \bibinfo{author}{A.~J. Shahani},
\newblock \bibinfo{title}{A side-by-side comparison of the solidification
  dynamics of quasicrystalline and approximant phases in the al--co--ni
  system},
\newblock \bibinfo{journal}{Acta Crystallographica Section A: Foundations and
  Advances} \bibinfo{volume}{75} (\bibinfo{year}{2019})
  \bibinfo{pages}{281--296}.
\bibitem[{Moniri et~al.(2019)Moniri, Xiao, and Shahani}]{moniri_mechanism_2019}
\bibinfo{author}{S.~Moniri}, \bibinfo{author}{X.~Xiao}, \bibinfo{author}{A.~J.
  Shahani},
\newblock \bibinfo{title}{The mechanism of eutectic modification by trace
  impurities},
\newblock \bibinfo{journal}{Scientific Reports} \bibinfo{volume}{9}
  (\bibinfo{year}{2019}) \bibinfo{pages}{1--13}.
\bibitem[{Moniri et~al.(2020)Moniri, Xiao, and
  Shahani}]{moniri_singlytwinned_2020}
\bibinfo{author}{S.~Moniri}, \bibinfo{author}{X.~Xiao}, \bibinfo{author}{A.~J.
  Shahani},
\newblock \bibinfo{title}{Singly-twinned growth of {Si} crystals upon chemical
  modification},
\newblock \bibinfo{journal}{Physical Review Materials} \bibinfo{volume}{4}
  (\bibinfo{year}{2020}) \bibinfo{pages}{063403}. \bibinfo{note}{Publisher:
  American Physical Society}.
\bibitem[{Reese et~al.(2021)Reese, Gladstein, Shevchenko, Xiao, Shahani, and
  Taub}]{reese_realtime_2021}
\bibinfo{author}{C.~W. Reese}, \bibinfo{author}{A.~Gladstein},
  \bibinfo{author}{P.~Shevchenko}, \bibinfo{author}{X.~Xiao},
  \bibinfo{author}{A.~J. Shahani}, \bibinfo{author}{A.~I. Taub},
\newblock \bibinfo{title}{Real-time visualization of particle evolution during
  reactive flux-assisted processing of aluminum melts},
\newblock \bibinfo{journal}{Scripta Materialia} \bibinfo{volume}{201}
  (\bibinfo{year}{2021}) \bibinfo{pages}{113978}.
\bibitem[{Deville et~al.(2009{\natexlab{a}})Deville, Maire, Lasalle, Bogner,
  Gauthier, Leloup, and Guizard}]{deville_situ_2009_1}
\bibinfo{author}{S.~Deville}, \bibinfo{author}{E.~Maire},
  \bibinfo{author}{A.~Lasalle}, \bibinfo{author}{A.~Bogner},
  \bibinfo{author}{C.~Gauthier}, \bibinfo{author}{J.~Leloup},
  \bibinfo{author}{C.~Guizard},
\newblock \bibinfo{title}{In situ x-ray radiography and tomography observations
  of the solidification of aqueous alumina particle suspensions—part {I}:
  Initial instants},
\newblock \bibinfo{journal}{Journal of the American Ceramic Society}
  \bibinfo{volume}{92} (\bibinfo{year}{2009}{\natexlab{a}})
  \bibinfo{pages}{2489--2496}.
\bibitem[{Deville et~al.(2009{\natexlab{b}})Deville, Maire, Lasalle, Bogner,
  Gauthier, Leloup, and Guizard}]{deville_situ_2009_2}
\bibinfo{author}{S.~Deville}, \bibinfo{author}{E.~Maire},
  \bibinfo{author}{A.~Lasalle}, \bibinfo{author}{A.~Bogner},
  \bibinfo{author}{C.~Gauthier}, \bibinfo{author}{J.~Leloup},
  \bibinfo{author}{C.~Guizard},
\newblock \bibinfo{title}{In situ x-ray radiography and tomography observations
  of the solidification of aqueous alumina particles suspensions. part {II}:
  Steady state},
\newblock \bibinfo{journal}{Journal of the American Ceramic Society}
  \bibinfo{volume}{92} (\bibinfo{year}{2009}{\natexlab{b}})
  \bibinfo{pages}{2497--2503}.
\bibitem[{Mathworks(2020)}]{MATLAB}
Mathworks, \bibinfo{title}{{MATLAB Image Processing Toolbox Reference
  (R2020b)}}, \bibinfo{organization}{The Mathworks, Inc.},
  \bibinfo{address}{Natick, Massachusetts}, \bibinfo{year}{2020}.
\bibitem[{Buffet et~al.(2010)Buffet, Nguyen-Thi, Bogno, Schenk,
  Mangelinck-No{\"e}l, Reinhart, Bergeon, Billia, and
  Baruchel}]{buffet2010measurement}
\bibinfo{author}{A.~Buffet}, \bibinfo{author}{H.~Nguyen-Thi},
  \bibinfo{author}{A.~Bogno}, \bibinfo{author}{T.~Schenk},
  \bibinfo{author}{N.~Mangelinck-No{\"e}l}, \bibinfo{author}{G.~Reinhart},
  \bibinfo{author}{N.~Bergeon}, \bibinfo{author}{B.~Billia},
  \bibinfo{author}{J.~Baruchel},
\newblock \bibinfo{title}{Measurement of solute profiles by means of
  synchrotron x-ray radiography during directional solidification of al-4 wt\%
  cu alloys},
\newblock in: \bibinfo{booktitle}{Materials Science Forum}, volume
  \bibinfo{volume}{649}, \bibinfo{organization}{Trans Tech Publ},
  \bibinfo{year}{2010}, pp. \bibinfo{pages}{331--336}.
\bibitem[{Soltani et~al.(2020)Soltani, Reinhart, Benoudia, Ngomesse, Zahzouh,
  and Nguyen-Thi}]{soltani2020impact}
\bibinfo{author}{H.~Soltani}, \bibinfo{author}{G.~Reinhart},
  \bibinfo{author}{M.~Benoudia}, \bibinfo{author}{F.~Ngomesse},
  \bibinfo{author}{M.~Zahzouh}, \bibinfo{author}{H.~Nguyen-Thi},
\newblock \bibinfo{title}{Impact of growth velocity on grain structure
  formation during directional solidification of a refined al-20 wt.\% cu
  alloy},
\newblock \bibinfo{journal}{Journal of Crystal Growth} \bibinfo{volume}{548}
  (\bibinfo{year}{2020}) \bibinfo{pages}{125819}.
\bibitem[{Tandjaoui et~al.(2013)Tandjaoui, Mangelinck-Noel, Reinhart, Billia,
  and Guichard}]{tandjaoui_twinning_2013}
\bibinfo{author}{A.~Tandjaoui}, \bibinfo{author}{N.~Mangelinck-Noel},
  \bibinfo{author}{G.~Reinhart}, \bibinfo{author}{B.~Billia},
  \bibinfo{author}{X.~Guichard},
\newblock \bibinfo{title}{Twinning occurrence and grain competition in
  multi-crystalline silicon during solidification},
\newblock \bibinfo{journal}{Comptes Rendus Physique} \bibinfo{volume}{14}
  (\bibinfo{year}{2013}) \bibinfo{pages}{141--148}.
\bibitem[{Andrew et~al.(2017)Andrew, Bhattiprolu, Butnaru, and
  Correa}]{andrew_usage_2017}
\bibinfo{author}{M.~Andrew}, \bibinfo{author}{S.~Bhattiprolu},
  \bibinfo{author}{D.~Butnaru}, \bibinfo{author}{J.~Correa},
\newblock \bibinfo{title}{The usage of modern data science in segmentation and
  classification: Machine learning and microscopy},
\newblock \bibinfo{journal}{Microscopy and Microanalysis} \bibinfo{volume}{23}
  (\bibinfo{year}{2017}) \bibinfo{pages}{156--157}.
\bibitem[{Volkenandt et~al.(2018)Volkenandt, Freitag, and
  Rauscher}]{volkenandt_machine_2018}
\bibinfo{author}{T.~Volkenandt}, \bibinfo{author}{S.~Freitag},
  \bibinfo{author}{M.~Rauscher},
\newblock \bibinfo{title}{Machine learning powered image segmentation},
\newblock \bibinfo{journal}{Microscopy and Microanalysis} \bibinfo{volume}{24}
  (\bibinfo{year}{2018}) \bibinfo{pages}{520--521}.
\bibitem[{Gürsoy et~al.(2014)Gürsoy, De~Carlo, Xiao, and
  Jacobsen}]{gursoy_tomopy_2014}
\bibinfo{author}{D.~Gürsoy}, \bibinfo{author}{F.~De~Carlo},
  \bibinfo{author}{X.~Xiao}, \bibinfo{author}{C.~Jacobsen},
\newblock \bibinfo{title}{{TomoPy}: a framework for the analysis of
  synchrotron tomographic data},
\newblock \bibinfo{journal}{Journal of Synchrotron Radiation}
  \bibinfo{volume}{21} (\bibinfo{year}{2014}) \bibinfo{pages}{1188--1193}.
\bibitem[{Vo et~al.(2018)Vo, Atwood, and Drakopoulos}]{vo_superior_2018}
\bibinfo{author}{N.~T. Vo}, \bibinfo{author}{R.~C. Atwood},
  \bibinfo{author}{M.~Drakopoulos},
\newblock \bibinfo{title}{Superior techniques for eliminating ring artifacts in
  x-ray micro-tomography},
\newblock \bibinfo{journal}{Optics Express} \bibinfo{volume}{26}
  (\bibinfo{year}{2018}) \bibinfo{pages}{28396--28412}.
\bibitem[{Dowd et~al.(1999)Dowd, Campbell, Marr, Nagarkar, Tipnis, Axe, and
  Siddons}]{dowd_developments_1999}
\bibinfo{author}{B.~A. Dowd}, \bibinfo{author}{G.~H. Campbell},
  \bibinfo{author}{R.~B. Marr}, \bibinfo{author}{V.~V. Nagarkar},
  \bibinfo{author}{S.~V. Tipnis}, \bibinfo{author}{L.~Axe},
  \bibinfo{author}{D.~P. Siddons},
\newblock \bibinfo{title}{Developments in synchrotron x-ray computed
  microtomography at the {National Synchrotron Light Source}},
\newblock in: \bibinfo{booktitle}{Developments in X-Ray Tomography {II}},
  volume \bibinfo{volume}{3772}, \bibinfo{publisher}{{SPIE}},
  \bibinfo{year}{1999}, pp. \bibinfo{pages}{224--236}.
  \DOIprefix\doi{10.1117/12.363725}.
\bibitem[{Kocks et~al.(1998)Kocks, Tomé, and Wenk}]{kocks_texture_1998}
\bibinfo{author}{U.~F. Kocks}, \bibinfo{author}{C.~N. Tomé},
  \bibinfo{author}{H.-R. Wenk}, \bibinfo{title}{{Texture and Anisotropy}:
  Preferred Orientations in Polycrystals and their Effect on Materials
  Properties}, \bibinfo{publisher}{Cambridge University Press},
  \bibinfo{year}{1998}.
\bibitem[{Engler and Randle(2009)}]{engler_introduction_2009}
\bibinfo{author}{O.~Engler}, \bibinfo{author}{V.~Randle},
  \bibinfo{title}{Introduction to Texture Analysis: Macrotexture, Microtexture,
  and Orientation Mapping, Second Edition}, \bibinfo{edition}{2} ed.,
  \bibinfo{publisher}{{CRC} Press}, \bibinfo{year}{2009}.
  \DOIprefix\doi{10.1201/9781420063660}.
\bibitem[{Bachmann et~al.(2010)Bachmann, Hielscher, and
  Schaeben}]{bachmann_texture_2010}
\bibinfo{author}{F.~Bachmann}, \bibinfo{author}{R.~Hielscher},
  \bibinfo{author}{H.~Schaeben},
\newblock \bibinfo{title}{Texture analysis with {MTEX} – free and open source
  software toolbox},
\newblock \bibinfo{journal}{Solid State Phenomena} \bibinfo{volume}{160}
  (\bibinfo{year}{2010}) \bibinfo{pages}{63--68}.
\bibitem[{Rohrer et~al.(2004)Rohrer, Saylor, Dasher, Adams, Rollett, and
  Wynblatt}]{rohrer_distribution_2004}
\bibinfo{author}{G.~S. Rohrer}, \bibinfo{author}{D.~M. Saylor},
  \bibinfo{author}{B.~E. Dasher}, \bibinfo{author}{B.~L. Adams},
  \bibinfo{author}{A.~D. Rollett}, \bibinfo{author}{P.~Wynblatt},
\newblock \bibinfo{title}{The distribution of internal interfaces in
  polycrystals},
\newblock \bibinfo{journal}{Zeitschrift für Metallkunde} \bibinfo{volume}{95}
  (\bibinfo{year}{2004}) \bibinfo{pages}{197--214}.
\bibitem[{Rowenhorst et~al.(2016)Rowenhorst, Gupta, Feng, and
  Spanos}]{rowenhorst_3d_2006}
\bibinfo{author}{D.~Rowenhorst}, \bibinfo{author}{A.~Gupta},
  \bibinfo{author}{C.~Feng}, \bibinfo{author}{G.~Spanos},
\newblock \bibinfo{title}{{3D} crystallographic and morphological analysis of
  coarse martensite: Combining {EBSD} and serial sectioning},
\newblock \bibinfo{journal}{Scripta Materialia} \bibinfo{volume}{55}
  (\bibinfo{year}{2016}) \bibinfo{pages}{11--16}.
\bibitem[{Larson et~al.(2013)Larson, Prosa, Ulfig, Geiser, and
  Kelly}]{larson2013local}
\bibinfo{author}{D.~J. Larson}, \bibinfo{author}{T.~Prosa},
  \bibinfo{author}{R.~M. Ulfig}, \bibinfo{author}{B.~P. Geiser},
  \bibinfo{author}{T.~F. Kelly},
\newblock \bibinfo{title}{Local electrode atom probe tomography},
\newblock \bibinfo{journal}{New York, US: Springer Science} \bibinfo{volume}{2}
  (\bibinfo{year}{2013}).
\bibitem[{Mjaavatten(2021)}]{mjaavatten_curvature}
\bibinfo{author}{A.~Mjaavatten}, \bibinfo{title}{Curvature of a 1d curve in a
  2d or 3d space}, \bibinfo{year}{2021}. \URLprefix
  \url{https://www.mathworks.com/matlabcentral/fileexchange/69452-curvature-of-a-1d-curve-in-a-2d-or-3d-space}.
\bibitem[{Mathiesen et~al.(2011)Mathiesen, Arnberg, Li, Meier, Schaffer,
  Snigireva, Snigirev, and Dahle}]{mathiesen_xray_2011}
\bibinfo{author}{R.~H. Mathiesen}, \bibinfo{author}{L.~Arnberg},
  \bibinfo{author}{Y.~Li}, \bibinfo{author}{V.~Meier}, \bibinfo{author}{P.~L.
  Schaffer}, \bibinfo{author}{I.~Snigireva}, \bibinfo{author}{A.~Snigirev},
  \bibinfo{author}{A.~K. Dahle},
\newblock \bibinfo{title}{X-ray videomicroscopy studies of eutectic {Al-Si}
  solidification in {Al-Si-Cu}},
\newblock \bibinfo{journal}{Metallurgical and Materials Transactions A}
  \bibinfo{volume}{42} (\bibinfo{year}{2011}) \bibinfo{pages}{170--180}.
\bibitem[{Reinhart et~al.(2020)Reinhart, Grange, Abou-Khalil, Mangelinck-Noël,
  Niane, Maguin, Guillemot, Gandin, and Nguyen-Thi}]{reinhart_impact_2020}
\bibinfo{author}{G.~Reinhart}, \bibinfo{author}{D.~Grange},
  \bibinfo{author}{L.~Abou-Khalil}, \bibinfo{author}{N.~Mangelinck-Noël},
  \bibinfo{author}{N.~T. Niane}, \bibinfo{author}{V.~Maguin},
  \bibinfo{author}{G.~Guillemot}, \bibinfo{author}{C.~A. Gandin},
  \bibinfo{author}{H.~Nguyen-Thi},
\newblock \bibinfo{title}{Impact of solute flow during directional
  solidification of a {Ni}-based alloy: In-situ and real-time x-radiography},
\newblock \bibinfo{journal}{Acta Materialia} \bibinfo{volume}{194}
  (\bibinfo{year}{2020}) \bibinfo{pages}{68--79}.
\bibitem[{Kalidindi(2015)}]{kalidindi_3_2015}
\bibinfo{author}{S.~R. Kalidindi},
\newblock \bibinfo{title}{Hierarchical materials informatics},
\newblock in: \bibinfo{editor}{S.~R. Kalidindi} (Ed.),
  \bibinfo{booktitle}{Hierarchical Materials Informatics},
  \bibinfo{publisher}{Butterworth-Heinemann}, \bibinfo{year}{2015}, pp.
  \bibinfo{pages}{75--110}. \DOIprefix\doi{10.1016/B978-0-12-410394-8.00003-5}.
\bibitem[{Cecen et~al.(2018)Cecen, Yabansu, and Kalidindi}]{cecen_new_2018}
\bibinfo{author}{A.~Cecen}, \bibinfo{author}{Y.~C. Yabansu},
  \bibinfo{author}{S.~R. Kalidindi},
\newblock \bibinfo{title}{A new framework for rotationally invariant two-point
  spatial correlations in microstructure datasets},
\newblock \bibinfo{journal}{Acta Materialia} \bibinfo{volume}{158}
  (\bibinfo{year}{2018}) \bibinfo{pages}{53--64}.
\bibitem[{Parisi and Plapp(2010)}]{parisi_defects_2010}
\bibinfo{author}{A.~Parisi}, \bibinfo{author}{M.~Plapp},
\newblock \bibinfo{title}{Defects and multistability in eutectic solidification
  patterns},
\newblock \bibinfo{journal}{{EPL} (Europhysics Letters)} \bibinfo{volume}{90}
  (\bibinfo{year}{2010}) \bibinfo{pages}{26010}.
\bibitem[{Hellman et~al.(2000)Hellman, Vandenbroucke, R{\"u}sing, Isheim, and
  Seidman}]{hellman2000analysis}
\bibinfo{author}{O.~C. Hellman}, \bibinfo{author}{J.~A. Vandenbroucke},
  \bibinfo{author}{J.~R{\"u}sing}, \bibinfo{author}{D.~Isheim},
  \bibinfo{author}{D.~N. Seidman},
\newblock \bibinfo{title}{Analysis of three-dimensional atom-probe data by the
  proximity histogram},
\newblock \bibinfo{journal}{Microscopy and Microanalysis} \bibinfo{volume}{6}
  (\bibinfo{year}{2000}) \bibinfo{pages}{437--444}.
\bibitem[{Fujii et~al.(1993)Fujii, Nakae, and Okada}]{fujii1993interfacial}
\bibinfo{author}{H.~Fujii}, \bibinfo{author}{H.~Nakae},
  \bibinfo{author}{K.~Okada},
\newblock \bibinfo{title}{Interfacial reaction wetting in the boron
  nitride/molten aluminum system},
\newblock \bibinfo{journal}{Acta Metallurgica et Materialia}
  \bibinfo{volume}{41} (\bibinfo{year}{1993}) \bibinfo{pages}{2963--2971}.
\bibitem[{Du et~al.(2003)Du, Chang, Huang, Gong, Jin, Xu, Yuan, Liu, He, and
  Xie}]{du_diffusion_2003}
\bibinfo{author}{Y.~Du}, \bibinfo{author}{Y.~A. Chang},
  \bibinfo{author}{B.~Huang}, \bibinfo{author}{W.~Gong},
  \bibinfo{author}{Z.~Jin}, \bibinfo{author}{H.~Xu}, \bibinfo{author}{Z.~Yuan},
  \bibinfo{author}{Y.~Liu}, \bibinfo{author}{Y.~He}, \bibinfo{author}{F.~Y.
  Xie},
\newblock \bibinfo{title}{Diffusion coefficients of some solutes in fcc and
  liquid al: critical evaluation and correlation},
\newblock \bibinfo{journal}{Materials Science and Engineering: A}
  \bibinfo{volume}{363} (\bibinfo{year}{2003}) \bibinfo{pages}{140--151}.
\bibitem[{Akamatsu and Faivre(2000)}]{akamatsu_traveling_2000}
\bibinfo{author}{S.~Akamatsu}, \bibinfo{author}{G.~Faivre},
\newblock \bibinfo{title}{Traveling waves, two-phase fingers, and eutectic
  colonies in thin-sample directional solidification of a ternary eutectic
  alloy},
\newblock \bibinfo{journal}{Physical Review E} \bibinfo{volume}{61}
  (\bibinfo{year}{2000}) \bibinfo{pages}{3757--3770}.
\bibitem[{Bogno et~al.(2011)Bogno, Reinhart, Buffet, Nguyen~Thi, Billia,
  Schenk, Mangelinck-Noël, Bergeon, and Baruchel}]{bogno_situ_2011}
\bibinfo{author}{A.~Bogno}, \bibinfo{author}{G.~Reinhart},
  \bibinfo{author}{A.~Buffet}, \bibinfo{author}{H.~Nguyen~Thi},
  \bibinfo{author}{B.~Billia}, \bibinfo{author}{T.~Schenk},
  \bibinfo{author}{N.~Mangelinck-Noël}, \bibinfo{author}{N.~Bergeon},
  \bibinfo{author}{J.~Baruchel},
\newblock \bibinfo{title}{In situ analysis of the influence of convection
  during the initial transient of planar solidification},
\newblock \bibinfo{journal}{Journal of Crystal Growth} \bibinfo{volume}{318}
  (\bibinfo{year}{2011}) \bibinfo{pages}{1134--1138}.
\bibitem[{Lee et~al.(2005)Lee, Liu, and Trivedi}]{lee_effect_2005}
\bibinfo{author}{J.~H. Lee}, \bibinfo{author}{S.~Liu},
  \bibinfo{author}{R.~Trivedi},
\newblock \bibinfo{title}{The effect of fluid flow on eutectic growth},
\newblock \bibinfo{journal}{Metallurgical and Materials Transactions A}
  \bibinfo{volume}{36A} (\bibinfo{year}{2005}) \bibinfo{pages}{3111}.
\bibitem[{McFadden et~al.(1984)McFadden, Rehm, Coriell, Chuck, and
  Morrish}]{mcfadden1984thermosolutal}
\bibinfo{author}{G.~McFadden}, \bibinfo{author}{R.~Rehm},
  \bibinfo{author}{S.~Coriell}, \bibinfo{author}{W.~Chuck},
  \bibinfo{author}{K.~Morrish},
\newblock \bibinfo{title}{Thermosolutal convection during directional
  solidification},
\newblock \bibinfo{journal}{Metallurgical Transactions A} \bibinfo{volume}{15}
  (\bibinfo{year}{1984}) \bibinfo{pages}{2125--2137}.
\bibitem[{Carlberg(1984)}]{carlberg_effect_1984}
\bibinfo{author}{T.~Carlberg},
\newblock \bibinfo{title}{The effect of convection upon off-eutectic composite
  growth of {Al-Cu} alloys},
\newblock \bibinfo{journal}{Journal of Crystal Growth} \bibinfo{volume}{66}
  (\bibinfo{year}{1984}) \bibinfo{pages}{106--120}.
\bibitem[{De~Wilde and Froyen(2005)}]{de_wilde_solutal_2005}
\bibinfo{author}{J.~De~Wilde}, \bibinfo{author}{L.~Froyen},
\newblock \bibinfo{title}{Solutal configuration during coupled two-phase
  {[$\alpha$-(Al) + $\theta$-Al\textsubscript{2}Cu]} planar univariant eutetic
  growth in {Al-Cu-(Ag, Si)} ternary eutetic alloys},
\newblock \bibinfo{journal}{Microgravity - Science and Technology}
  \bibinfo{volume}{16} (\bibinfo{year}{2005}) \bibinfo{pages}{40--44}.
\bibitem[{Li et~al.(2007)Li, Ren, and Fautrelle}]{li2007spiral}
\bibinfo{author}{X.~Li}, \bibinfo{author}{Z.~Ren},
  \bibinfo{author}{Y.~Fautrelle},
\newblock \bibinfo{title}{The spiral growth of lamellar eutectics in a high
  magnetic field during the directional solidification process},
\newblock \bibinfo{journal}{Scripta materialia} \bibinfo{volume}{56}
  (\bibinfo{year}{2007}) \bibinfo{pages}{505--508}.
\bibitem[{Li et~al.(2010)Li, Ren, Fautrelle, Zhang, and
  Esling}]{li2010morphological}
\bibinfo{author}{X.~Li}, \bibinfo{author}{Z.~Ren},
  \bibinfo{author}{Y.~Fautrelle}, \bibinfo{author}{Y.~Zhang},
  \bibinfo{author}{C.~Esling},
\newblock \bibinfo{title}{Morphological instabilities and alignment of lamellar
  eutectics during directional solidification under a strong magnetic field},
\newblock \bibinfo{journal}{Acta Materialia} \bibinfo{volume}{58}
  (\bibinfo{year}{2010}) \bibinfo{pages}{1403--1417}.
\bibitem[{Van~Alkemade(1893)}]{van1893graphical}
\bibinfo{author}{A.~V.~R. Van~Alkemade},
\newblock \bibinfo{title}{Graphical treatment of some thermodynamic problems
  with equilibrium states of salt solutions with solid phases},
\newblock \bibinfo{journal}{Z. Phys. Chem} \bibinfo{volume}{11}
  (\bibinfo{year}{1893}) \bibinfo{pages}{289--327}.
\bibitem[{Rumball(1968)}]{rumball1968cellular}
\bibinfo{author}{W.~Rumball},
\newblock \bibinfo{title}{Cellular growth in eutectic systems},
\newblock \bibinfo{journal}{Metallurgia} \bibinfo{volume}{78}
  (\bibinfo{year}{1968}) \bibinfo{pages}{141--145}.
\bibitem[{Lahiri and Choudhury(2015)}]{lahiri_effect_2015}
\bibinfo{author}{A.~Lahiri}, \bibinfo{author}{A.~Choudhury},
\newblock \bibinfo{title}{Effect of surface energy anisotropy on the stability
  of growth fronts in multiphase alloys},
\newblock \bibinfo{journal}{Transactions of the Indian Institute of Metals}
  \bibinfo{volume}{68} (\bibinfo{year}{2015}) \bibinfo{pages}{1053--1057}.
\bibitem[{Lahiri and Choudhury(2017)}]{lahiri_revisiting_2017}
\bibinfo{author}{A.~Lahiri}, \bibinfo{author}{A.~Choudhury},
\newblock \bibinfo{title}{Revisiting {Jackson-Hunt} calculations: Unified
  theoretical analysis for generic multi-phase growth in a multi-component
  system},
\newblock \bibinfo{journal}{Acta Materialia} \bibinfo{volume}{133}
  (\bibinfo{year}{2017}) \bibinfo{pages}{316--332}.
\bibitem[{De~Wilde et~al.(2005)De~Wilde, Froyen, Witusiewicz, and
  Hecht}]{de2005two}
\bibinfo{author}{J.~De~Wilde}, \bibinfo{author}{L.~Froyen},
  \bibinfo{author}{V.~Witusiewicz}, \bibinfo{author}{U.~Hecht},
\newblock \bibinfo{title}{Two-phase planar and regular lamellar coupled growth
  along the univariant eutectic reaction in ternary alloys: An analytical
  approach and application to the {Al-Cu-Ag} system},
\newblock \bibinfo{journal}{Journal of applied physics} \bibinfo{volume}{97}
  (\bibinfo{year}{2005}) \bibinfo{pages}{113515}.
\bibitem[{Witusiewicz et~al.(2021)Witusiewicz, Hecht, Fernandez, Rodriguez, and
  Ezquerro}]{witusiewicz_insitu_2021}
\bibinfo{author}{V.~T. Witusiewicz}, \bibinfo{author}{U.~Hecht},
  \bibinfo{author}{J.~Fernandez}, \bibinfo{author}{J.~Rodriguez},
  \bibinfo{author}{J.~M. Ezquerro},
\newblock \bibinfo{title}{In-situ observation of eutectic growth during
  directional solidification of succinonitrile - {(D)}camphor- neopentyl glycol
  alloys under imposed velocity transients},
\newblock \bibinfo{journal}{Acta Materialia} \bibinfo{volume}{203}
  (\bibinfo{year}{2021}) \bibinfo{pages}{116469}.
\bibitem[{{\c{S}}erefoǧlu et~al.(2012){\c{S}}erefoǧlu, Bottin-Rousseau,
  Akamatsu, and Faivre}]{cserefoglu2012dynamics}
\bibinfo{author}{M.~{\c{S}}erefoǧlu}, \bibinfo{author}{S.~Bottin-Rousseau},
  \bibinfo{author}{S.~Akamatsu}, \bibinfo{author}{G.~Faivre},
\newblock \bibinfo{title}{Dynamics of rod eutectic growth patterns in confined
  geometry},
\newblock in: \bibinfo{booktitle}{IOP Conference Series: Materials Science and
  Engineering}, volume \bibinfo{volume}{27-1}, \bibinfo{organization}{IOP
  Publishing}, \bibinfo{year}{2012}, p. \bibinfo{pages}{012030}.
\bibitem[{Parisi(2008)}]{parisi_stability_2008}
\bibinfo{author}{A.~Parisi},
\newblock \bibinfo{title}{Stability of lamellar eutectic growth},
\newblock \bibinfo{journal}{Acta Materialia} \bibinfo{volume}{56}
  (\bibinfo{year}{2008}) \bibinfo{pages}{1348--1357}.
\bibitem[{Rayleigh(1878)}]{rayleigh1878instability}
\bibinfo{author}{L.~Rayleigh},
\newblock \bibinfo{title}{On the instability of jets},
\newblock \bibinfo{journal}{Proceedings of the London mathematical society}
  \bibinfo{volume}{1} (\bibinfo{year}{1878}) \bibinfo{pages}{4--13}.
\bibitem[{Bottin-Rousseau et~al.(2022)Bottin-Rousseau, Witusiewicz, Hecht,
  Fernandez, Laveron-Simavilla, and
  Akamatsu}]{bottin-rousseau_coexistence_2022}
\bibinfo{author}{S.~Bottin-Rousseau}, \bibinfo{author}{V.~T. Witusiewicz},
  \bibinfo{author}{U.~Hecht}, \bibinfo{author}{J.~Fernandez},
  \bibinfo{author}{A.~Laveron-Simavilla}, \bibinfo{author}{S.~Akamatsu},
\newblock \bibinfo{title}{Coexistence of rod-like and lamellar eutectic growth
  patterns},
\newblock \bibinfo{journal}{Scripta Materialia} \bibinfo{volume}{207}
  (\bibinfo{year}{2022}) \bibinfo{pages}{114314}.
\bibitem[{Akamatsu et~al.(2007)Akamatsu, Bottin-Rousseau, Perrut, Faivre,
  Witusiewicz, and Sturz}]{akamatsu2007real}
\bibinfo{author}{S.~Akamatsu}, \bibinfo{author}{S.~Bottin-Rousseau},
  \bibinfo{author}{M.~Perrut}, \bibinfo{author}{G.~Faivre},
  \bibinfo{author}{V.~Witusiewicz}, \bibinfo{author}{L.~Sturz},
\newblock \bibinfo{title}{Real-time study of thin and bulk eutectic growth in
  succinonitrile--(d) camphor alloys},
\newblock \bibinfo{journal}{Journal of crystal growth} \bibinfo{volume}{299}
  (\bibinfo{year}{2007}) \bibinfo{pages}{418--428}.
\bibitem[{Santala and
  Glaeser(2006)}]{santala_surfaceenergyanisotropyinduced_2006}
\bibinfo{author}{M.~K. Santala}, \bibinfo{author}{A.~M. Glaeser},
\newblock \bibinfo{title}{Surface-energy-anisotropy-induced orientation effects
  on rayleigh instabilities in sapphire},
\newblock \bibinfo{journal}{Surface Science} \bibinfo{volume}{600}
  (\bibinfo{year}{2006}) \bibinfo{pages}{782--792}.
\bibitem[{Bottin-Rousseau et~al.(2021)Bottin-Rousseau, Medjkoune, Senninger,
  Carroz, Soucek, Hecht, and Akamatsu}]{bottinrousseau_lockedlamellar_2021}
\bibinfo{author}{S.~Bottin-Rousseau}, \bibinfo{author}{M.~Medjkoune},
  \bibinfo{author}{O.~Senninger}, \bibinfo{author}{L.~Carroz},
  \bibinfo{author}{R.~Soucek}, \bibinfo{author}{U.~Hecht},
  \bibinfo{author}{S.~Akamatsu},
\newblock \bibinfo{title}{Locked-lamellar eutectic growth in thin
  {Al–Al\textsubscript{2}Cu} samples: In situ directional solidification and
  crystal orientation analysis},
\newblock \bibinfo{journal}{Journal of Crystal Growth} \bibinfo{volume}{570}
  (\bibinfo{year}{2021}) \bibinfo{pages}{126203}.
\bibitem[{Garmong and Rhodes(1974{\natexlab{a}})}]{garmong_interfacial_1974}
\bibinfo{author}{G.~Garmong}, \bibinfo{author}{C.~G. Rhodes},
\newblock \bibinfo{title}{Interfacial structure of {Al}-{CuAl}\textsubscript{2}
  eutectic composites},
\newblock \bibinfo{journal}{Acta Metallurgica} \bibinfo{volume}{22}
  (\bibinfo{year}{1974}{\natexlab{a}}) \bibinfo{pages}{1373--1382}.
\bibitem[{Garmong and Rhodes(1974{\natexlab{b}})}]{garmong_structure_1974}
\bibinfo{author}{G.~Garmong}, \bibinfo{author}{C.~G. Rhodes},
\newblock \bibinfo{title}{The structure of interphase boundaries in
  {Al}-{CuAl}\textsubscript{2} curved eutectic crystals},
\newblock \bibinfo{journal}{Metallurgical and Materials Transactions B}
  \bibinfo{volume}{5} (\bibinfo{year}{1974}{\natexlab{b}})
  \bibinfo{pages}{2507--2513}.
\bibitem[{Kaya and Smith(1992)}]{kaya_transmission_1992}
\bibinfo{author}{M.~Kaya}, \bibinfo{author}{R.~W. Smith},
\newblock \bibinfo{title}{Transmission electron microscope study of a
  directionally solidified {Cu-MgCu\textsubscript{2}} eutectic},
\newblock \bibinfo{journal}{Journal of Materials Science} \bibinfo{volume}{27}
  (\bibinfo{year}{1992}) \bibinfo{pages}{2258--2266}.
\bibitem[{Biswas et~al.(2010)Biswas, Siegel, and
  Seidman}]{biswas_simultaneous_2010}
\bibinfo{author}{A.~Biswas}, \bibinfo{author}{D.~J. Siegel},
  \bibinfo{author}{D.~N. Seidman},
\newblock \bibinfo{title}{Simultaneous segregation at coherent and semicoherent
  heterophase interfaces},
\newblock \bibinfo{journal}{Physical Review Letters} \bibinfo{volume}{105}
  (\bibinfo{year}{2010}) \bibinfo{pages}{076102}.
\bibitem[{Asta et~al.(2009)Asta, Beckermann, Karma, Kurz, Napolitano, Plapp,
  Purdy, Rappaz, and Trivedi}]{asta_solidification_2009}
\bibinfo{author}{M.~Asta}, \bibinfo{author}{C.~Beckermann},
  \bibinfo{author}{A.~Karma}, \bibinfo{author}{W.~Kurz},
  \bibinfo{author}{R.~Napolitano}, \bibinfo{author}{M.~Plapp},
  \bibinfo{author}{G.~Purdy}, \bibinfo{author}{M.~Rappaz},
  \bibinfo{author}{R.~Trivedi},
\newblock \bibinfo{title}{Solidification microstructures and solid-state
  parallels: Recent developments, future directions},
\newblock \bibinfo{journal}{Acta Materialia} \bibinfo{volume}{57}
  (\bibinfo{year}{2009}) \bibinfo{pages}{941--971}.
\bibitem[{Rappaz and Kurtuldu(2015)}]{rappaz_thermodynamic_2015}
\bibinfo{author}{M.~Rappaz}, \bibinfo{author}{G.~Kurtuldu},
\newblock \bibinfo{title}{Thermodynamic aspects of homogeneous nucleation
  enhanced by icosahedral short range order in liquid fcc-type alloys},
\newblock \bibinfo{journal}{{JOM}} \bibinfo{volume}{67} (\bibinfo{year}{2015})
  \bibinfo{pages}{1812--1820}.

\end{thebibliography}

\clearpage
\newpage

\section*{Figures}

\begin{figure}[h!]
\includegraphics[width=\textwidth]{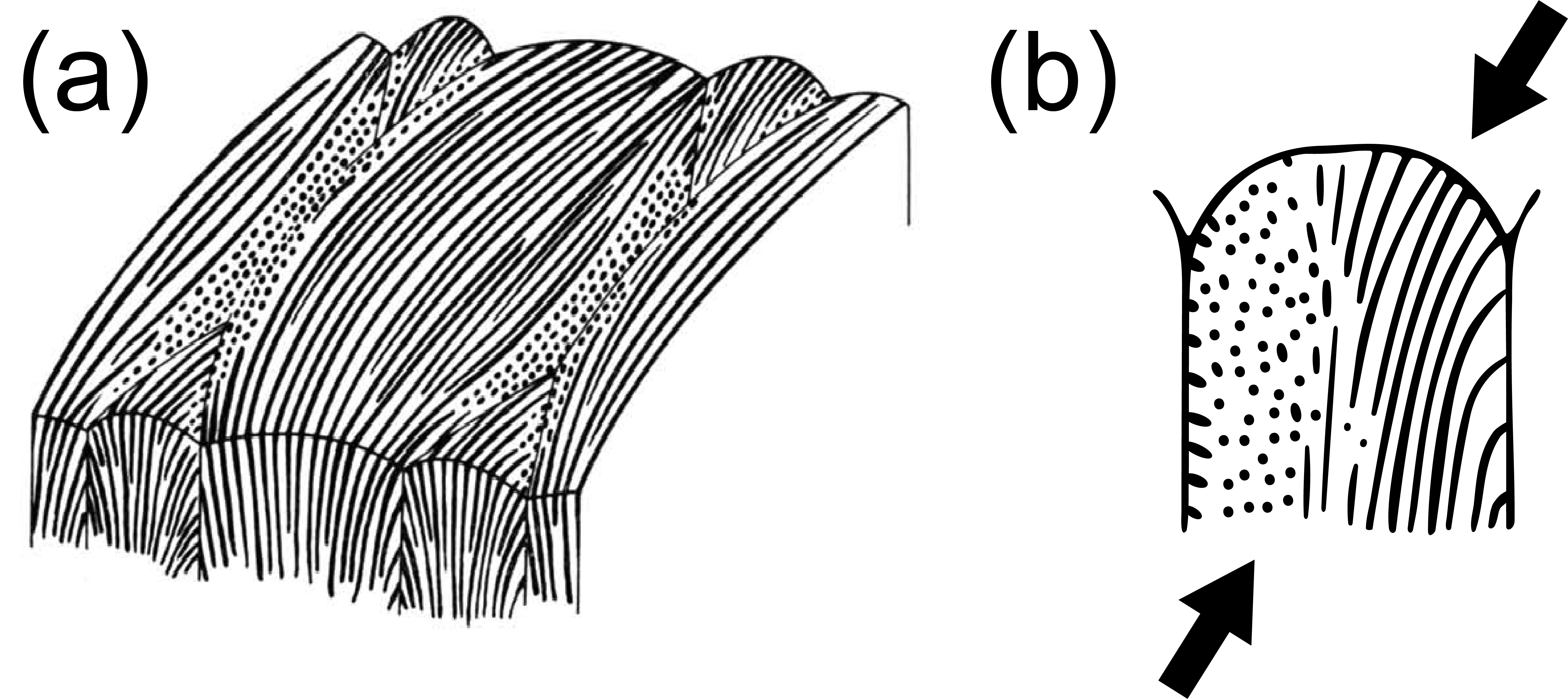}
\caption{\textit{Proposed structure of eutectic colonies}. (a) According to Chadwick \cite{chadwick_modification_1962}, constitutional undercooling due to impurities induces not only a colony structure but also triggers rod formation at the edges of the colony. This is where the impurity concentration is thought to be higher, and hence also the constitutional undercooling. (b) Hunt \cite{hunt_lamella_1966} proposed instead that the eutectic is forced to accommodate the curvature of the solid-liquid interface in cells, such that rods are predicted when the local growth direction is away from the low\mbox{-}energy, lamellar plane. See text for details. Adapted with permission from Refs.~\cite{chadwick_modification_1962, hunt_lamella_1966}.}
\label{F1}
\end{figure}

\begin{figure}[h!]
\includegraphics[width=\textwidth]{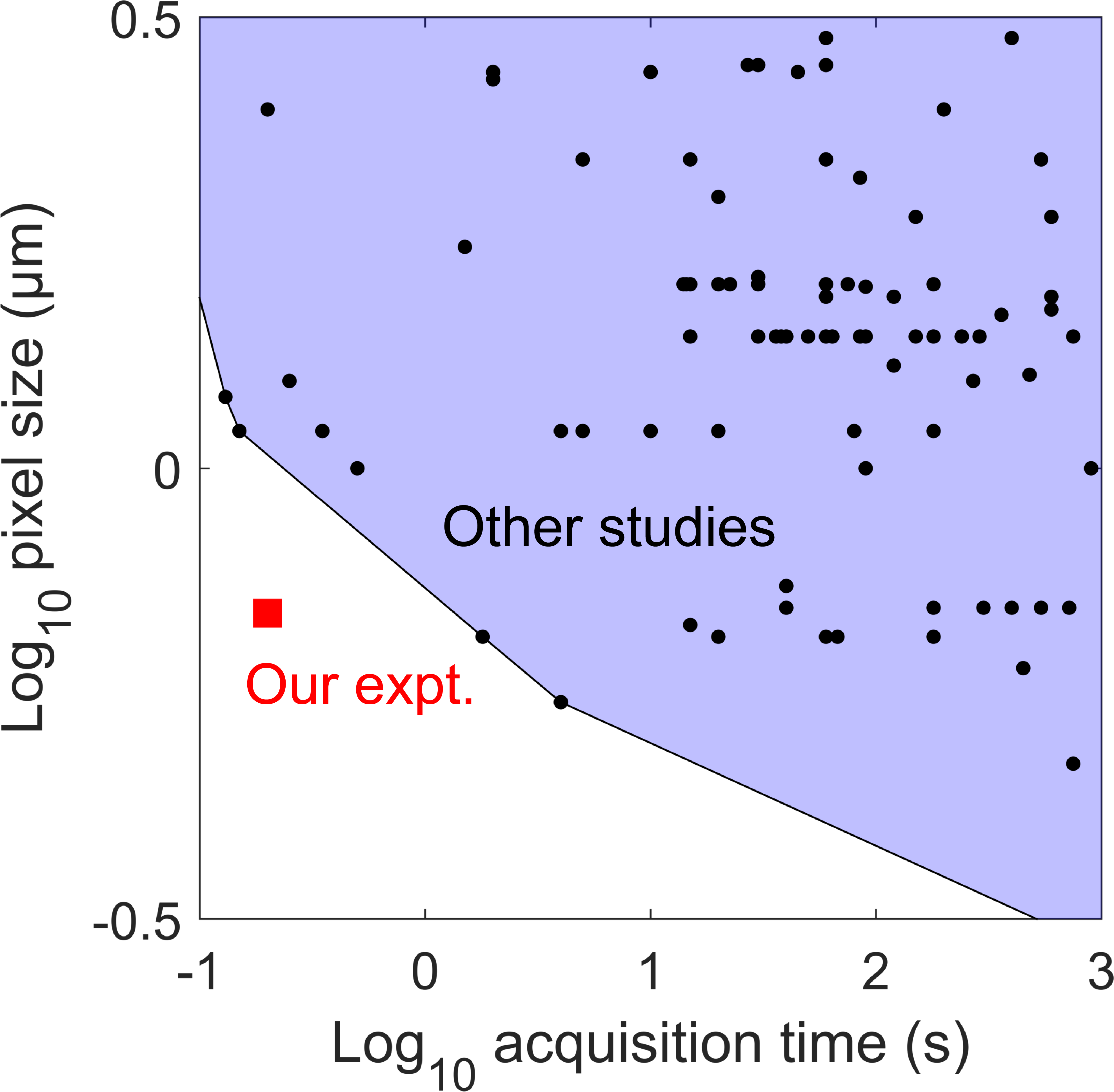}
\caption{\textit{State-of-the-art in 4D imaging}. The present experiment (red) lies outside the space\mbox{-}time manifold (blue) based on a survey of previous works that use \textit{in~situ} 4D X-ray microtomography (black) \cite{shahani_characterization_2020}. In the classical sense, the acquisition time represents not the frame rate, but instead the time to collect a 3D dataset (tomogram). Dividing line calculated from a convex hull of other studies. The study of eutectic solidification at this spatial and temporal resolution is made possible by our novel data fusion approach.}
\label{F2} 
\end{figure}

\begin{figure}[h!]
\includegraphics[width=\textwidth]{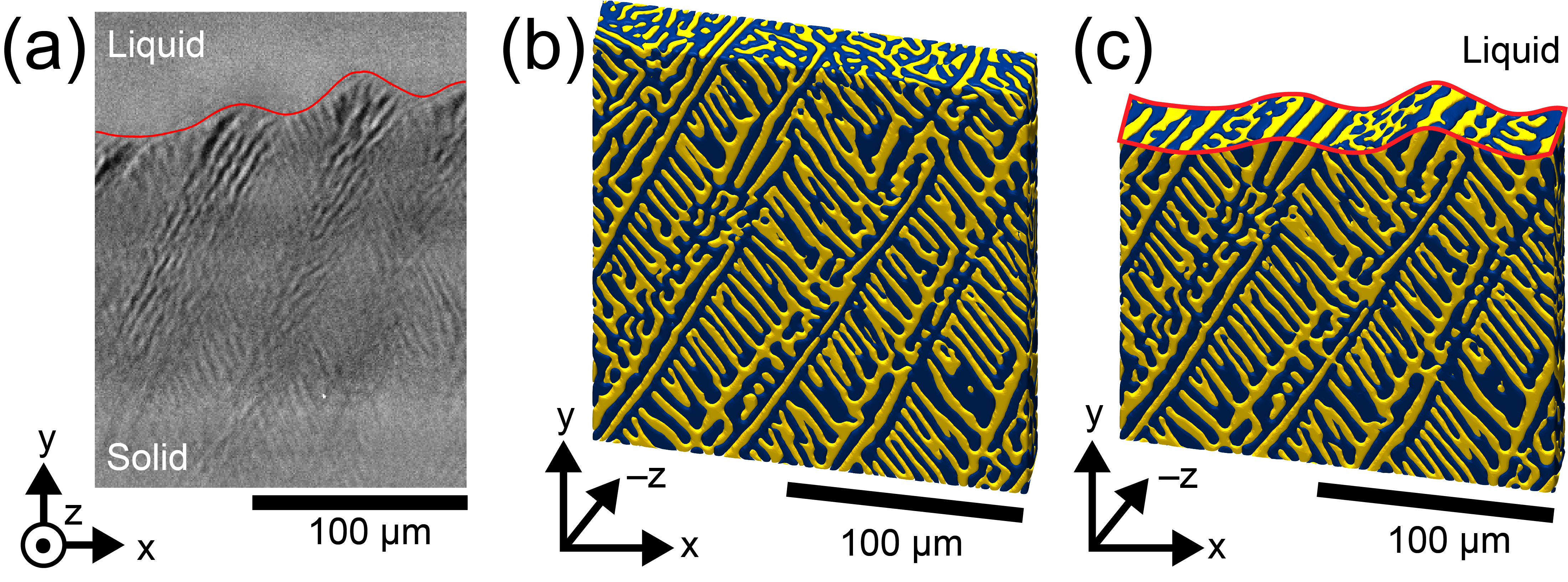}
\caption{\textit{Pseudo\mbox{-}4D approach}, incorporating (a) \textit{in~situ} X\mbox{-}ray projection radiographs to capture the moving solid\mbox{-}liquid interface (red line) and (b) \textit{ex~situ} tomographs (reconstructed volumes), to resolve the microstructure evolution in 4D. Al is rendered in blue and Al\textsubscript{2}Cu in yellow. The coordinates $x$ and $y$ are in the plane of the detector while $z$ is in\mbox{-}line with the X\mbox{-}ray beam. (c) We interpolate the location of the solid\mbox{-}liquid interface (red line) within the tomographic volume to obtain the pseudo\mbox{-}4D dataset.}
\label{F3}
\end{figure}

\begin{figure}[h!]
\includegraphics[width=\textwidth]{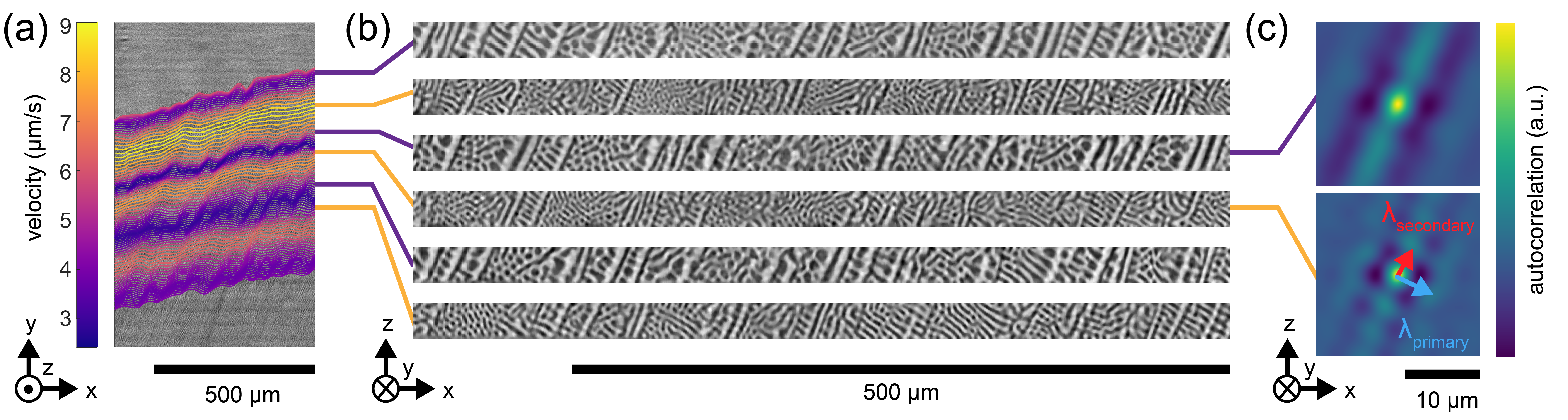}
\caption{\textit{Analysis of X\mbox{-}radiographs}. The eutectic\mbox{-}liquid interface of the eutectic colonies oscillates from 3 to \SI{9} {\micro\meter/\second} within the X\mbox{-}radiography field\mbox{-}of\mbox{-}view. Panel (a) represents a spatiotemporal diagram of eutectic\mbox{-}liquid interfaces from 100 time steps, separated by an interval of 1~s. The corresponding location of the solid\mbox{-}liquid interface within the X\mbox{-}ray tomography data is displayed in (b), depicted from a bird’s eye view (see coordinate system). We visualize the interface at those timesteps where the velocity is maximum and minimum (see lines). A microstructural adjustment to the velocity changes is visually apparent. Al is the darker phase in the cross\mbox{-}sections while Al\textsubscript{2}Cu is brighter. \textcolor{black}{The lamellar spacing $\lambda$ from these cross\mbox{-}sections can be determined by (c) auto-correlation (or self-convolution) functions, which depict the real\mbox{-}space periodicity. We calculate $\lambda$ along two directions, termed ``primary" and ``secondary."} }
\label{F4}
\end{figure}

\begin{figure}[h!]
\includegraphics[width=\textwidth]{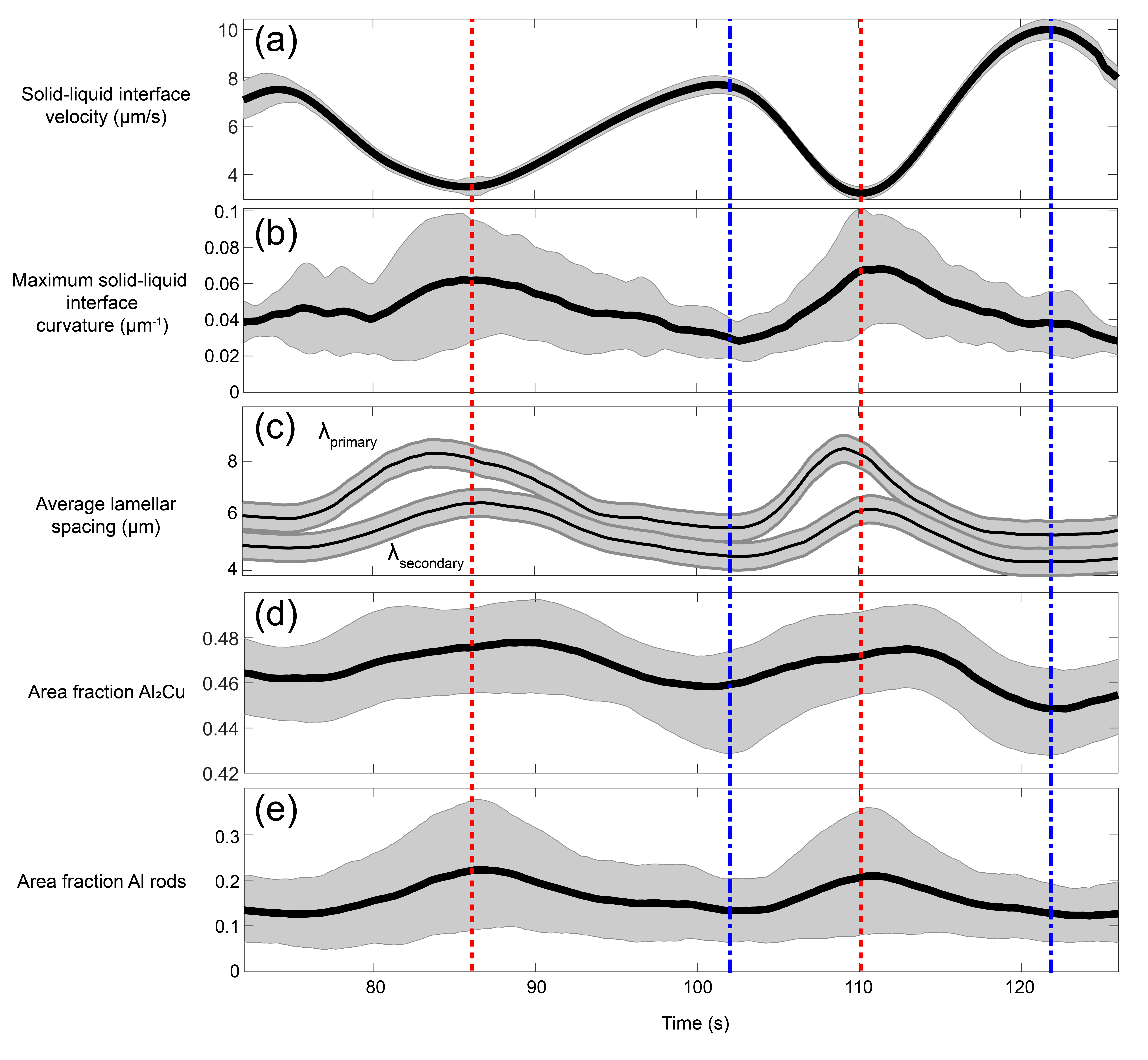}
\caption{\textit{Time\mbox{-}resolved statistics on solidification dynamics}. The eutectic\mbox{-}liquid interface trace in the X\mbox{-}radiography images enables us to (a) quantify the interface velocity, which oscillates in the range 3\mbox{-}\SI{9}{\micro\meter/\second} (see also \textbf{Fig.~\ref{F4}}) and (b) quantify the maximum interfacial curvature (magnitude only). A low velocity corresponds to a high curvature in the eutectic\mbox{-}liquid interface and \textit{vice versa}. After fusing X\mbox{-}radiography and tomography datasets, we can now calculate the following microstructural attributes within the eutectic\mbox{-}liquid interfaces: (c) average lamellar spacing \textcolor{black}{of the primary and secondary lamellae}, (d) area fraction of Al\textsubscript{2}Cu, and (e) area fraction of Al rods. The features in (c\mbox{-}e) vary periodically in time and out\mbox{-}of\mbox{-}phase with the velocity curve in (a). Dotted lines corresponding to the local minimum velocity (in red) and local maximum velocity (blue) are added to guide the reader's eye. Grey bands represent uncertainty intervals. See text for details.}
\label{F5} 
\end{figure}

\begin{figure}[h!]
\includegraphics[width=\textwidth]{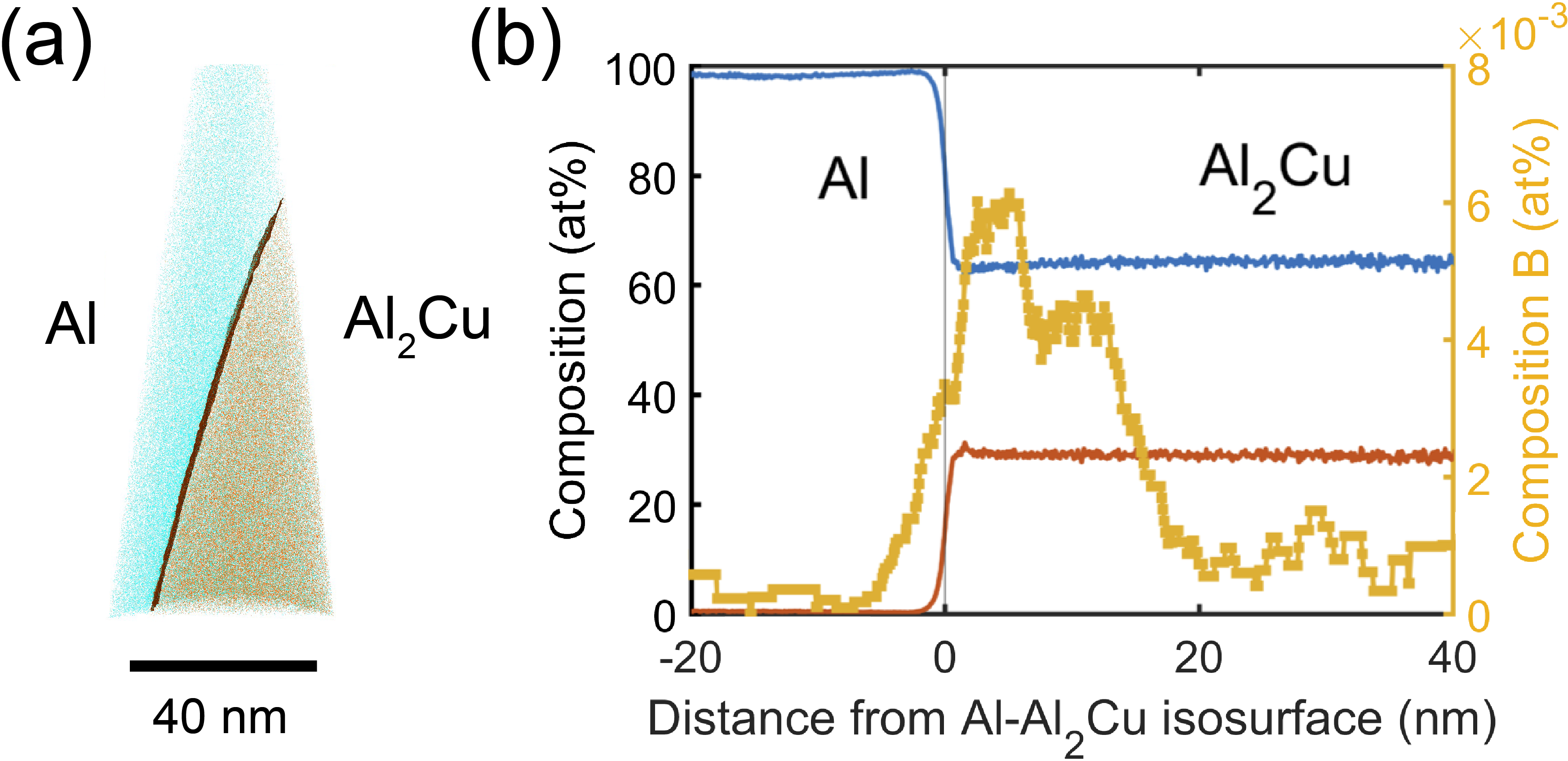}
\caption{\textcolor{black}{\textit{Atom probe tomography of Al\mbox{-}Al\textsubscript{2}Cu interface}. Panel (a) shows the reconstructed tip showing the volume of material \textcolor{black}{extracted from the foil sample}. We place an isosurface at the inflection point of the major element concentration profile to represent the Al\mbox{-}Al\textsubscript{2}Cu interface. Panel (b) shows the composition profile at distances away from the isosurface, revealing an elevated concentration of B near the interface.}}
\label{F13}
\end{figure}

\begin{figure}[h!]
\includegraphics[width=\textwidth]{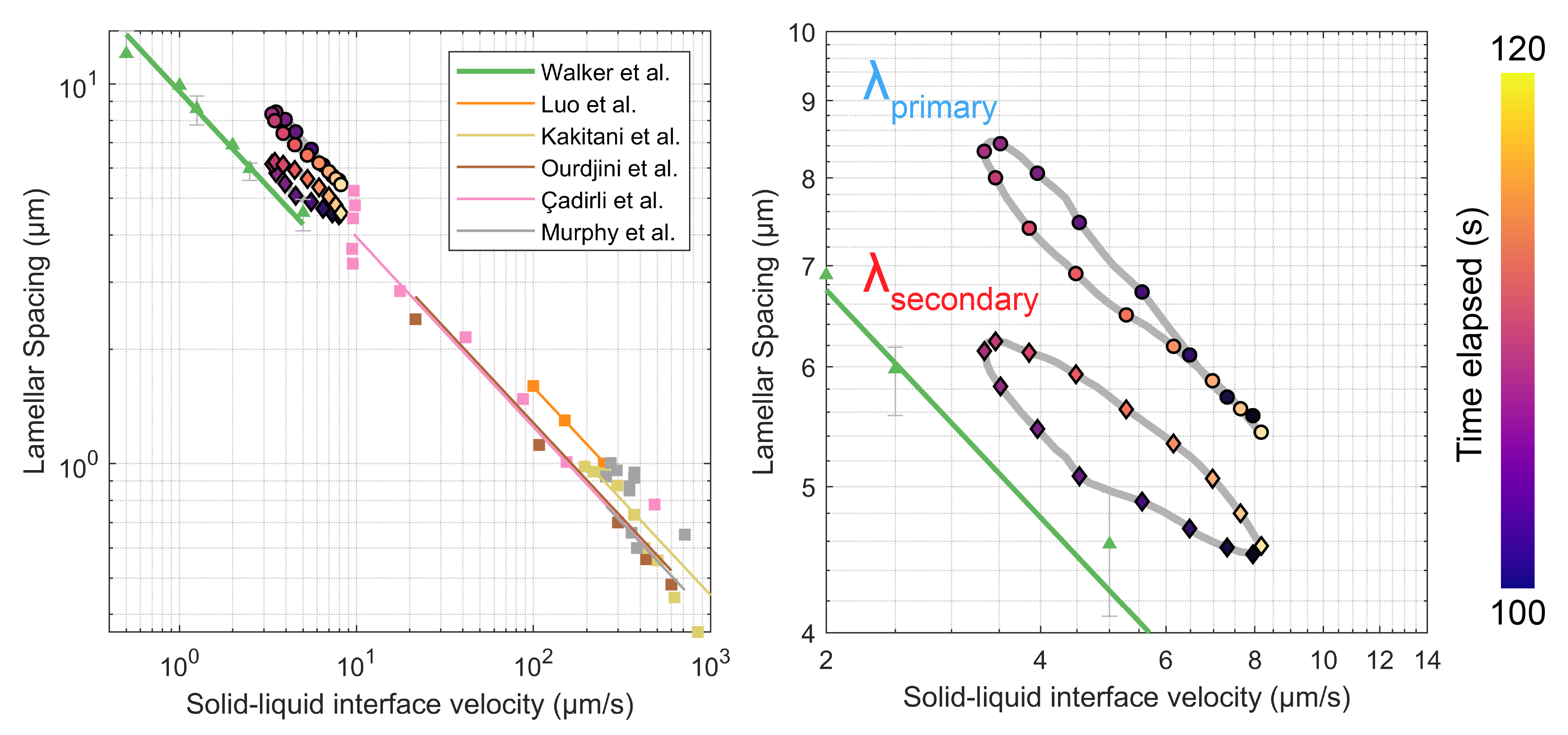}
\caption{\textit{Spacing-velocity scaling behavior}. The \textcolor{black}{average lamellar spacing of the primary and secondary lamella}  shows good agreement with previous studies of this eutectic alloy \cite{,ourdjini_eutectic_1994,cadirli_directional_1999,walker_eutectic_2007,murphy_combined_2013,luo_research_2019,kakitani_thermal_2019} despite the fact that the solidification proceeds under non\mbox{-}steady state conditions and with smoothly curved eutectic\mbox{-}liquid interfaces. The difference in the two curves can be reconciled by accounting for the difference in crystallographic orientations of these solid-solid interfaces, see \textbf{Fig.\mbox{ }\ref{F10}}.}
\label{F6} 
\end{figure}

\begin{figure}[h!]
\includegraphics[width=\textwidth]{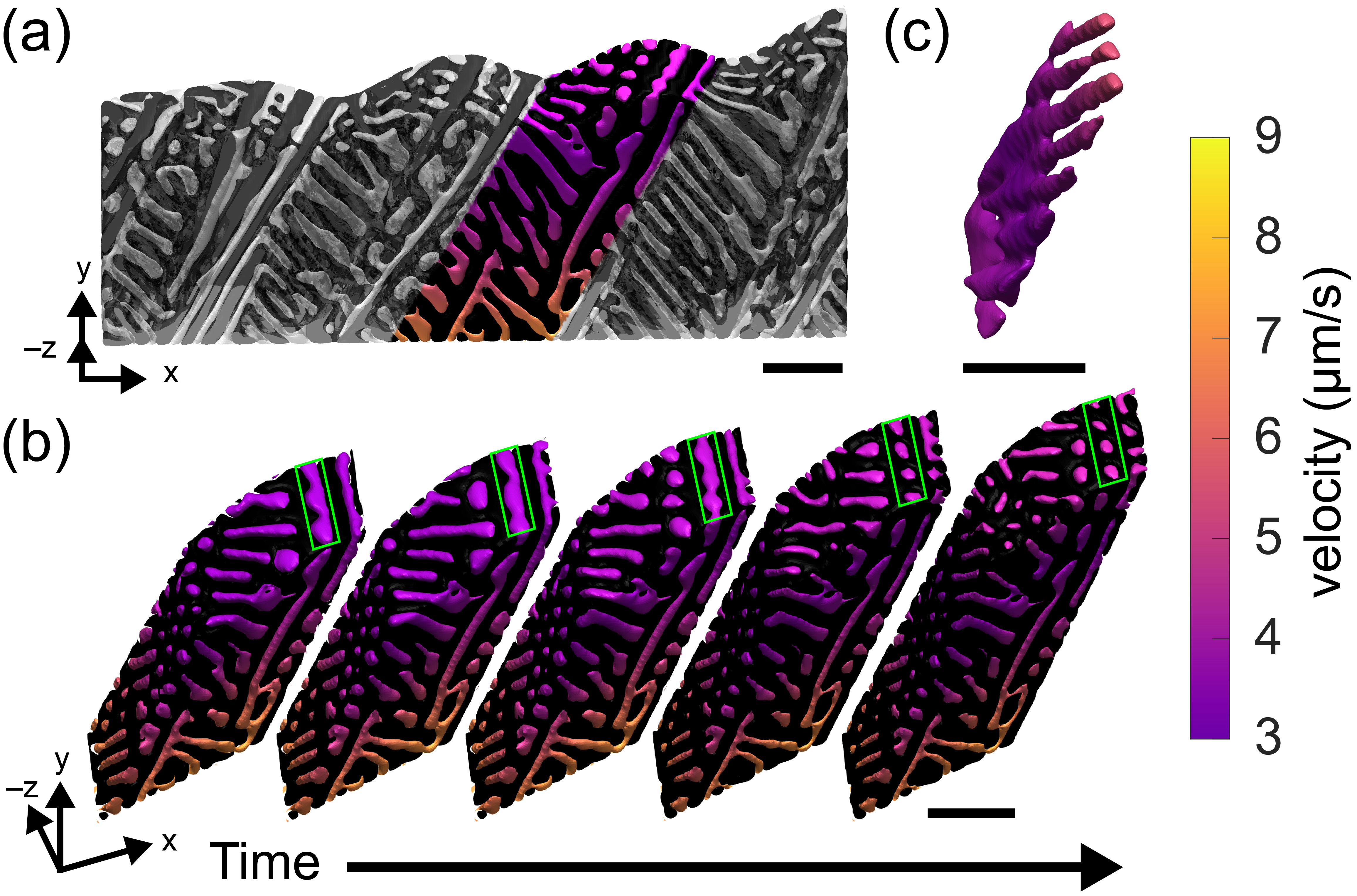}
\caption{\textit{Mechanism of lamella\mbox{-}to\mbox{-}rods transition}: (a) shows a region of the tomographic volume from which we extract a colony\mbox{-}of\mbox{-}interest. The colony is slanted in the $x$\mbox{-}$y$ plane owing to the tilt of the solid\mbox{-}solid interfaces with respect to the macroscopic growth direction. Panel (b) renders this colony in 3D every 1 s to show the evolution of Al lamella to rods (see green boxes). A zoomed\mbox{-}in view is given in (c), which shows the transition within a single Al lamella. Al\textsubscript{2}Cu is depicted in black and Al in color corresponding to the velocity. Scale bar is \SI{20}{\micro\meter} for all panels.}
\label{F7}
\end{figure}

\begin{figure}[h!]
\includegraphics[width=\textwidth]{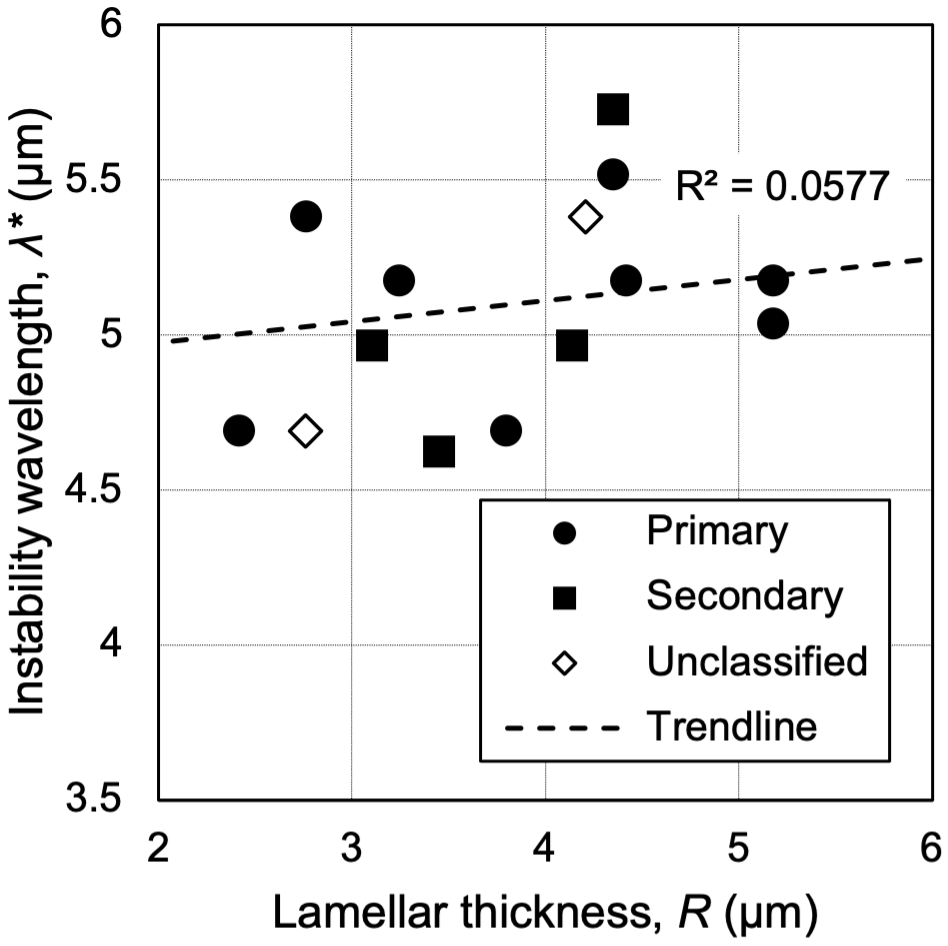}
\caption{\textcolor{black}{\textit{Relation between lamellar thickness and instability wavelength.} The latter is equivalent to the rod spacing at the lamella-to-rod transition. We measure both quantities within a 2~s interval starting at 87.6~s. In this temporal window, we detect 14~lamellae that breakdown to form rods, 12 of which have ``primary" and ``secondary" orientations (as visualized in Fig.~\ref{F10}(c)), and two of which have undetermined or unclassified orientations. The instability wavelength is uncorrelated to lamellar thickness ($R^2 = 0.0577$). Instead, the average wavelength of 5.1~\textpm~0.3~\SI{}{\micro \meter} matches reasonably well the smallest stable spacing of \SI{4.9}{\micro \meter} for an interfacial velocity of \SI{3.8}{\micro \meter/s} \cite{walker_eutectic_2007}, predicted using the Jackson\mbox{-}Hunt model~\cite{Jackson_Hunt_1966}.}}
\label{FX}
\end{figure}

\begin{figure}[h!]
\includegraphics[width=\textwidth]{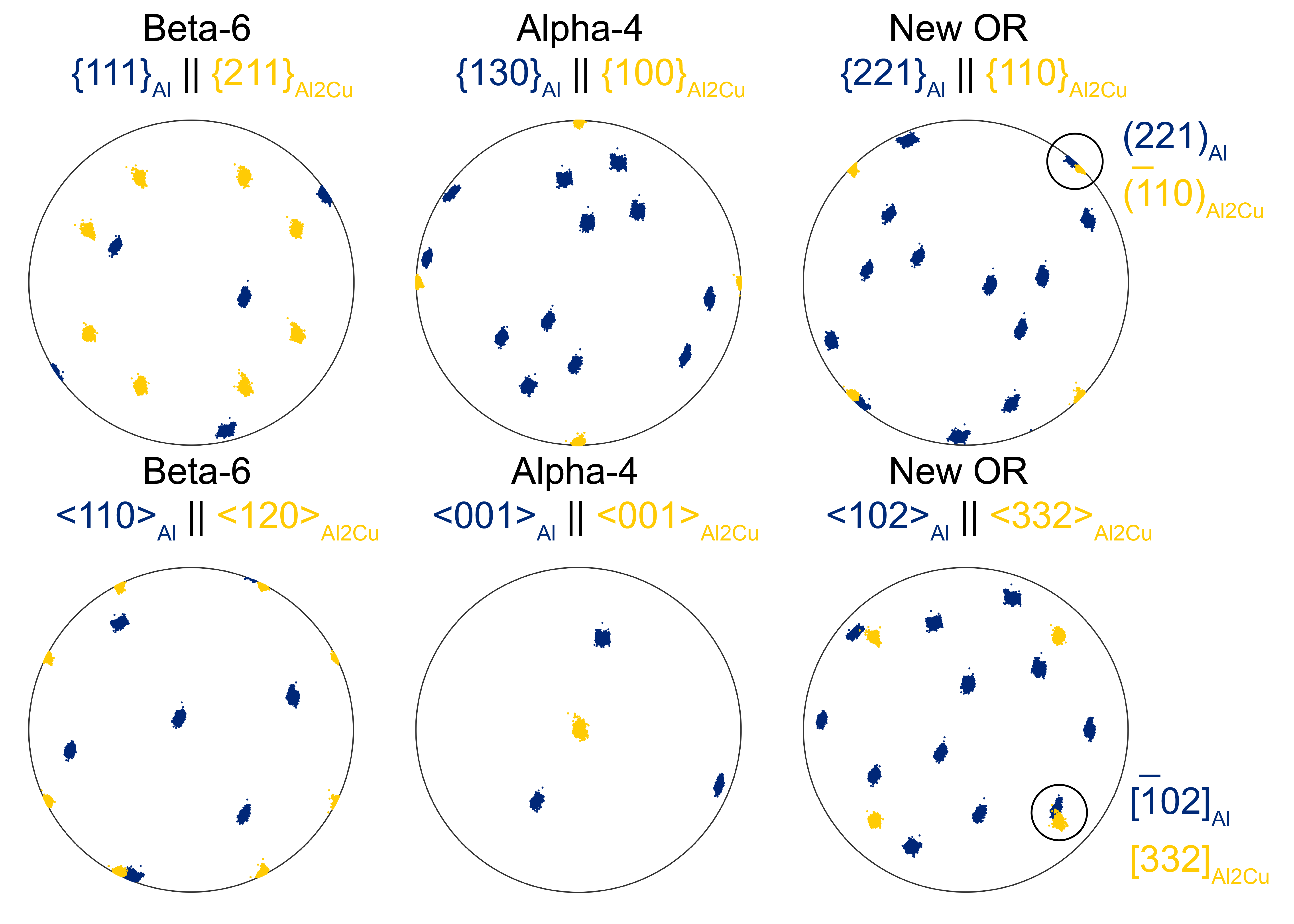}
\caption{\textit{Orientation relationships between eutectic phases}. Al and Al\textsubscript{2}Cu crystals do not match with the ``Beta\mbox{-}6" nor ``Alpha\mbox{-}4" ORs but instead show high misorientations of 17.95\textdegree and 5.42\textdegree, and 49.03\textdegree and 11.69\textdegree, for the plane and direction, respectively. See the pole figures in the left and center column, respectively. We determined a lower misorientation of 3.34\textdegree~between $(211)_{\textrm{Al}}$ // $(\bar{1}10)_{\textrm{Al\textsubscript{2}Cu}}$ and 6.88\textdegree~between $[\bar{1}02]_{\textrm{Al}} // [332]_{\textrm{Al\textsubscript{2}Cu}}$. See pole figures in the right column. All are centered along $(001)_{\textrm{Al\textsubscript{2}Cu}}$. Poles of Al are shown in blue and those of Al\textsubscript{2}Cu are in yellow.}
\label{F8}
\end{figure}

\begin{figure}[h!]
\includegraphics[width=\textwidth]{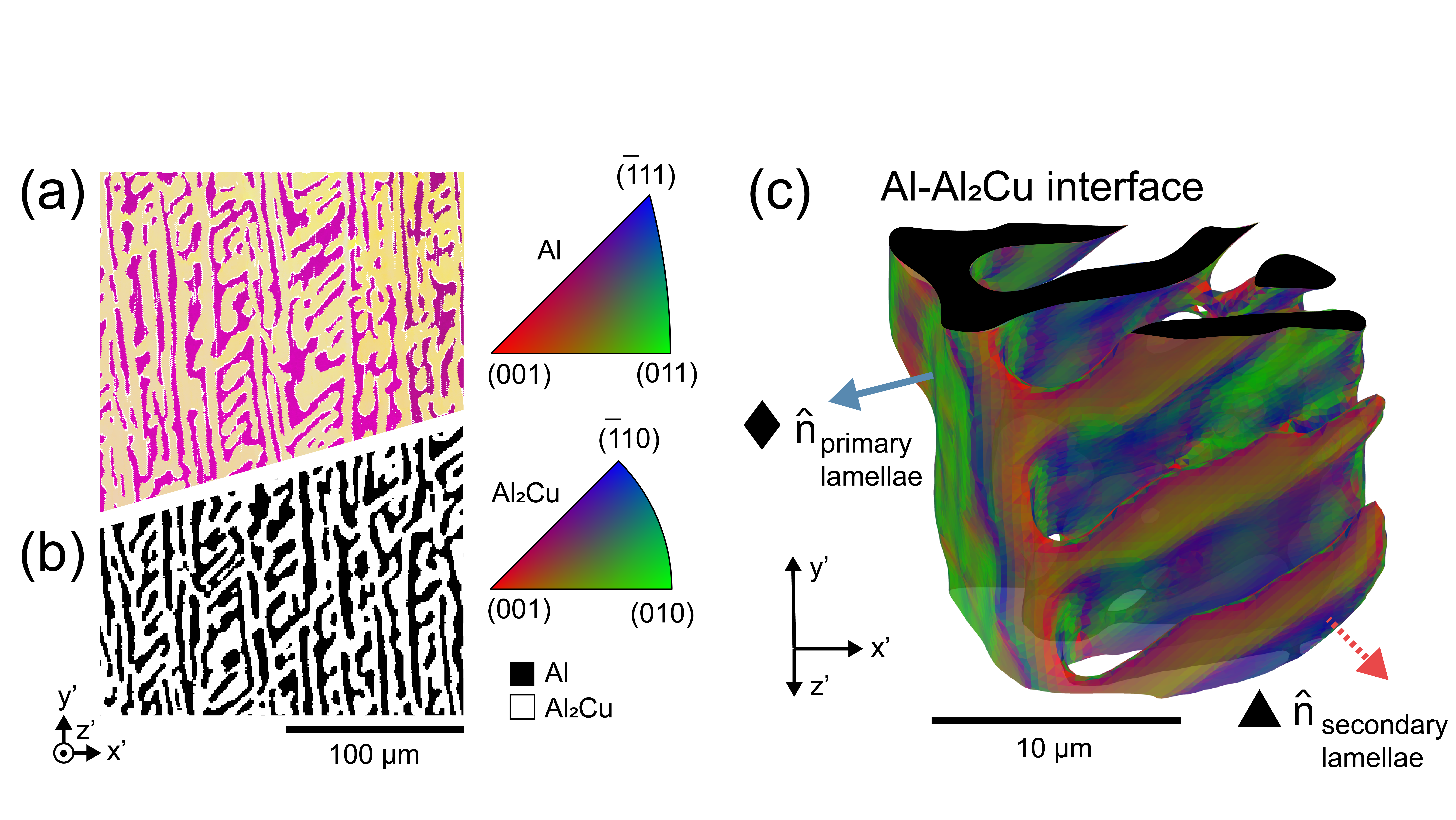}
\caption{\textit{Conversion of specimen to crystallographic frame}. (a) Crystallographic orientations were determined with EBSD. The data indicates that the eutectic is in the form of two interpenetrating single crystals. Pink orientations correspond to $(\bar{6}57)$ of Al\textsubscript{2}Cu while the yellow orientations are $(\bar{1}37)$ of Al. See standard triangles. (b) The segmented tomographic volume is aligned to the same EBSD frame of reference ($x$', $y$', $z$') with Al in white and Al\textsubscript{2}Cu in black. When combined together in (c), the normal vectors along the solid\mbox{-}solid interfaces in the X\mbox{-}ray tomography volume can be rotated into crystallographic frame. Interfaces in (c) are colored with respect to the Al phase, see standard triangle. We identify two predominant lamellar orientations, which we denote as ``primary" and ``secondary" during analysis. }
\label{F9}
\end{figure}

\begin{figure}[h!]
\includegraphics[width=\textwidth]{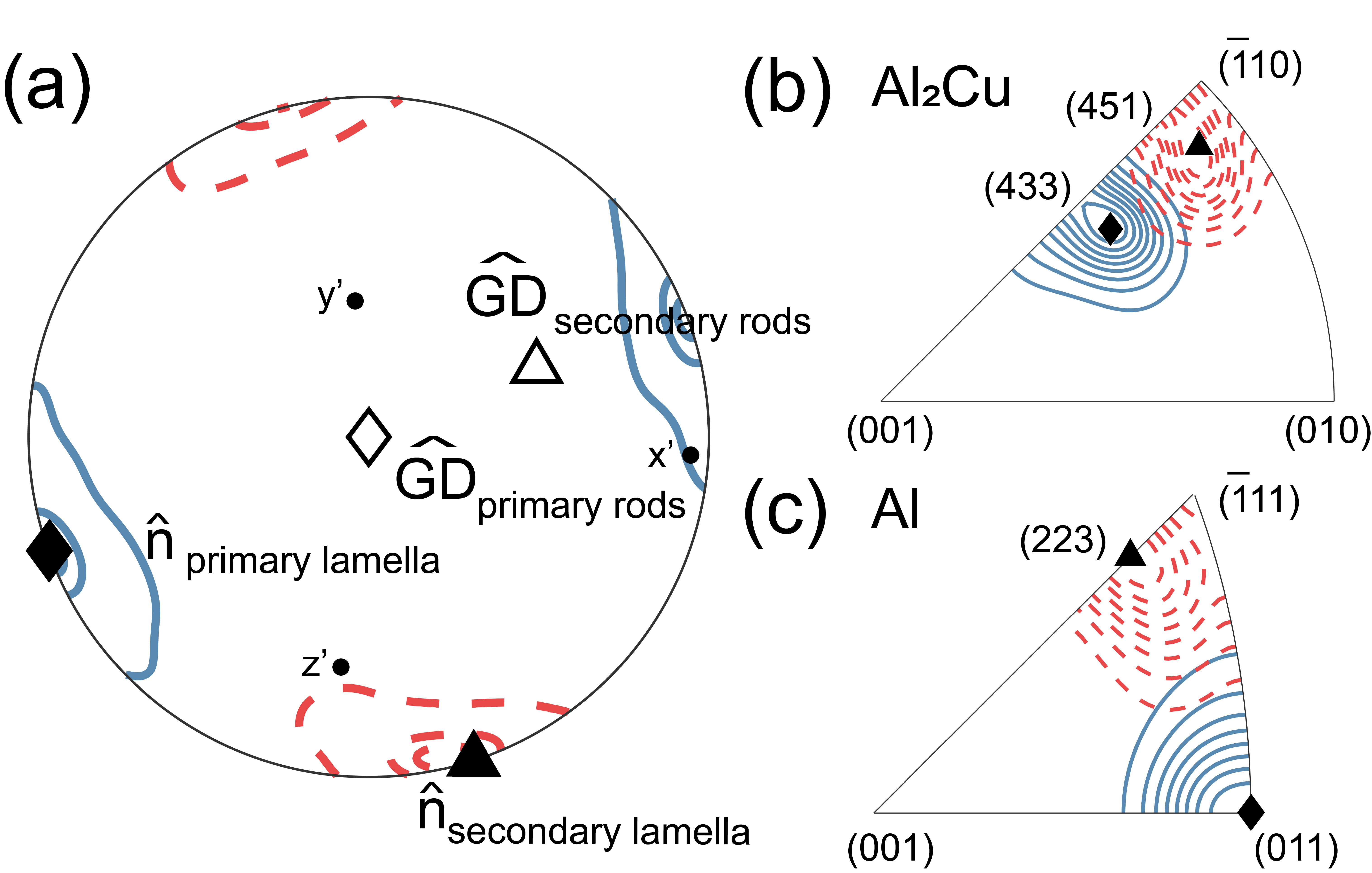}
\caption{\textit{Orientations of solid\mbox{-}solid interfaces}. (a) Stereographic projection of normal vectors computed from tomographic dataset. Zone axis is cross product between primary and secondary lamellar orientations (this coincides with the growth direction (GD) of the primary rods). The two lamellae are nearly orthogonal (\textit{i.e.}, 80 $\pm$ 2\textdegree~ apart on great circle). Specimen coordinates $x$', $y$', and $z$' shown for perspective. GD of secondary rods is 30\textdegree away from that of primary rod and 53\textdegree from the primary lamellar normal. (b\mbox{-}c) IPFs showing the distribution of the primary lamellar normal vectors (solid blue lines) is centered around (011) Al and (433) Al\textsubscript{2}Cu while that of the secondary lamellae (dotted red lines) is centered around (223) Al and $(\bar{4}51)$ Al\textsubscript{2}Cu. In all panels, blue contours correspond to primary directions and dashed red contours the secondary directions, retrieved from our mapping of the X\mbox{-}ray tomography data to the crystal frames.}
\label{F10}
\end{figure}

\begin{figure}[h!]
\includegraphics[width=\textwidth]{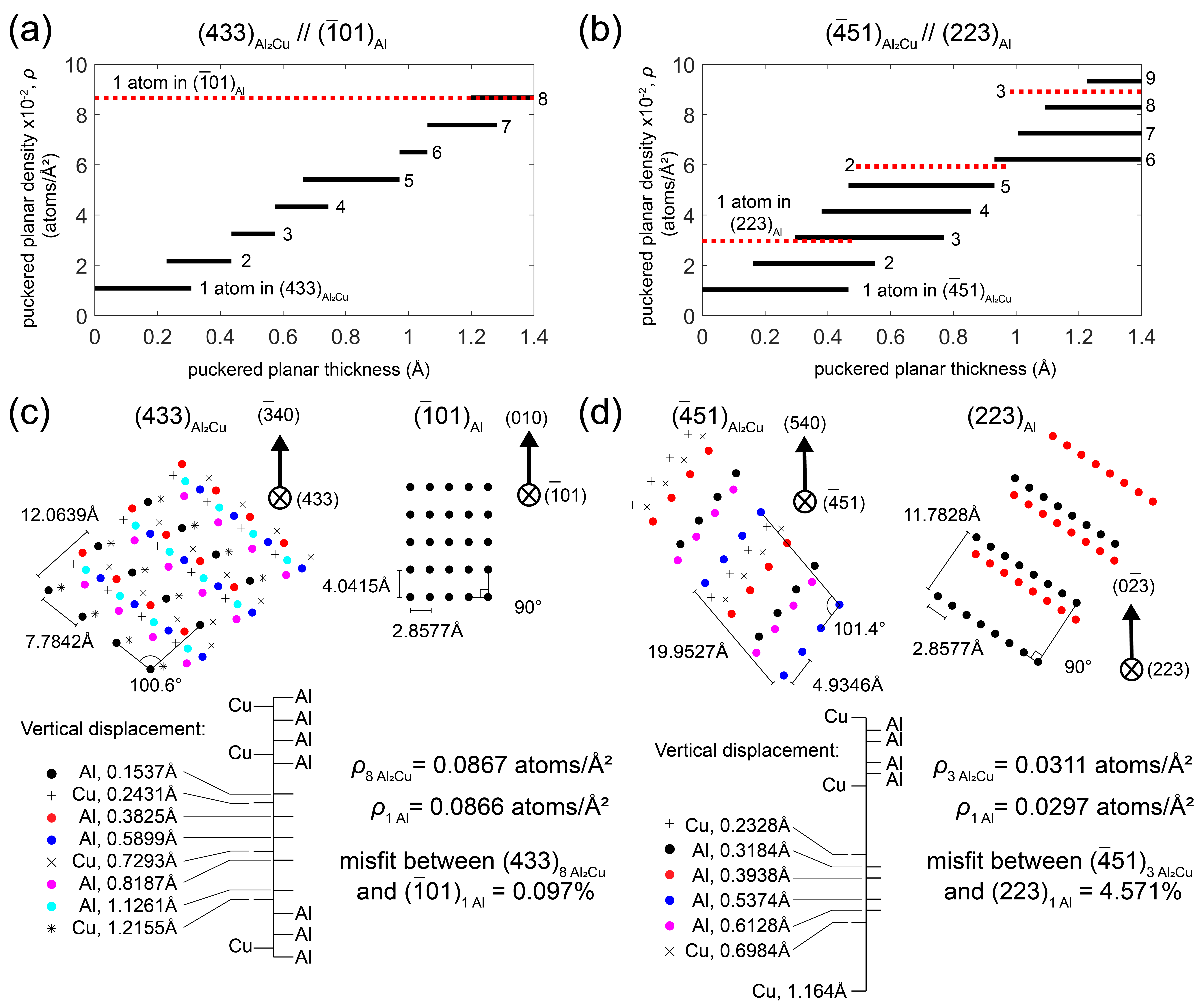}
\caption{\textit{Misfit of solid\mbox{-}solid interfaces}. Panels (a\mbox{-}b) show the effect of an interfacial thickness on the planar density for primary and secondary lamellae, respectively. If we consider the planar density of the interphase planes to contain atoms within a certain atomic layer thickness, there is a minimum misfit for the primary interphase boundary of 0.01\% for 8 atoms/unit area of Al\textsubscript{2}Cu (black line) and 1 atom/unit area of Al (dashed red line). We find also a minimum misfit between the secondary interphase boundary of 4.57\% for 3 atoms/unit area of Al\textsubscript{2}Cu and 1 atom/unit area of Al. (c\mbox{-}d) show projections of these two interfaces, accounting for puckered planes that minimize misfit between Al and Al\textsubscript{2}Cu. Here, the misfit is defined as the relative difference in atomic planar density $\rho$ of the two solid phases.}
\label{F11}
\end{figure}

\clearpage
\newpage

\section*{Supplemental Information}

\renewcommand{\thefigure}{S\arabic{figure}}
\setcounter{figure}{0}

\renewcommand{\thetable}{S\arabic{table}}
\setcounter{table}{0}

\begin{table}[ht]
\captionsetup{font=normalsize,format=hang}
\centering
\caption{\textit{Alloy composition}, retrieved from inductively coupled plasma mass spectroscopy (ICP\mbox{-}MS). Measurement obtained from as\mbox{-}cast ingot, \textit{i.e.}, prior to directional solidification.}
\begin{tabular}[t]{cc}
Element&(wt \%)\\
\hline
Cr&$<$0.01\\
Cu&33.7\\
Fe&$<$0.01\\
Mg&$<$0.01\\
Mn&$<$0.01\\
Ni&$<$0.01\\
Pb&$<$0.01\\
Si&0.02\\
Sn&$<$0.01\\
Ti&$<$0.01\\
Zn&0.01\\
Al&Remainder\\
\end{tabular}
\label{tableSI}
\end{table}%

\noindent
\renewcommand{\theVideo}{V\arabic{Video}}
\setcounter{Video}{0}

\begin{Video}
\captionsetup{font=normalsize,justification=raggedright,format=hang}
\caption{\textit{X-radiography images}, combined as a video. Playback speed is equal to acquisition speed (5 frames per second, fps). }
\label{V_S1}  
\end{Video} 

\begin{Video}
\captionsetup{font=normalsize,justification=raggedright,format=hang,singlelinecheck=false}
\caption{\textit{Processed X\mbox{-}radiography images}, combined as a video. Playback speed (5fps).                   }
\label{V_S2}   
\end{Video}

\begin{Video}
\captionsetup{font=normalsize,justification=raggedright,format=hang}
\caption{\textit{Full bird’s\mbox{-}eye\mbox{-}view of the solid liquid interface}, corresponding to the (\textit{x-z}) plane, see \textbf{Fig.~\ref{F4}(b)} for all timesteps analyzed in \textbf{Fig.~\ref{F5}}. The dimensions are \SI{22} {\micro\meter} \texttimes \SI{1312} {\micro\meter}. }
\label{V_S3}   
\end{Video}

\begin{Video}
\captionsetup{font=normalsize,justification=raggedright,format=hang}
\caption{\textit{3D render of the eutectic colony} in \textbf{Fig.~\ref{F7}(b)} showing the rod-to-lamella transition. Al\textsubscript{2}Cu in black while Al is colored by velocity, see \textbf{Fig.~\ref{F7}} for color\mbox{-}bar and scale\mbox{-}bar). }
\label{V_S4}   
\end{Video}

\begin{Video}
\captionsetup{font=normalsize,justification=raggedright,format=hang}
\caption{\textit{3D render of the coplanar rod\mbox{-}to\mbox{-}lamella transition}, showing the single Al lamella in \textbf{Fig.~\ref{F7}(c)} (Al phase colored by velocity while Al\textsubscript{2}Cu is transparent, see \textbf{Fig.~\ref{F7}} for color\mbox{-}bar and scale\mbox{-}bar). }
\label{V_S5}   
\end{Video}

\begin{figure}[h!]
\includegraphics[width=\textwidth]{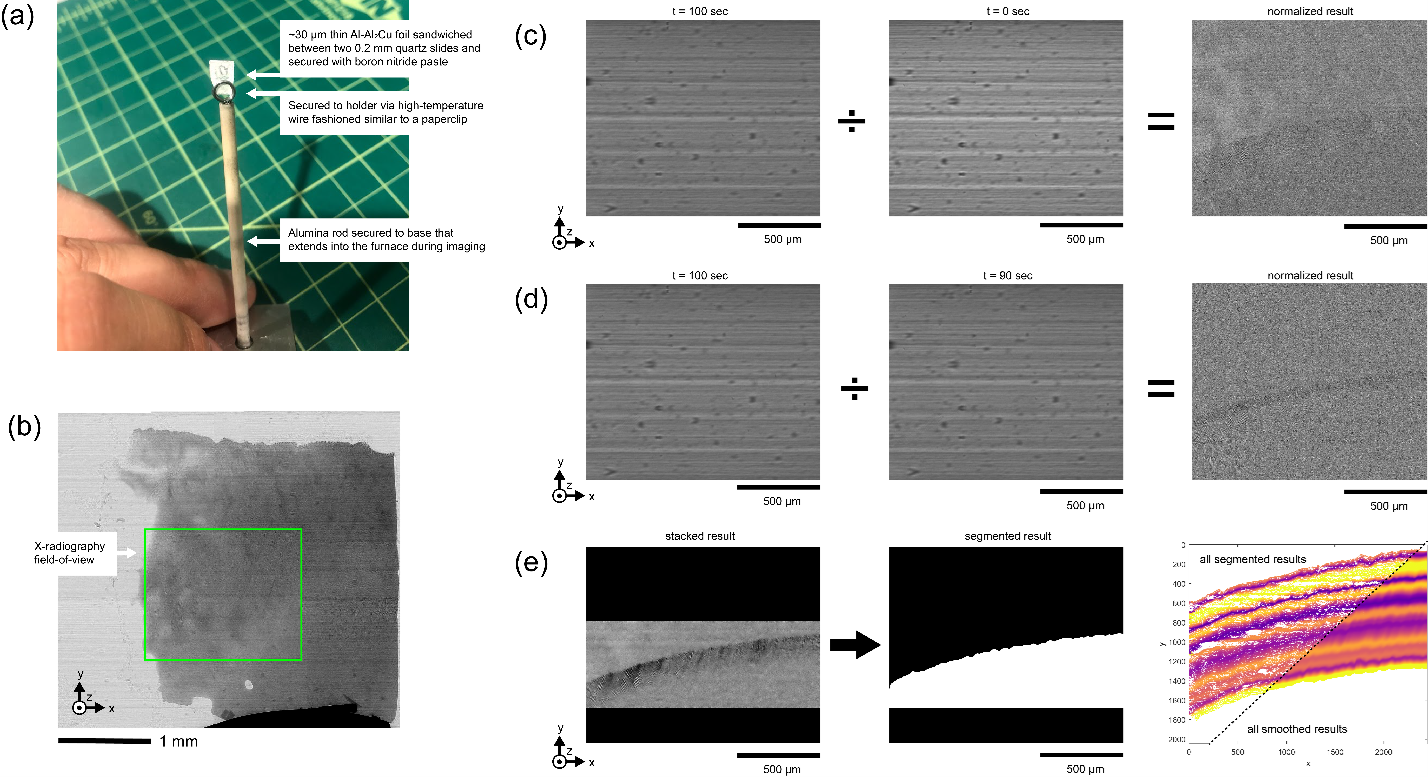}
\caption{\textit{Sample and data processing workflow for X\mbox{-}radiography.} Panel (a) shows a photo of the encased foil sample attached to the holder. The $\sim$30 \SI{} {\micro\meter} thin Al\mbox{-}Al\textsubscript{2}Cu foil was sandwiched between two 0.2 mm quartz slides and secured with boron nitride paste. This assembly was secured to a makeshift holder made of high\mbox{-}temperature wire fashioned similarly to a paperclip. This holder was secured to an alumina holder that was attached to the base of the kinematic mount at the beamline. This set up ensured we could insert the sample into the furnace, position it in the path of the beam, and acquire the \textit{ex situ} tomogram. Panel (b) shows a montage of the sample and the X\mbox{-}radiography field\mbox{-}of\mbox{-}view is boxed in green for a sense of scale. Panel (c) shows the normalization procedure whereby a specific frame (taken at $t$ = 100 s) is normalized by an image with 100\% liquid in the  field\mbox{-}of\mbox{-}view ($t$ = 0 s), in order to enable identification of the eutectic and liquid phases after removing detector artifacts. Panel (d) shows the normalization procedure whereby a specific frame ($t$ = 100 s) is normalized by a previous frame ($t$ = 90 s) to better enhance the eutectic\mbox{-}liquid interface. Although contrast is much improved, segmentation did not yield promising results until a series of normalized frames ($t$ = 100s/99s + 100s/98s + … + 100s/62s + 100s/63s) were cropped, stacked together, then filtered with a non\mbox{-}local means filter (search window size of 41). The result is shown in panel (e), left. Following segmentation \textit{via} the commercially available semantic segmentation (machine learning) routine (Zeiss Zen Blue, Intellesis module), the result is shown in panel (e), center. A spatiotemporal diagram of the eutectic\mbox{-}liquid interfaces extracted by this procedure are shown in panel (e), right; the spatially smoothed traces are depicted at far right. }
\label{SI_1}
\end{figure}

\begin{figure}[h!]
\includegraphics[width=\textwidth]{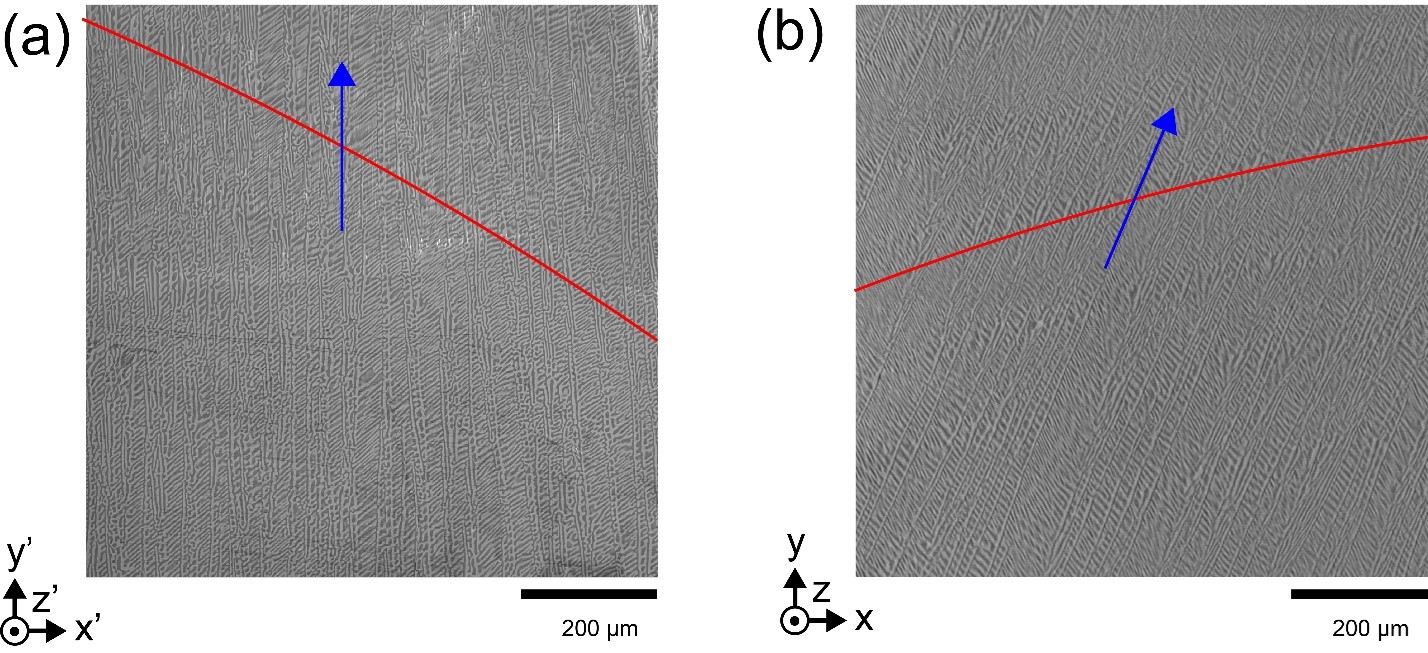}
\caption{\textit{Alignment of the tomography frame\mbox{-}of\mbox{-}reference to the EBSD frame\mbox{-}of\mbox{-}reference.} Panel (a) is the Scanning Electron Microscope (SEM) image of the Al\mbox{-}Al\textsubscript{2}Cu foil after the \textit{in situ} experiment where the EBSD scan was obtained. Here the EBSD frame of reference is $x'$, $y'$ and $z'$. Panel (b) is a select cross\mbox{-}sectional slice of the Al\mbox{-}Al\textsubscript{2}Cu foil from the reconstructed tomogram. Here the X\mbox{-}ray frame of reference is $x$, $y$ and $z$. The red line in both indicate the band of region of lower velocity (coarser eutectic) and the blue arrow in both indicate the direction of the primary lamella traces. Both annotations are added to guide the reader’s eye. We align the X\mbox{-}ray data to the EBSD frame\mbox{-}of\mbox{-}reference following an affine transformation (specifically, a rotation 22\textdegree~counterclockwise about $z // z’$, then a mirror operation across $y’$). This operation brings the two frame\mbox{-}of\mbox{-}reference into coincidence.}
\label{SI_2}
\end{figure}

\begin{figure}[h!]
\begin{center}
\includegraphics[width=0.5\textwidth]{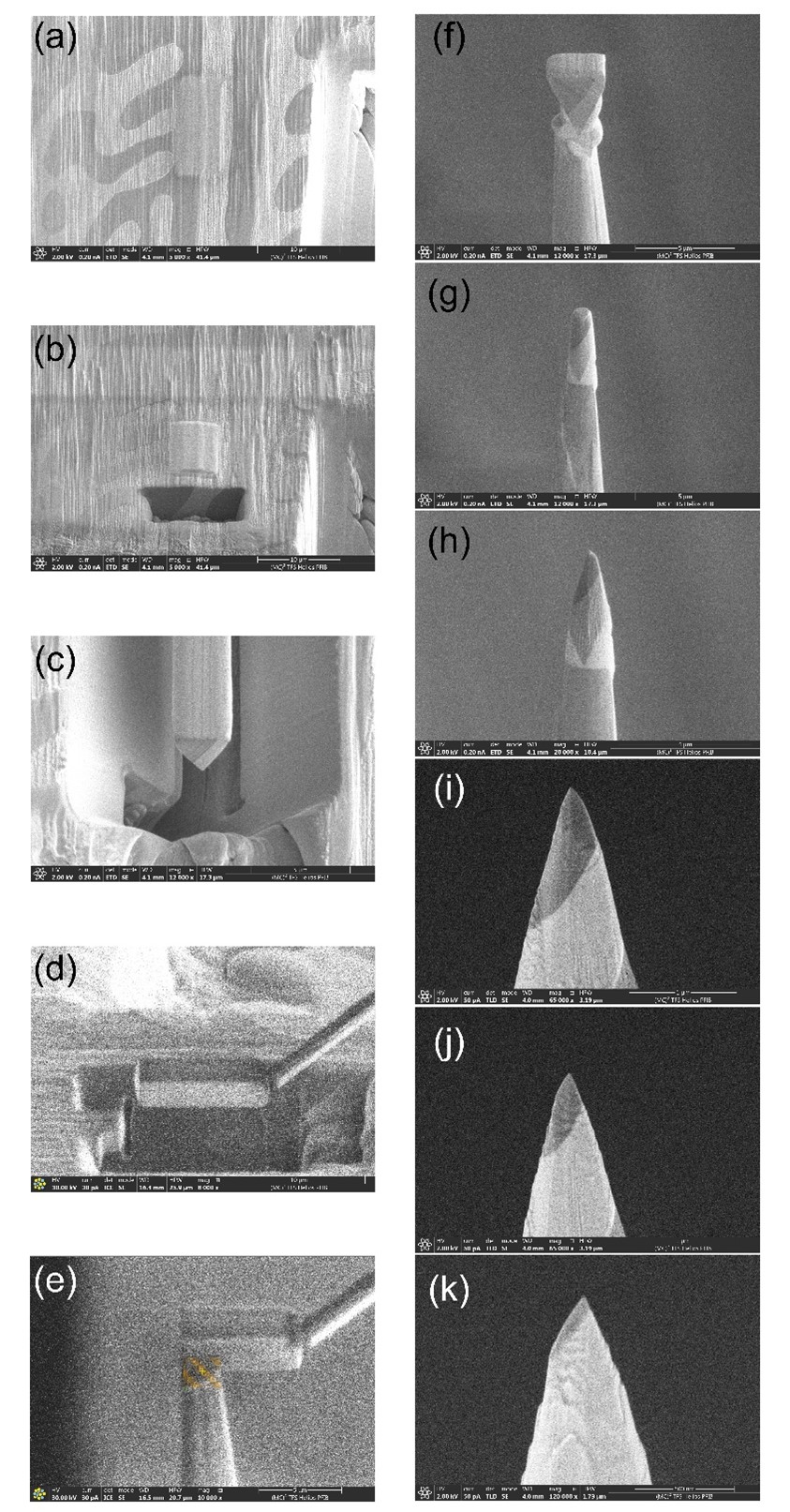}
\end{center}
\caption{\textit{Sample tip preparation for atom probe tomography (APT).} The sample tips were excavated from the volume of material with the Thermo Fisher Helios G4 PFIB UXe. The heterophase Al\mbox{-}Al\textsubscript{2}Cu boundary cuts through this volume. The deposited Pt/W cap in panel (a) shows the targeted region. Cross sectional views in panels (b\mbox{-}c) show the heterophase boundary is completely captured by the lift\mbox{-}out. Panel (d) shows snapshots of the lift\mbox{-}out from the excavation region and in panel (e), its attachment to a Mo transfer post. Panels (f\mbox{-}k) shows snapshots during the annular pattern thinning procedure to achieve the sharp tip geometry necessary for APT.}
\label{SI_3}
\end{figure}

\begin{figure}[h!]
\includegraphics[width=\textwidth]{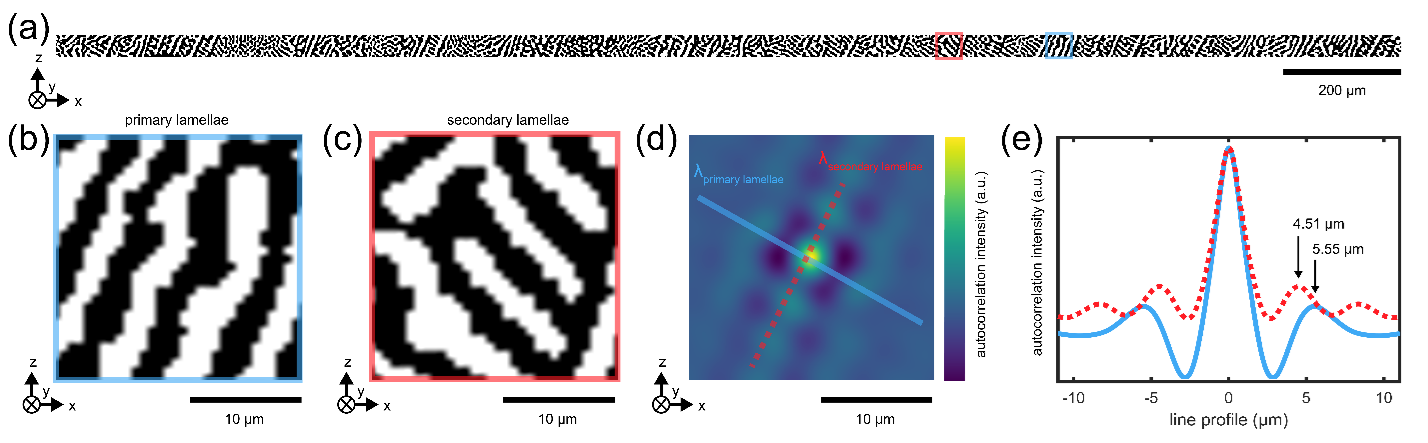}
\caption{\textit{Procedure for determining the average eutectic spacing} by autocorrelation (also termed as a two\mbox{-}point correlation, self\mbox{-}convolution, or Patterson function). We generate the real space periodicity within the eutectic\mbox{-}liquid interface by multiplying the 2D fast Fourier transform (FFT) with its conjugate and then taking the inverse Fourier transform of the product. Panel (a) is the segmented, transverse view of the eutectic\mbox{-}liquid interface. The lamellae we define as ``primary” and ``secondary” are boxed in blue and red in (b) and (c), respectively. We generate in (d) the autocorrelation function from which we can trace in (e) line profiles to obtain the lamellar spacing for both the primary and secondary lamellae, here \SI{5.55} {\micro\meter} and \SI{4.51} {\micro\meter}, respectively.}
\label{SI_4}
\end{figure}

\begin{figure}[h!]
\includegraphics[width=\textwidth]{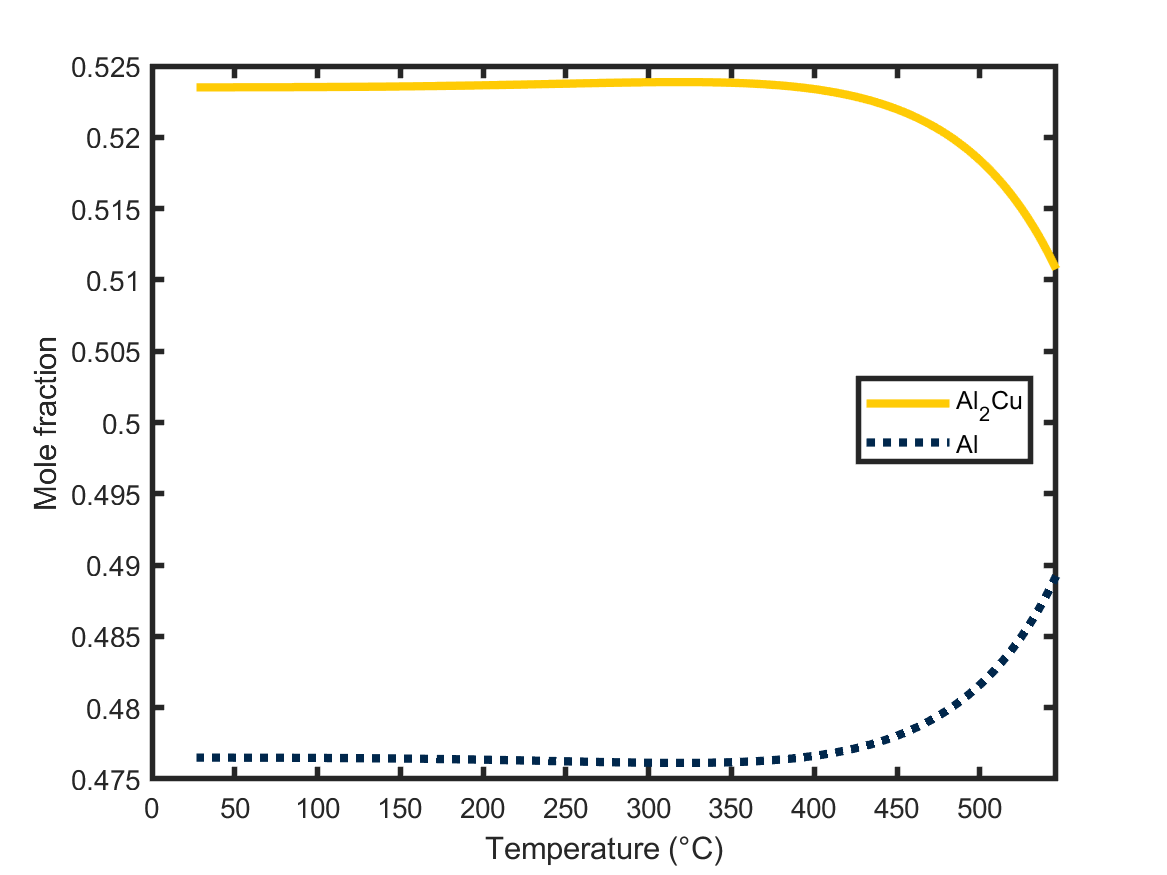}
\caption{\textit{Equilibrium concentration changes during cooling.} We use ThermoCalc to estimate the equilibrium concentration of each solid eutectic phase during cooling and predict a 2.4\% increase in Al\textsubscript{2}Cu concentration during cooling to substantiate the observed concentration gradient of Cu at the heterophase boundary captured by APT.}
\label{SI_5}
\end{figure}

\begin{figure}[h!]
\includegraphics[width=\textwidth]{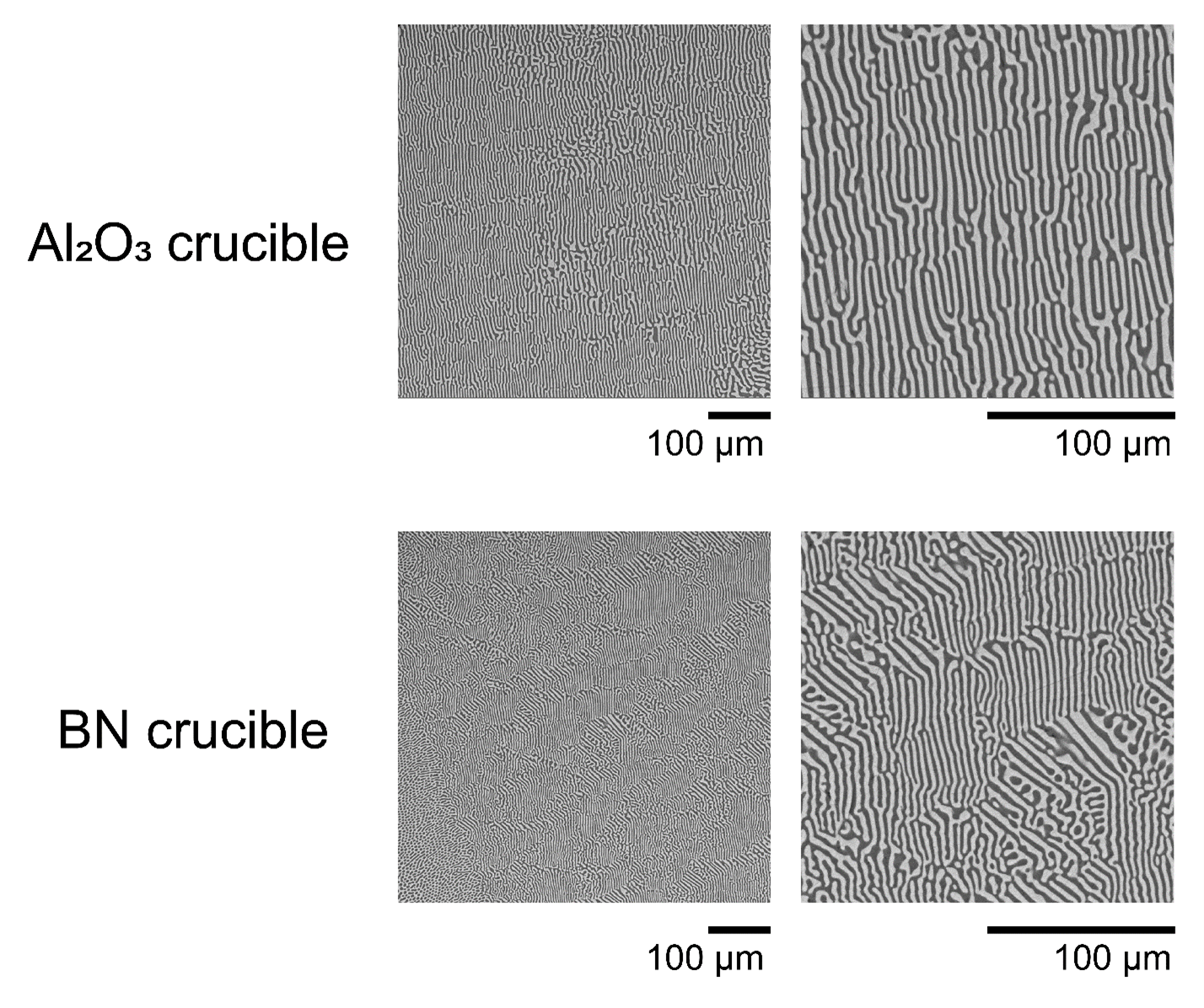}
\caption{\textit{Influence of trace Boron on eutectic microstructure.} We solidified two otherwise identical Al\mbox{-}33wt\%Cu samples of eutectic composition, the first using an Al\textsubscript{2}O\textsubscript{3} crucible (top row) and the second using a BN crucible (the source of B, bottom row). The transverse cross sections taken from the same distance (3 cm) along the solidification direction show different microstructures, which supports the idea of an underlying chemical effect. Specifically, there is a dominance of a single, primary lamella orientation in the sample solidified with Al\textsubscript{2}O\textsubscript{3} while colonies and secondary lamella orientations appear with BN crucible. The latter resembles the cross\mbox{-}sections in the pseudo\mbox{-}4D dataset, \textit{cf.} \textbf{Figs.~\ref{F4} and \ref{F7}}. We collected these micrographs with a scanning electron microscope imaged using back-scatter electrons to show phase contrast: the bright phase is Al.}
\label{SI_6}
\end{figure}

\begin{figure}[h!]
\includegraphics[width=\textwidth]{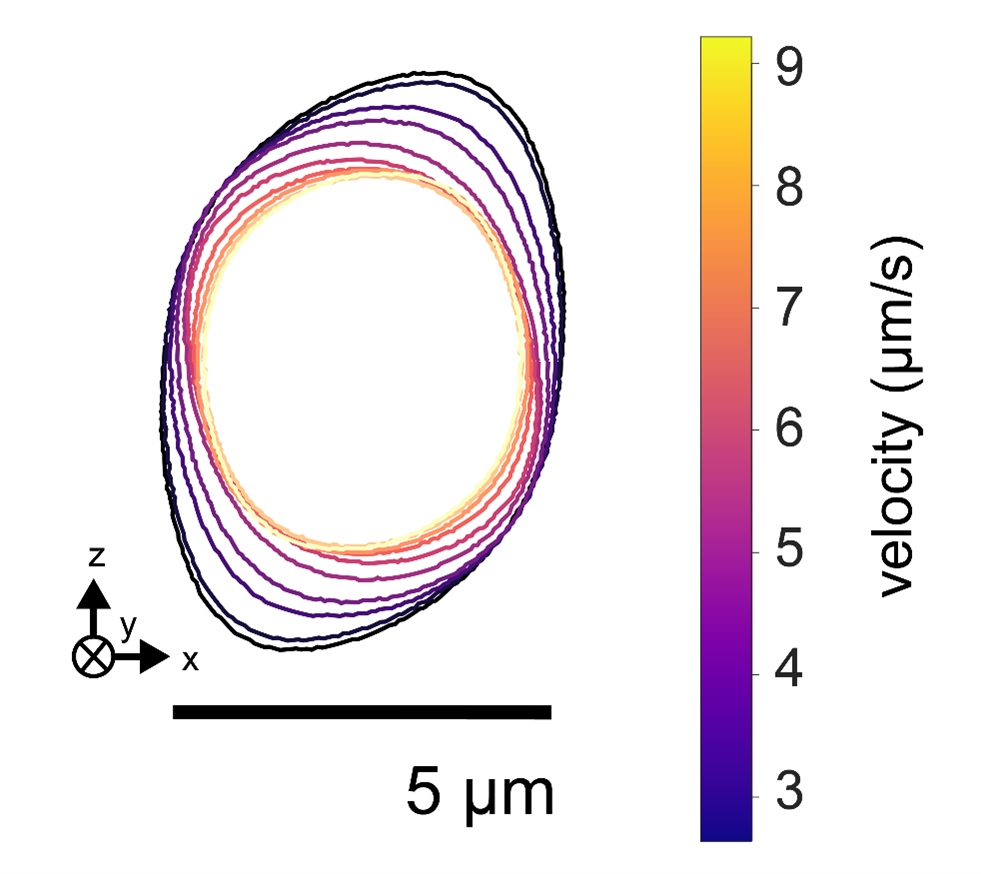}
\caption{\textit{Anisotropy of Al rods.} We can evaluate the morphology of the coplanar Al rods in detail by segmenting the autocorrelation image in \textbf{Fig.~\ref{SI_4}(d)}. Note these rods have eccentricity ~0.7\mbox{-}0.9 (corresponding to aspect ratios from ~1\mbox{-}1.5) and are not circular. Instead, they show an elongation along the primary lamella direction (see again \textbf{Fig.~\ref{SI_4}}), which hints at a resistance to morphological instability due to a low\mbox{-}energy solid\mbox{-}solid boundary.}
\label{SI_7}
\end{figure}

\begin{figure}[h!]
\includegraphics[width=\textwidth]{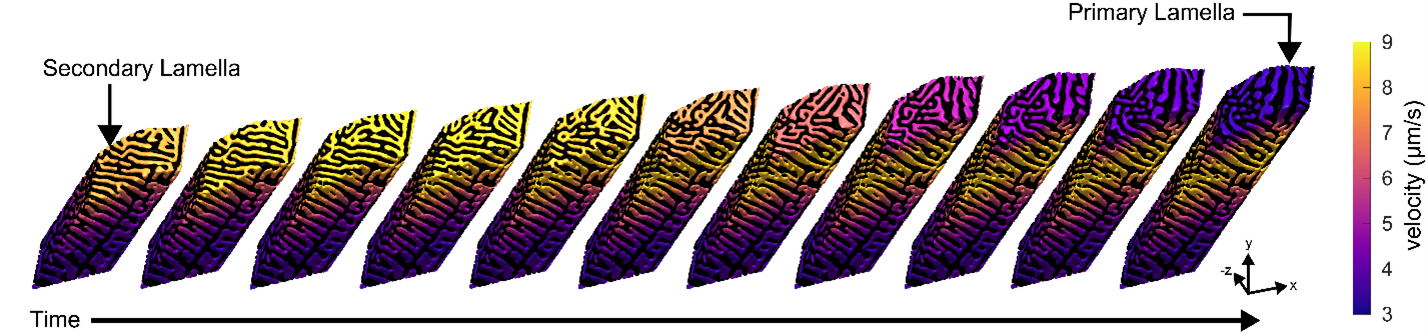}
\caption{\textit{Lamella-to-lamella transition.} Sequence of time\mbox{-}resolved, 3D renders depicted 1 s apart. The data demonstrate that primary lamellae (orthogonal to sample surfaces) predominate at low velocities, corresponding to a greater impurity concentration in the boundary layer and presumably also a greater B segregation to the lamellar interfaces. At high velocities, secondary lamellae take over. See arrows for examples of each type of lamella. Al phase is colored according to eutectic\mbox{-}liquid interfacial velocity (see color\mbox{-}bar) while Al\textsubscript{2}Cu is black.}
\label{SI_8}
\end{figure}

\end{document}